\documentclass[10pt,aps,preprintnumbers,prd,noshowpacs,nofootinbib,noshowkeys,floatfix,superscriptaddress]{revtex4}
\usepackage[dvips]{graphics,graphicx}
\usepackage[colorlinks=true,linktocpage=true,linkcolor=blue,citecolor=blue]{hyperref}
\usepackage[usenames,dvipsnames]{color}
\usepackage{amsmath, amssymb,oldgerm}
\usepackage{multirow}
\usepackage{xcolor}
\usepackage[normalem]{ulem}  
\usepackage{braket}
\usepackage{slashed}
\usepackage[mathscr]{eucal}
\usepackage{cancel}
\usepackage{tikz}                   
\usepackage{pgfplots}				

\definecolor{goethe-blau}{cmyk}{1.0,0.2,0.0,0.4}
\definecolor{hellgrau}{cmyk}{0.04,0.04,0.05,0.02}
\definecolor{sandgrau}{cmyk}{0.12,0.09,0.13,0.0}
\definecolor{dunkelgrau}{cmyk}{0.25,0.25,0.30,0.75}
\definecolor{emo-rot}{cmyk}{0.04,1.0,0.8,0.07}
\definecolor{purple}{cmyk}{0.08,1.0,0.3,0.36}
\definecolor{senfgelb}{cmyk}{0.01,0.25,1.0,0.05}
\definecolor{gruen}{cmyk}{0.62,0.4,0.87,0.09}
\definecolor{magenta}{cmyk}{0.08,0.86,0.12,0.12}
\definecolor{orange}{cmyk}{0.0,0.7,1.0,0.04}
\definecolor{sonnengelb}{cmyk}{0.0,0.12,0.95,0.0}
\definecolor{helles-gruen}{cmyk}{0.4,0.17,0.81,0.07}
\definecolor{lichtblau}{cmyk}{0.8,0.0,0.06,0.04}

\newcommand{\Tr}{\text{Tr}}
\newcommand{\n}{\nonumber}

\newcommand{\mC}{\mathfrak{C}}

\newcommand{\ms}{\mathfrak{s}}

\newcommand{\beq}{\begin{equation}}
\newcommand{\eeq}{\end{equation}}

\newcommand{\eqs}{Eqs.~}
\newcommand{\eq}{Eq.~}

\newcommand{\inranglesm}{\rangle_\text{in}}

\newcommand{\outlanglesm}{{}_\text{out}\langle}

\newcommand{\taum}{\tau}
\newcommand{\proj}{\Delta}

\newcommand{\mn}{\mathfrak{n}}
\newcommand{\mh}{\mathfrak{h}}
\newcommand{\map}{\mathfrak{p}}

\newcommand{\tc}{\mathfrak{K}}
\newcommand{\mft}{\mathfrak{p}}
\newcommand{\mfu}{\mathfrak{n}}
\newcommand{\mfv}{\mathfrak{z}}
\newcommand{\mfw}{\mathfrak{q}}
\newcommand{\tct}{\mathfrak{g}}
\newcommand{\lmur}{{\langle\mu\rangle}}
\newcommand{\ns}{\text{NS}}
\newcommand{\nr}{\text{nr}}

\newcommand{\mfp}{\text{mfp}}
\newcommand{\hydro}{\text{hydro}}
\newcommand{\brm}{-}

\newcommand{\bbb}{\textcolor{black}}

\begin{document}

\title{Relativistic second-order dissipative spin hydrodynamics from the method of moments}
\author{Nora Weickgenannt}

\affiliation{Institute for Theoretical Physics, Goethe University,
Max-von-Laue-Str.\ 1, D-60438 Frankfurt am Main, Germany}

\author{David Wagner}

\affiliation{Institute for Theoretical Physics, Goethe University,
Max-von-Laue-Str.\ 1, D-60438 Frankfurt am Main, Germany}

\author{Enrico Speranza}

\affiliation{Illinois Center for Advanced Studies of the Universe and Department of Physics, University of Illinois at Urbana-Champaign, Urbana, IL 61801, USA}

\author{Dirk H.\ Rischke}

\affiliation{Institute for Theoretical Physics, Goethe University,
Max-von-Laue-Str.\ 1, D-60438 Frankfurt am Main, Germany}
\affiliation{Helmholtz Research Academy Hesse for FAIR, Campus Riedberg, Max-von-Laue-Str.\ 12, D-60438 Frankfurt am Main, Germany}

\begin{abstract}
We derive relativistic second-order dissipative fluid-dynamical equations of motion for 
massive spin-1/2 particles from kinetic theory using the method of moments. Besides the 
usual conservation laws for charge, energy, and momentum, such a theory of relativistic 
dissipative spin hydrodynamics features an equation of motion for the rank-3 spin tensor, 
which follows from the conservation of total angular momentum. Extending the 
conventional method of moments for spin-0 particles, we expand the spin-dependent 
distribution 
function near local equilibrium in terms of moments of the momentum and spin variables. 
We work to next-to-leading order in the Planck constant $\hbar$. 
As shown in previous work, at this order in $\hbar$ the Boltzmann equation for spin-1/2 
particles features a nonlocal collision term. From the Boltzmann equation, we then obtain an 
infinite set of equations of motion for the irreducible moments of the deviation of the 
single-particle distribution function from local equilibrium. In order to close this system of 
moment equations, a truncation procedure is needed. We employ the ``14+24-moment 
approximation", where ``14" corresponds to the components of the charge current and the 
energy-momentum tensor and ``24" to the components of the spin tensor, which completes 
the derivation of the equations of motion of second-order dissipative spin hydrodynamics. 
For applications to heavy-ion phenomenology, we also determine dissipative corrections to 
the Pauli-Lubanski vector.
\end{abstract}



\maketitle

\section{Introduction}

The derivation of a theory of relativistic hydrodynamics when spin degrees of freedom are 
dynamical variables coupled to the fluid, often referred to as ``relativistic spin hydrodynamics", has 
recently attracted a lot of attention~\cite{Florkowski:2017ruc,Florkowski:2017dyn,Hidaka:2017auj,Florkowski:2018myy,Weickgenannt:2019dks,Bhadury:2020puc,Weickgenannt:2020aaf,Shi:2020htn,Speranza:2020ilk,Bhadury:2020cop,Singh:2020rht,Bhadury:2021oat,Peng:2021ago,Sheng:2021kfc,Sheng:2022ssd,Hu:2021pwh,Hu:2022lpi,Fang:2022ttm,Wang:2022yli,Montenegro:2018bcf,Montenegro:2020paq,Gallegos:2021bzp,Hattori:2019lfp,Fukushima:2020ucl,Li:2020eon,She:2021lhe,Wang:2021ngp,Wang:2021wqq,Hongo:2021ona}. One of the main motivations to 
develop such a theory comes from the physics of the quark-gluon plasma (QGP) created in 
nuclear collisions. In this case, the vorticity of the hot and dense matter triggers hadron spin 
polarization in the final state \cite{Liang:2004ph,Voloshin:2004ha,Betz:2007kg,Becattini:2007sr}. 
This mechanism resembles the time-honored Barnett effect \cite{Barnett:1935}, which shows the 
interplay between a classical property of the system, the rotation, with the spin, which is a 
quantum property of matter. Experimental evidence of these phenomena comes from the analysis 
carried out in Refs.\ \cite{STAR:2017ckg,Adam:2018ivw,ALICE:2019aid,Mohanty:2021vbt}, where it was shown 
that hadrons emitted in noncentral nuclear collisions are indeed spin-polarized.
Theoretical models have successfully managed to describe the global-polarization data (i.e., the 
polarization along the direction of angular momentum of the collision) 
\cite{Becattini:2007sr,Becattini:2013vja,Becattini:2013fla,Becattini:2015ska,Becattini:2016gvu,Karpenko:2016jyx,Pang:2016igs,Xie:2017upb}. 
However, the explanation of the longitudinal-polarization data (i.e., the polarization along the 
beam direction) is still an open question \cite{Becattini:2017gcx,Becattini:2020ngo,Florkowski:2019qdp,Florkowski:2019voj,Zhang:2019xya,Becattini:2019ntv,Xia:2019fjf,Wu:2019eyi,Sun:2018bjl,Liu:2019krs,Florkowski:2021wvk}, see also important recent developments in Refs.\ \cite{Liu:2021uhn,Fu:2021pok,Becattini:2021suc,Becattini:2021iol}. 
Since the spacetime evolution of the QGP is very accurately described by relativistic 
hydrodynamics \cite{Heinz:2013th,Florkowski:2017olj}, it is natural to extend conventional 
relativistic hydrodynamics to incorporate the dynamics of spin. This novel theory, besides being of 
fundamental interest by itself as it connects quantum properties of matter with hydrodynamics, 
may provide an important tool towards a deeper understanding of relativistic strong-interaction 
matter under extreme conditions.

The basic idea of relativistic spin hydrodynamics, as put forward in Ref.\ \cite{Florkowski:2017ruc}, is 
that, in addition to the usual hydrodynamic quantities such as the energy-momentum tensor, one 
introduces the rank-3 spin tensor and studies its evolution using additional equations of motion 
constructed from the conservation of the total angular momentum of the system. 
Over the past few years, different methods to derive relativistic spin hydrodynamics have been 
applied: kinetic theory~\cite{Florkowski:2017ruc,Florkowski:2017dyn,Hidaka:2017auj,Florkowski:2018myy,Weickgenannt:2019dks,Bhadury:2020puc,Weickgenannt:2020aaf,Shi:2020htn,Speranza:2020ilk,Bhadury:2020cop,Singh:2020rht,Bhadury:2021oat,Peng:2021ago,Sheng:2021kfc,Sheng:2022ssd,Hu:2021pwh,Hu:2022lpi,Fang:2022ttm,Wang:2022yli}, 
an effective action~\cite{Montenegro:2018bcf,Montenegro:2020paq,Gallegos:2021bzp}, 
an entropy-current analysis~\cite{Hattori:2019lfp,Fukushima:2020ucl,Li:2020eon,She:2021lhe,Wang:2021ngp,Wang:2021wqq},
holographic duality~\cite{Gallegos:2020otk,Garbiso:2020puw,Cartwright:2021qpp}, and linear-response theory 
\cite{Montenegro:2020paq,Hongo:2021ona}. Despite these formidable efforts, an agreement on how to formulate
a theory of relativistic dissipative spin hydrodynamics has not yet been reached. An important 
issue in deriving this theory is that the definitions of the energy-momentum and spin tensors are  
not unique: their form is fixed only up to so-called ``pseudo-gauge transformations", which do not 
change the global charges (i.e., the global energy, momentum, and angular 
momentum)~\cite{Hehl:1976vr,Speranza:2020ilk}. The physical implications of various choices 
of energy-momentum and spin tensors have been investigated in different works and this topic is still 
intensely debated~\cite{Becattini:2018duy,Speranza:2020ilk,Fukushima:2020ucl,Li:2020eon,Buzzegoli:2021wlg,Das:2021aar,Daher:2022xon}. 
In Ref.\ \cite{Weickgenannt:2020aaf}, it was proposed that in the Hilgevoord-Wouthuysen (HW) 
pseudo-gauge choice \cite{HILGEVOORD19631} nonlocal collisions serve as a source term in the 
equation of motion of the spin tensor, providing a physical interpretation of polarization through rotation 
in a manifestly relativistic kinetic and hydrodynamic framework.

One of the most powerful ways to derive conventional relativistic hydrodynamics is using the method of 
moments starting from the Boltzmann equation [see, e.g., Refs.\ \cite{Denicol:2012es,Denicol:2012cn} 
and refs.\ therein]. In this approach, the single-particle distribution function is expanded in momentum 
space around its local-equilibrium value in terms of a series of irreducible Lorentz tensors formed from the
particle four-momentum. In order to study deviations from equilibrium, a consistent power-counting 
scheme is needed. Usually in the context of deriving hydrodynamics from kinetic theory, such a 
power counting is constructed by comparing the mean free path $\lambda_\mathrm{mfp}$ of particle
scattering with the length scale $L_\mathrm{hydro}$ associated with gradients of the hydrodynamical 
variables, the ratio of the two being the Knudsen number 
$\mathrm{Kn} \equiv \lambda_\mathrm{mfp}/L_\mathrm{hydro}$. In spin kinetic theory, however, another 
scale, $\Delta$, enters via the nonlocal collision term 
\cite{Weickgenannt:2020aaf,Weickgenannt:2021cuo}, allowing to mutually transfer spin and orbital 
angular momentum. For a consistent power-counting scheme, it turns out that 
$\Delta/\ell_\mathrm{vort} \sim \mathrm{Kn}$, where $\ell_\mathrm{vort}$ is the length scale associated 
with the fluid vorticity. For $\Delta \ll \lambda_\mathrm{mfp}$, this means that $\ell_\mathrm{vort}$ is not of the 
order $L_\mathrm{hydro}$, like typical gradients of hydrodynamical quantities, but can
be much smaller 
[for a related discussion, see Ref.\ \cite{Li:2020eon}]. 

In this paper we extend the method of moments to include spin dynamics. 
This requires the extension of ordinary phase space by spin degrees of freedom. Here, we
choose a description in terms of a spin four-vector $\ms^\mu$, which is normalized and
orthogonal to the particle four-momentum $p^\mu$.
Starting from the quantum kinetic theory with nonlocal collisions developed in Refs.\
\cite{Weickgenannt:2019dks,Weickgenannt:2020aaf,Weickgenannt:2021cuo} [see also the related 
works \cite{Yang:2020hri,Wang:2020pej,Sheng:2021kfc}], we expand the single-particle 
distribution function in terms of irreducible moments formed by $p^\mu$ and $\ms^\mu$.
After deriving the equations of motion for the spin moments, we employ a truncation to close the 
system of equations. For the truncation we use the HW pseudo-gauge and choose the 
``14+24-moment approximation", which extends the usual 14-moment approximation 
\cite{Denicol:2012es} by 24 additional moments related to the components of the spin tensor. 
In this way, we derive for the first time a second-order dissipative theory of relativistic spin 
hydrodynamics. 

The paper is organized as follows. In Sec.\ \ref{sec:kinetic} we briefly review the kinetic theory 
developed in Refs.\ \cite{Weickgenannt:2020aaf,Weickgenannt:2021cuo}. In Sec.\ \ref{sec:currents} 
we summarize the equations of motion of spin hydrodynamics for the conserved quantities in the HW 
pseudo-gauge. The extended power-counting scheme mentioned above is subject of 
Sec.\ \ref{sec:scheme}. In Sec.\ \ref{moments} we
generalize the method of moments as used in Ref.~\cite{Denicol:2012cn} to include
spin degrees of freedom. In order to define the distribution function in local equilibrium, 
one needs to impose matching conditions, which are discussed in Sec.\ \ref{sec:match}. 
The equations of motion for the spin moments are derived in Sec.\ \ref{geomsec}. 
In Sec.\ \ref{sec:lin} the linearized collision term is expressed in terms of the spin moments. 
In order to obtain a closed set of equations of motion we employ the 14+24-moment approximation in 
Sec.\ \ref{1424}. Furthermore, we calculate the relaxation times for the spin moments and compare 
them with those related to the usual dissipative quantities. In Sec.\ \ref{PLsec}, in order to 
establish a connection with the phenomenology of heavy-ion collisions, we give the expression for the 
Pauli-Lubanski vector, which is the observable used to quantify the particle spin polarization.
Finally, in Sec.\ \ref{NSLspin} we also present the Navier-Stokes limit of the
second-order equations of motion, before concluding this work with a summary and
an outlook.

We use the following notation and conventions, $a\cdot b=a^\mu b_\mu$,
$a_{[\mu}b_{\nu]}\equiv a_\mu b_\nu-a_\nu b_\mu$, $a_{(\mu}b_{\nu)}\equiv a_\mu b_\nu
+a_\nu b_\mu$, $g_{\mu \nu} = \mathrm{diag}(+,-,-,-)$,
$\epsilon^{0123} = - \epsilon_{0123} = 1$, and repeated indices are summed over. 
The dual of any rank-2 tensor $A^{\mu\nu}$ is defined as 
$\tilde{A}^{\mu\nu}\equiv \epsilon^{\mu\nu\alpha\beta}A_{\alpha\beta}$.

\section{Kinetic theory with spin}
\label{sec:kinetic}

In this section we give a brief review of the kinetic theory for massive spin-1/2 particles developed in  
Refs.~\cite{Weickgenannt:2020aaf,Weickgenannt:2021cuo}, which will be used to derive 
hydrodynamical equations of motion in the following sections. 
All information about the microscopic theory is contained in the spin-dependent distribution function
$f(x,p,\ms)$, which depends on space-time coordinate $x^\mu$, four-momentum $p^\mu$,
and the spin vector $\ms^\mu$ and is uniquely defined in terms of the Wigner function for spinor fields,
see Refs.~\cite{Weickgenannt:2020aaf,Weickgenannt:2021cuo} for details. 
Its dynamics is described by the generalized Boltzmann equation
\begin{equation}
p\cdot \partial f=\mC[f]\;, \label{boltz}
\end{equation}
where $\mC[f]$ is the collision term. As shown in 
Refs.~\cite{Weickgenannt:2020aaf,Weickgenannt:2021cuo} this collision term contains
a nonlocal part, which allows to convert vorticity into spin. Neglecting a contribution from
pure spin exchange without momentum exchange (which will be justified below), it reads explicitly
\begin{eqnarray}
{\mC}[f] &  =& \int d\Gamma_1 d\Gamma_2 d\Gamma^\prime\,    
{\mathcal{W}}\,  
[f(x+\Delta_1,p_1,\ms_1)f(x+\Delta_2,p_2,\ms_2)
-f(x+\Delta,p,\ms)f(x+\Delta^\prime,p^\prime,\ms^\prime)]\; ,
\label{finalcollisionterm} 
\end{eqnarray}
where the integration measure 
\begin{equation}
  d\Gamma  \equiv d^4p\, \delta(p^2-m^2) d S(p)
\end{equation}
denotes integration over the extended phase space, with
\begin{equation} \label{dSp}
dS(p) \equiv \frac{\sqrt{p^2}}{\sqrt{3}\pi}  d^4\ms\,  \delta(\ms\cdot\ms+3)\delta(p\cdot \ms)\;.
\end{equation}

The transition rate $\mathcal{W}$ in Eq.\ (\ref{finalcollisionterm}) is defined as
\begin{eqnarray}
{\mathcal{W}}&\equiv& \delta^{(4)}(p+p^\prime-p_1-p_2)\,
 \frac{1}{8} \sum_{s,r}   h_{s r} (p,\ms)  
 \sum_{s',r',s_1,s_2,r_1,r_2} h_{s^\prime r^\prime}(p^\prime, \ms^\prime) \,  
 h_{s_1 r_1}(p_1, \ms_1)
  \, h_{s_2 r_2}(p_2, \ms_2) \n \\
    &&\times \langle{p,p^\prime;r,r^\prime|t|p_1,p_2;s_1,s_2}\rangle
  \langle{p_1,p_2;r_1,r_2|t^\dagger|p,p^\prime;s,s^\prime}\rangle \;,
  \label{local_col_GLW_after}
\end{eqnarray}
with
\begin{equation}
h_{s r}(p,\ms) \equiv \delta_{s r}+\ms \cdot  n_{s r}(p)\; ,
\end{equation} 
where
\begin{equation}
n^\mu_{sr}(p)\equiv\frac{1}{2m} \bar{u}_s(p)\gamma^5\gamma^\mu u_r(p)\; .
\end{equation}
The scattering matrix element for a general  interaction 
$\rho\equiv-(1/\hbar) \partial \mathcal{L}_I/(\partial \bar{\psi})$, where $\mathcal{L}_I$ is the 
interaction Lagrangian
and $\bar{\psi}$ is the Dirac-adjoint fermion spinor, is defined as \cite{DeGroot:1980dk}
\begin{align}
&\langle{p,p^\prime;r,r^\prime|t|p_1,p_2;s_1,s_2}\rangle\equiv
-\sqrt{\frac{(2\pi\hbar)^{7}}{2}}\, \bar{u}_r(p) \outlanglesm p^\prime;r^\prime|\rho(0)| 
p_1,p_2;s_1,s_2\inranglesm\; .
 \label{turho}
\end{align}
In \eq\eqref{finalcollisionterm} the nonlocality of the collision term is given by the spatial separations
\begin{equation}
\label{deltanon}
\Delta^\mu\equiv -\frac{\hbar}{2m(p\cdot\hat{t}+m)}\, 
\epsilon^{\mu\nu\alpha\beta}p_\nu \hat{t}_\alpha \ms_{\beta}\;,
\end{equation}
where $\hat{t}^\mu$ is the time-like unit vector which is equal to $(1,\boldsymbol{0})$ 
in the frame where $p^\mu$ is measured. The Boltzmann equation \eqref{boltz} is the starting point to 
derive dissipative equations of motion for spin hydrodynamics.

\section{Equations of motion of spin hydrodynamics}
\label{sec:currents}

The dynamical quantities in spin hydrodynamics are the charge current $N^\mu$, the 
energy-momentum tensor $T^{\mu\nu}$, and the spin tensor $S^{\lambda,\mu\nu}$. It should be noted 
that the form of these quantities depends on the choice of the pseudo-gauge. In this paper, we choose 
the so-called Hilgevoord-Wouthuysen (HW) pseudo-gauge \cite{HILGEVOORD19631}, which 
corresponds to a frame where the spin of a particle is measured in its rest frame. 
As will become clear later in Sec.\ \ref{1424}, the dynamical moments depend on the choice of
pseudo-gauge, which hence affects the evolution of the system. 
Since the HW spin tensor is conserved in equilibrium 
[see discussion in Ref.\ \cite{Weickgenannt:2020aaf}], we expect that it evolves on the
same time scales as the charge current and the energy-momentum tensor, i.e., on
hydrodynamic time scales. 

In kinetic theory the form of the 
charge current, as well as the energy-momentum and the spin tensor 
can be obtained from the Wigner-function formalism, employing a power-series
expansion in $\hbar$ 
\cite{Weickgenannt:2019dks,Weickgenannt:2020aaf,Speranza:2020ilk}. 
In the following, we will work up to first order in $\hbar$, such that the currents have the form
\begin{subequations} \label{conscurr}
\begin{align}
N^\mu &= \left\langle p^\mu \right\rangle\;,\\
T^{\mu\nu}&= \left\langle p^\mu p^\nu \right\rangle+\hbar\, T^{\mu\nu}_\mathrm{i}\;, 
\label{conscurr_t}\\
S^{\lambda,\mu\nu}&= \frac12\left\langle p^\lambda  \Sigma_{\ms}^{\mu\nu}\right\rangle 
-\frac{\hbar}{4m^2}\partial^{[\nu} \left\langle p^{\mu]} p^\lambda \right\rangle\;. 
\end{align}
\end{subequations}
Here we defined
\begin{equation}
\langle \cdots \rangle\equiv \int d\Gamma\, (\cdots) f(x,p,\ms)\;,
\end{equation}
and the dipole-moment tensor
\begin{equation}
\Sigma_\ms^{\mu\nu}\equiv -\frac{1}{m} \epsilon^{\mu\nu\alpha\beta} p_\alpha \ms_\beta\;. 
\end{equation}
The interaction contribution $\hbar \,T_\mathrm{i}^{\mu\nu}$ in \eq\eqref{conscurr_t} 
is of second order in $\hbar$ (see below) and 
hence will be neglected in the equations of motion for $T^{\mu\nu}$. However, its
antisymmetric part contributes to first order in $\hbar$ to 
the equation of motion of the spin tensor, see Eq.\ (\ref{eoms_s}).
This antisymmetric part arises from nonlocal collisions, which are responsible for the conversion of 
orbital to spin angular momentum.
The equations of motion of spin hydrodynamics read~\cite{Weickgenannt:2020aaf}
\begin{subequations} \label{eoms_nts}
\begin{align}
\partial_\mu N^\mu&=0\;, \label{eom_n}\\
\partial_\mu T^{\mu\nu}&=0\;, \label{eom_t}\\
\partial_\lambda S^{\lambda,\mu\nu}&=\frac12 \int d\Gamma\, \Sigma_\ms^{\mu\nu} \mC[f] 
\equiv T_\mathrm{i}^{[\nu\mu]}\; . \label{eoms_s}
\end{align}
\end{subequations}
By explicitly performing a pseudo-gauge transformation from the canonical to the HW
energy-momentum tensor, we observe that $T^{\mu \nu} \sim \int d\Gamma
v^\mu p^\nu$, with some vector $v^\mu$ \cite{Weickgenannt:2022jes}. Combining this
with Eq.\ (\ref{eoms_s}) and the conservation of total angular momentum 
$J^{\mu\nu}\equiv\Delta^{[\mu} p^{\nu]}+(\hbar/2) \Sigma_\ms^{\mu\nu}$ in a microscopic collision
we obtain
\begin{equation}
\hbar\, T_\mathrm{i}^{\mu\nu}=\int d\Gamma \Delta^\mu p^\nu \mC[f]\;. \label{ti}
\end{equation}
We now show that this is of second order in $\hbar$. Namely, when 
expanding the distribution functions in the collision term (\ref{finalcollisionterm}) 
in a Taylor series around $x$, we recover to lowest order the standard local collision term.
Under the integral in Eqs.\ (\ref{eoms_s}) or (\ref{ti}), respectively, this contribution vanishes
[the local collision term conserves spin or orbital angular momentum separately, see discussion
in Ref.\ \cite{Weickgenannt:2020aaf}]. The next term in the Taylor series gives rise
to the nonlocal collision term, which does not separately conserve spin or orbital angular momentum
and is of linear order in the shifts (\ref{deltanon}).
These shifts are of first order in $\hbar$, and together with the prefactor
$\sim \Delta^\mu p^\nu$ in Eq.\ (\ref{ti}) we obtain 
$\hbar T_\mathrm{i}^{\mu\nu} \sim \mathcal{O}(\hbar^2)$.

It is convenient to decompose the quantities in \eq\eqref{conscurr} with respect to the fluid velocity 
$u^\mu$. In this work, the latter is defined as the normalized timelike eigenvector of 
$T_\mathrm{ni}^{\mu\nu}\equiv T^{\mu\nu}-\hbar\, T_\mathrm{i}^{\mu\nu}
= \langle p^\mu p^\nu \rangle$ with eigenvalue $\epsilon$,
\begin{equation}
T_\mathrm{ni}^{\mu\nu} u_\nu= \epsilon u^\mu\; . \label{landau}
\end{equation}
In other words, we choose the Landau frame (with respect to $T_\mathrm{ni}^{\mu\nu}$). 
All momenta appearing in the microscopic expressions for the hydrodynamic quantities in 
\eq\eqref{conscurr} are decomposed into parts parallel and orthogonal to the fluid velocity,
\begin{equation} \label{p_decomp}
p^\mu = E_p u^\mu + p^{\langle \mu\rangle}\;,
\end{equation}
with $E_p\equiv p\cdot u$ and $p^{\langle\mu\rangle}\equiv \proj^{\mu\nu}p_\nu$, where 
$\proj^{\mu\nu}\equiv g^{\mu\nu}-u^\mu u^\nu$ is the projector onto the
three-space orthogonal to the fluid velocity. 
Furthermore, products of two momenta are split into parallel, orthogonal, and traceless orthogonal 
parts, making use of the traceless projector $\proj^{\mu\nu}_{\alpha\beta}\equiv 
(1/2)\proj^{(\mu}_\alpha \proj^{\nu)}_\beta-(1/3) \proj^{\mu\nu}\proj_{\alpha\beta}$ and the notation 
$p^{\langle\mu} p^{\nu\rangle}\equiv \proj^{\mu\nu}_{\alpha\beta}p^\alpha p^\beta$. 
The tensor decompositions of $N^\mu$, $T^{\mu \nu}_\mathrm{ni}$, and $S^{\lambda, \mu \nu}$
then take the form
\begin{subequations}
\begin{align}
N^\mu ={}& n u^\mu + n^\mu\;,\\
T_\mathrm{ni}^{\mu\nu}={}& \epsilon u^\mu u^\nu -\Delta^{\mu\nu}(P_0+\Pi)+\pi^{\mu\nu}\; ,\\
S^{\lambda,\mu\nu}={}&u^\lambda \tilde{\mathfrak{N}}^{\mu\nu}
+\Delta^\lambda_{\, \alpha}   \tilde{\mathfrak{P}}^{\alpha\mu\nu}
+ u_{(\alpha}\tilde{\mathfrak{H}}^{\lambda)\mu\nu\alpha}+\tilde{\mathfrak{Q}}^{\lambda\mu\nu}
-\frac{\hbar}{4m^2}\partial^{[\nu}\left[\epsilon u^{\mu]} u^\lambda-\Delta^{\mu]\lambda}(P_0+\Pi)
+\pi^{\mu]\lambda}\right]\; . \label{spitendeco}
\end{align}
\end{subequations}
Here we defined the usual hydrodynamic currents, which are given by the particle density 
$n\equiv\langle E_p\rangle$, the particle diffusion current $n^\mu\equiv  \langle p^{\lmur}\rangle$, 
the energy density $\epsilon\equiv \langle E_p^2\rangle$, the thermodynamic pressure $P_0$, 
the bulk viscous pressure $\Pi$ with $P_0+\Pi\equiv -(1/3)\langle \proj^{\mu\nu} p_\mu p_\nu\rangle$, 
and the shear-stress tensor $\pi^{\mu\nu}\equiv \langle p^{\langle\mu} p^{\nu\rangle}\rangle$.  
In addition, the following new quantities associated with spin transport occur, 
\begin{subequations}
\begin{align}
\tilde{\mathfrak{N}}^{\mu\nu}& \equiv -\frac{1}{2m}\epsilon^{\mu\nu\alpha\beta}u_\alpha 
\langle E_p^2\, \ms_\beta\rangle\;, \\
 \tilde{\mathfrak{P}}^{\alpha\mu\nu}& \equiv -\frac{1}{6m}\epsilon^{\alpha\mu\nu\beta} \langle 
 \Delta^{\rho\sigma} p_\rho p_\sigma\, \ms_\beta\rangle\;,\\
\tilde{ \mathfrak{H}}^{\lambda\mu\nu\alpha}& \equiv-\frac{1}{2m}\epsilon^{\mu\nu\alpha\beta}
\langle E_p p^{\langle\lambda\rangle}  \ms_\beta\rangle\;, \\
 \tilde{\mathfrak{Q}}^{\lambda\mu\nu}& \equiv -\frac{1}{2m}\epsilon^{\mu\nu\alpha\beta}
 \langle p^{\langle\lambda} p_{\alpha\rangle}\, \ms_\beta\rangle\;,
\end{align}
\end{subequations}
which are dual to the \emph{spin-energy tensor}
\begin{subequations}\label{nphqdef}
\begin{equation}
{\mathfrak{N}}^{\mu\nu}\equiv -\frac{1}{2m}u^{\mu} \langle E_p^2\, \ms^{\nu}\rangle\;,
\end{equation}
the \emph{spin-pressure tensor}  
\begin{equation}
{\mathfrak{P}}^{\mu}\equiv 
-\frac{1}{6m} \langle \Delta^{\rho\sigma} p_\rho p_\sigma\, \ms^\mu\rangle\;,
\end{equation}
the \emph{spin-diffusion tensor}
\begin{equation}
{ \mathfrak{H}}^{\lambda\mu}\equiv
-\frac{1}{2m}\langle  E_p p^{\langle\lambda\rangle}  \ms^{\mu}\rangle\;,
\end{equation}
and the \emph{spin-stress tensor}
\begin{equation}
 {\mathfrak{Q}}^{\lambda\mu\nu}\equiv 
 -\frac{1}{2m}\langle p^{\langle\mu} p^{\nu\rangle}\, \ms^{\lambda}\rangle\;. 
\end{equation}
\end{subequations}
We remark that the 24 degrees of freedom of the spin tensor in \eq\eqref{spitendeco} are distributed 
as follows: 3 from the spin-energy tensor, 3 from the spin-pressure tensor, 9 from the spin-diffusion 
tensor, and 9 from the spin-stress tensor. Although \eqs\eqref{nphqdef} in principle contain more than 
these degrees of freedom, as we will see later, certain components will be fixed by the matching 
conditions and constraints, such that the number of dynamical components in our framework reduces 
to 24. As in standard (spin-averaged) dissipative hydrodynamics, the system of equations of motion 
\eqref{eoms_nts} is not sufficient to determine all 14+24=38 dynamical degrees of freedom of the 
system. In the remainder of this paper we will derive additional equations of motion for the dissipative 
currents from the Boltzmann equation \eqref{boltz} using the method of moments, and thus close the 
system of equations of motion \cite{Denicol:2012cn}.

\section{Power-counting scheme}
\label{sec:scheme}

In this section we introduce a novel power-counting scheme, which allows to extend the 
concept of local equilibrium in the presence of spin and nonlocal collisions, 
and expand the distribution function $f(x,p,\ms)$ around this equilibrium state. 
In kinetic theory, local equilibrium is defined by the condition 
that the collision term vanishes. However, in Ref.~\cite{Weickgenannt:2020aaf}, it was found that the 
nonlocal part of the collision term vanishes only in \emph{global equilibrium}. This means that
the single-particle distribution function assumes the equilibrium form
\begin{equation}
f_{\mathrm{eq}}(x,p,\ms)=\frac{1}{(2\pi\hbar)^3} \exp \left(-\beta_0 u\cdot p+\alpha_0
+\frac\hbar4 \Omega_{\mu\nu}\Sigma_\ms^{\mu\nu}\right)\;, \label{fl_eq}
\end{equation}
where $\beta_0 \equiv 1/T$ is the inverse temperature,
$\alpha_0 \equiv \beta_0 \mu$, with $\mu$ being the chemical potential, and
$\Omega_{\mu \nu}$ the so-called spin potential, and the following \emph{global-equilibrium conditions}
are fulfilled,
\begin{subequations} \label{global_eq_cond}
\begin{align}
\partial_\mu \alpha_0 & = 0\;, \label{gec_1} \\
\partial_{(\mu} \beta_0 u_{\nu)} & = 0\;, \label{gec_2} \\
\Omega_{\mu \nu} & = \varpi_{\mu \nu} \equiv - \frac{1}{2}\, \partial_{[\mu}\beta_0 u_{\nu]}\;,
\label{gec_3}
\end{align}
\end{subequations}
where $\varpi_{\mu \nu}$ is the so-called \emph{thermal vorticity}. 

However, having to abandon the concept of local equilibrium in the presence of spin and
nonlocal collisions seems to be too restrictive. On the one hand, the local part of the
collision term vanishes also in \emph{local equilibrium}, without imposing the global-equilibrium
conditions (\ref{global_eq_cond}), just as in conventional kinetic theory. On the other hand, 
the nonlocal collision term captures physics on a length scale $\sim \Delta \sim \hbar/m$,
i.e., on the order of the Compton wavelength of a particle. This is
typically smaller or on the order of the range of the interaction $\ell_{\mathrm{int}}$, which is 
usually assumed to be much smaller than the mean free path $\lambda_{\mathrm{mfp}}$, such that the
particles can be treated as free between collisions. Finally,
in order to derive hydrodynamics from kinetic theory, it is assumed that
hydrodynamic quantities vary over a scale $L_{\hydro}$ which is much larger
than the mean free path, i.e., an expansion in powers of the Knudsen number 
Kn $\equiv \lambda_{\mfp}/L_{\hydro}$ is applicable. 
Thus, the scales in the problem are ordered as follows,
\begin{equation}
\Delta \lesssim \ell_{\mathrm{int}}\ll \lambda_{\mfp} \ll L_{\hydro}\;. \label{scaleord}
\end{equation}
We expect that physics on the scale $\Delta$ should not have a major influence on
what happens on the hydrodynamic scale $L_{\hydro}$. Thus, we should be able
to extend the concept of local equilibrium to situations where terms of order $\Delta/L_\mathrm{hydro}$ can 
be neglected. This requires a novel power-counting scheme, which will be introduced in
the following.

We start by defining the hydrodynamic scale $L_{\hydro}$ as
\begin{equation}
\frac1m p\cdot\partial f_{0p} \sim \frac{1}{L_{\hydro}} f_{0p}\;, \label{lhyd}
\end{equation}
where
\begin{equation} \label{fle_0}
f_{0p}\equiv \frac{1}{(2\pi\hbar)^3} e^{-\beta_0 u\cdot p+\alpha_0}
\end{equation}
is the local-equilibrium distribution function (\ref{fl_eq}) to zeroth order in $\hbar$. 
Equation \eqref{lhyd} yields
\begin{subequations} \label{local_eq_cond}
\begin{align}
\partial_\mu \alpha_0 &\sim \mathcal{O}(L_\text{hydro}^{-1})\;,\\
\frac{1}{\beta_0}\partial_{(\mu}\beta_0 u_{\nu)}&\sim
\mathcal{O}(L_\text{hydro}^{-1})\label{parbusymord}\;.
\end{align}
\end{subequations}
These conditions relax the more restrictive global-equilibrium conditions
(\ref{gec_1}) and (\ref{gec_2}) to situations where local equilibrium is established.
If $L_\mathrm{hydro} \rightarrow \infty$, or in other words hydrodynamic gradients vanish, 
global equilibrium is recovered. 
Note that only the symmetric part of $\partial_\mu \beta_0 u_\nu$ enters
the local-equilibrium conditions (\ref{local_eq_cond}). The antisymmetric part, which
is equal to the thermal vorticity (\ref{gec_3}), does not appear. In fact, this part is not
even constrained by the global-equilibrium conditions (\ref{global_eq_cond}), because
there exist global-equilibrium states with arbitrarily large thermal vorticity \cite{Becattini:2012tc}.
This fact will become important below, as it will allow us to deviate from the standard 
power-counting of gradients of hydrodynamic quantities. 

We now decompose Eq.\ (\ref{parbusymord}) with respect to the fluid velocity $u^\mu$. 
To this end, we define $\nabla^\mu\equiv \proj^\mu_\nu \partial^\nu$ and 
$\dot{A}\equiv u\cdot \partial A \equiv dA/d\tau$, as well as the expansion scalar 
$\theta\equiv \nabla\cdot u$, the shear tensor 
$\sigma^{\mu\nu}\equiv \nabla^{\langle\mu} u^{\nu\rangle}$, and the fluid vorticity 
$\omega^{\mu\nu}\equiv (1/2)\nabla^{[\mu} u^{\nu]}$.
Contracting \eq\eqref{parbusymord} with $u^\mu u^\nu$ yields 
\begin{equation} \label{25}
\frac{\dot{\beta}_0}{\beta_0} \sim\mathcal{O}(L_\text{hydro}^{-1})\;.
\end{equation} 
Furthermore, we obtain by contracting with 
$\proj_{\alpha\beta}^{\mu\nu}$ and $\proj^{\mu\nu}$, respectively,
\begin{subequations}
\begin{align}
\sigma_{\alpha\beta} \sim \mathcal{O}(L_\text{hydro}^{-1})\;,\\
\theta \sim \mathcal{O}(L_\text{hydro}^{-1})\;.
\end{align}
\end{subequations}
Contracting with $\proj^\mu_\alpha u^\nu$ gives
\begin{equation}
\frac{1}{\beta_0} \nabla_\alpha \beta_0 +\dot{u}_\alpha \sim \mathcal{O}(L_\text{hydro}^{-1})\;. 
\label{hgru}
\end{equation}
While in principle only the sum of $(1/\beta_0)\nabla_\alpha \beta_0$ and $\dot{u}_\alpha$
is of order $\mathcal{O}(L_\text{hydro}^{-1})$, we will consider situations where both
are independently of this order of magnitude. 
This is valid when being sufficiently far away from the boundary of a 
rigidly rotating system close to equilibrium.

Now consider the thermal vorticity $\varpi_{\mu \nu}$, cf.\ Eq.\ (\ref{gec_3}). 
As discussed above, this quantity does
not enter Eq.\ (\ref{lhyd}), and it can be arbitrarily large, even in global equilibrium.
Contracting $\varpi_{\mu \nu}$ with $\Delta^\mu_\alpha u^\nu$, we obtain
\begin{equation}
\frac{1}{\beta_0} \nabla_\alpha \beta_0 -\dot{u}_\alpha \sim \mathcal{O}(L_\text{hydro}^{-1})\;,
\label{hgru2}
\end{equation}
where we used the fact that both $(1/\beta_0)\nabla_\alpha \beta_0$ and $\dot{u}_\alpha$
are of order $\mathcal{O}(L_\text{hydro}^{-1})$.
However, contracting $\varpi_{\mu \nu}$ with $\proj^\mu_\alpha \proj^\nu_\beta$ 
and dividing by $\beta_0$ yields (up to a sign) the fluid vorticity $\omega_{\alpha \beta}$. 
We assume that this quantity is associated with a different scale, which we call $\ell_\mathrm{vort}$,
\begin{equation}\label{omega_vort}
\omega_{\alpha\beta}  \sim \mathcal{O}(\ell_{\mathrm{vort}}^{-1})\;.
\end{equation}
This assumption forms the basis of the novel power-counting scheme introduced here
[for a related discussion, see Ref.~\cite{Li:2020eon}]. A priori,
$\ell_\mathrm{vort}$ can be arbitrarily small, even in global equilibrium. We will later on restrict
it in order to neglect terms of higher order in $\hbar$.

The mean free path $\lambda_\mathrm{mfp}$ is related to the collision term (\ref{finalcollisionterm}) by
\begin{equation}
\frac1m \mathfrak{C}[f] \sim \frac{1}{\lambda_{\mfp}}f\;.
\end{equation}
However, the nonlocal part of the collision term is proportional to the scale $\Delta$, cf.\ 
Eq.\ (\ref{deltanon}), which 
characterizes the nonlocality of the collision. It is also a microscopic scale and should not be larger 
than the interaction range, cf.\ Eq.\ (\ref{scaleord}). Furthermore, it is important to note that both 
$\Delta$ and the polarization are of order $\hbar$. We consider here a situation where polarization is 
only generated by nonlocal collisions, i.e., there is no initial polarization.
For the semiclassical expansion to apply we need
\begin{equation}
\Delta\, \partial f \sim \frac{ \hbar}{m}\, \partial f \ll f\; . \label{semi}
\end{equation}
Comparison to \eq\eqref{lhyd} shows that we have to require that
\begin{equation}
\Delta\, \partial f \ll \frac{L_{\hydro}}{m} p\cdot \partial f\;,
\end{equation}
which implies
\begin{equation}
\Delta \ll  L_{\hydro}\;, \label{coco}
\end{equation}
which is consistent with \eq\eqref{scaleord}. However, the
gradient in \eq\eqref{semi}, when acting on the local-equilibrium
distribution function (\ref{fle_0}), also generates a term proportional to
the vorticity. Considering Eq.\ (\ref{omega_vort}), we therefore have to demand that
\begin{equation}
\Delta \ll \ell_\mathrm{vort}\;, \label{cococo}
\end{equation}
i.e., $\ell_\mathrm{vort}$ can no longer be arbitrarily small, such as in a global-equilibrium situation with
arbitrarily fast rotation.
However, $\ell_\mathrm{vort}$ can be smaller than $L_{\hydro}$ and does not even need to be larger 
than the mean free path.

We now consider a situation in which $\ell_\mathrm{vort}\ll L_{\hydro}$ such that
\begin{equation}
\frac{\Delta}{\ell_\mathrm{vort}}\sim \frac{\lambda_{\mfp}}{L_{\hydro}} \equiv \mathrm{Kn}\;. \label{samepc}
\end{equation}
In principle, it would not be necessary to require that $\Delta/\ell_\mathrm{vort}$ is of order
Kn, i.e., we could have introduced another quantity related to this ratio. This, however,
is not necessary for our purposes.

We will now show that the distribution function
\begin{equation}
f_{\mathrm{eq}}(x,p,\ms)=f_{0p}\left( 1+\frac\hbar4 \Omega_{\mu\nu}\Sigma_\ms^{\mu\nu}\right)
+\mathcal{O}(\hbar^2) \label{fle}
\end{equation}
leads to a vanishing collision term in \eq\eqref{finalcollisionterm}, if one neglects 
terms of order $\Delta / L_\mathrm{hydro} \ll \lambda_\mathrm{mfp}/L_\mathrm{hydro} =$ Kn, 
where we have used that $\Delta \lesssim \ell_\mathrm{int} \ll \lambda_\mathrm{mfp}$. 
Here, $\Omega_{\mu\nu} = -\Omega_{\nu \mu}$ is the 
Lagrange multiplier of the \textit{total} angular momentum, and not just of the spin angular momentum.
This means that $\beta_0 u^\mu$ contains a contribution from the rotational motion of the fluid, 
or in other words, that $\Omega^{\mu \nu}$ also enters $\beta_0 u^\mu$, 
\begin{equation} \label{beta_b_om}
\beta_0 u^\mu = b^\mu + \Omega^{\mu\nu}x_\nu\;,
\end{equation}
where $b^\mu$ is the Lagrange multiplier for the linear momentum of the fluid.  

We now expand each distribution function in \eq\eqref{finalcollisionterm} to linear order in 
$\Delta$ and insert \eq\eqref{fle}, see Ref.~\cite{Weickgenannt:2020aaf} for details. 
In the terms linear in $\Delta$, the derivatives of the distribution functions lead to terms 
proportional to $\partial_\nu\beta_0 u_\mu$.
According to Eq.\ (\ref{parbusymord}), the symmetric part of the latter gives rise to terms 
of order $\Delta/L_{\hydro}$, or with Eq.\ (\ref{beta_b_om}),
\begin{equation}
\frac{1}{\beta_0}\, \Delta^\lambda \partial_{\lambda} b_{\mu} \sim 
\frac{1}{\beta_0} \, \Delta^\lambda x^\nu \partial_{\lambda }
\Omega_{\mu\nu} \sim \mathcal{O}(\Delta/L_{\hydro})\;.
\end{equation}
This is much smaller than the leading dissipative corrections, which are of first order in Knudsen 
number, and will be neglected in our extended concept of local equilibrium.

However, the antisymmetric part of $\partial_\nu\beta_0 u_\mu$ has to be kept, because it can be
nonzero even in global equilibrium (for instance, for a globally rotating system). Requiring that the 
nonlocal collision term vanishes up to corrections of order $\mathcal{O}(\Delta/L_{\hydro})$ 
leads to the condition
\begin{equation} \label{special_cond}
\frac{1}{\beta_0}\, \partial^\nu \beta_0 u^\mu 
=\frac{1}{\beta_0}\,  \Omega^{\mu\nu}+\mathcal{O}(L_{\hydro}^{-1})\;,
\end{equation}
i.e., the spin potential is equal to the thermal vorticity up to terms
of order $\mathcal{O}(L_{\hydro}^{-1})$, which vanish in global equilibrium. This is
then consistent with Eqs.\ (\ref{gec_2}) and (\ref{gec_3}).

As an antisymmetric rank-2 tensor, the spin potential contains six independent parameters.
It is convenient to decompose $\Omega^{\mu\nu}$ as
\begin{equation} \label{Omega_decomp}
\Omega^{\mu\nu}=\epsilon^{\mu\nu\alpha\beta} u_\alpha \omega_{0\beta}+ u^{[\mu} \kappa_0^{\nu]}\;,
\end{equation}
with
\begin{equation} \label{kappa_0}
 \kappa_0^\mu\equiv -\Omega^{\mu\nu}u_\nu  
\end{equation}
and
\begin{equation}
 \omega_0^\mu\equiv \frac12\epsilon^{\mu\nu\alpha\beta}u_\nu\Omega_{\alpha\beta}\;.
\end{equation}
Since $\kappa_0 \cdot u = 0$ (because $\Omega^{\mu \nu} = - \Omega^{\nu \mu}$)
and $\omega_0 \cdot u = 0$, both $\kappa_0^\mu$ and $\omega_0^\mu$ contain
three independent parameters.

Multiplying Eq.\ (\ref{special_cond}) with $u_\nu$, from Eq.\ (\ref{kappa_0}) we see that
\begin{equation}
\frac{\kappa_0^\mu}{\beta_0}  \sim \mathcal{O}(L_{\hydro}^{-1})\;,
\end{equation}
where we have used Eq.\ (\ref{25}) and the fact that $\dot{u}^\mu \sim \mathcal{O}(L_\mathrm{hydro}^{-1})$.
Multiplying Eq.\ (\ref{Omega_decomp}) with $\Delta^{\lambda}_{\; \mu}$ and antisymmetrizing
the resulting equation in the indices $(\lambda, \nu)$, we then derive
\begin{equation}
\beta_0\omega^{\mu\nu}= \epsilon^{\mu\nu\alpha\beta} u_\alpha \omega_{0\beta}
+\mathcal{O}( L_{\hydro}^{-1})\;.
\end{equation}
With Eq.\ (\ref{omega_vort}) it follows that
\begin{equation}
\frac{\omega_0^\mu}{\beta_0}  \sim \mathcal{O}(\ell_\mathrm{vort}^{-1}) 
\gg \frac{\kappa_0^\mu}{\beta_0} \;.
\end{equation}

\section{Expansion around equilibrium}
\label{moments}

In this section we discuss the expansion of the distribution function around local equilibrium,
using the method of moments. We will generalize the approach of Ref.~\cite{Denicol:2012cn}
to also include spin degrees of freedom. Our starting point is the decomposition
\begin{equation}
    f_{p\ms}\equiv f_{\mathrm{eq}}+\delta f_{p\ms} \;, \label{fsplit1}
\end{equation}
with $f_{\mathrm{eq}}$ from Eq.\ (\ref{fle}) and
\begin{equation} \label{delta_fps}
 \delta f_{p\ms}\equiv f_{0p}\left(\phi_p+\ms\cdot \zeta_p\right)\;.
\end{equation}
The spin-independent part $\phi_p$ has the same form as in Ref.\ \cite{Denicol:2012cn}, i.e.,
\begin{equation} \label{phi_p}
\phi_p \equiv \sum_{l=0}^\infty \lambda_p^{\langle \mu_1 \cdots \mu_l \rangle}
p_{\langle\mu_1} \cdots p_{\mu_l\rangle}\;,
\end{equation}
where
\begin{equation}
\lambda_p^{\langle \mu_1 \cdots \mu_l \rangle} \equiv 
\sum_{n=0}^{N_l}\mathcal{H}_{pn}^{(l)} \rho_n^{\mu_1 \cdots\mu_l} \;.
\end{equation}
Here, $p_{\langle\mu_1} \cdots p_{\mu_l\rangle}$ are the irreducible tensors in momentum 
space and
\begin{equation}
\rho_n^{\mu_1 \cdots\mu_l} \equiv \left\langle E_p^n p^{\langle\mu_1} \cdots p^{\mu_l\rangle}
\right\rangle_\delta
\end{equation}
are the spin-independent irreducible moments of the deviation of the single-particle distribution 
function from local equilibrium, with
\begin{equation}
\langle\cdots\rangle_\delta\equiv \langle\cdots\rangle-\langle\cdots\rangle_{\mathrm{eq}}\;,
\end{equation}
where
\begin{equation}
\langle \cdots \rangle_{\mathrm{eq}} \equiv \int d\Gamma\, (\cdots) f_{\mathrm{eq}}(x,p,\ms)\; .
\end{equation}
The function $ \mathcal{H}_{pn}^{(l)}$ in Eq.\ (\ref{phi_p}) is defined as
\begin{equation}
 \mathcal{H}_{pn}^{(l)}=\frac{w^{(l)}}{l!}\sum_{m=n}^{N_l} a_{mn}^{(l)} \mathcal{P}_{pm}^{(l)}\;, \label{Hcoeffdef}
\end{equation}
where 
\begin{equation}
\mathcal{P}_{pn}^{(l)}\equiv \sum_{r=0}^n a_{nr}^{(l)} E_p^r \label{polynomialE}
\end{equation}
are orthogonal polynomials in energy, the coefficients $a_{nr}^{(l)}$ of which are 
determined such that
\begin{equation} \label{orthrel_poly}
2\int dP\, \frac{w^{(l)}}{(2l+1)!!}\left(\proj^{\alpha\beta}p_\alpha p_\beta\right)^l f_{0p} 
\mathcal{P}_{pm}^{(l)} \mathcal{P}_{pn}^{(l)}=\delta_{mn}\;,
\end{equation}
where we defined $dP \equiv d^3p/(2 p_0)$.
The normalization in Eq.\ (\ref{Hcoeffdef}) is determined as $w^{(l)} = (-1)^l/I_{2l,l}$,
where 
\begin{equation}
I_{nq}(\alpha_0,\beta_0)\equiv {\frac{1}{(2q+1)!!}} \left\langle E_p^{n-2q}(-\proj^{\alpha\beta} 
p_\alpha p_\beta)^q \right\rangle_\mathrm{eq} \label{thermdynint}
\end{equation}
are standard thermodynamic integrals.

Extending the approach of Ref.\ \cite{Denicol:2012cn} to spin degrees of freedom requires us
to introduce the four-vector $\zeta^\mu_p$ in Eq.\ (\ref{delta_fps}), which has an
expansion in terms of the irreducible tensors in momentum space,
\begin{equation}
 \zeta_p^\mu= \sum_{l=0}^{\infty}\eta_p^{\mu,\langle \mu_1\cdots \mu_l\rangle}
 p_{\langle\mu_1} \cdots p_{ \mu_l\rangle}\;. \label{zetaetaexpand}
\end{equation}
Without loss of generality, we may assume that $\zeta_p^\mu$ is orthogonal to $p^\mu$,
$p \cdot \zeta_p = 0$, since any part parallel to $p^\mu$ would vanish in Eq.\ (\ref{delta_fps})
anyway because of the constraint $\ms \cdot p =0$.
Using $p\cdot \zeta_p^\mu=0$ we obtain
\begin{equation}
    u\cdot\zeta_p=-\frac{1}{E_p} p_{\lmur} \zeta_p^\lmur.
\end{equation}
Therefore, the expansion \eqref{zetaetaexpand} takes the form
\begin{equation}
 \zeta_p^\mu= \left(g^\mu_{\nu}-\frac{p_{\langle\nu\rangle}}{E_p} u^\mu \right) \sum_{l=0}^{\infty}\eta_p^{\langle\nu\rangle,\langle \mu_1\cdots \mu_l\rangle}
 p_{\langle\mu_1} \cdots p_{ \mu_l\rangle} \;. \label{zetaetaexpand2}
\end{equation}
The coefficients $\eta_p^{\lmur,\langle \mu_1 \cdots \mu_l\rangle}$ are further expanded in terms of
polynomials in energy
\begin{equation}
 \eta_p^{\lmur,\langle \mu_1 \cdots \mu_l\rangle}= \sum_{n\in \mathbb{S}_l}
 d_n^{\mu,\langle \mu_1 \cdots \mu_l\rangle}  \mathcal{P}^{(l)}_{pn}\;,
\label{etapolyexpand}
\end{equation}
with 
\begin{equation}
{d}_n^{\mu,\langle \mu_1\cdots \mu_l\rangle} = \brm
\frac{w^{(l)}}{l!} \left\langle \mathcal{P}_{pn}^{(l)}\ms^\lmur p^{\langle \mu_1} \cdots 
p^{\mu_l\rangle}\right\rangle_\delta   \;.  \label{lpsppr}
\end{equation}
In Eq.\ (\ref{etapolyexpand}), $\mathbb{S}_l \subset \mathbb{N}_0$ is the set of indices 
of the spin moments which will be considered as dynamical degrees of freedom. We will
specify $\mathbb{S}_l$ for any given $l$ further below.
In order to prove Eq.\ (\ref{lpsppr}), insert $\delta f_{p\ms}$ from Eq.\ (\ref{delta_fps}) with
Eqs.\ (\ref{zetaetaexpand}) and (\ref{etapolyexpand}) on the right-hand side, and use 
\begin{equation}
\int dS(p) \, \ms^\mu = 0\;, \;\;\; \int dS(p) \, \ms^\mu \ms^\nu = -2 \left(g^{\mu \nu}
- \frac{p^\mu p^\nu}{p^2} \right)\;,
\end{equation}
the fact that $p \cdot \zeta_p =0$, as well as the orthogonality relations (\ref{orthrel_poly}) 
and (\ref{orthrel}).

Defining the spin moments
\begin{equation}
 \tau_n^{\mu,\mu_1 \cdots \mu_l}\equiv \langle E_p^n\, \ms^\mu p^{\langle \mu_1}\cdots 
 p^{\mu_l\rangle}\rangle_\delta\; , \label{spinmom}
\end{equation}
we obtain
\begin{equation}
  \eta_p^{\lmur,\langle \mu_1 \cdots \mu_l\rangle}= \brm 
  \sum_{n \in \mathbb{S}_l} \mathcal{H}^{(l)}_{pn} 
  \tau_n^{\langle\mu\rangle,\mu_1\cdots \mu_l}\;. \label{etatauexpand}
\end{equation}
Thus, the distribution function (\ref{fsplit1}) can be written as
\begin{equation}
 f_{p\ms}=f_{0p}\left\{ 1+\frac\hbar4\Omega_{\mu\nu}\Sigma_\ms^{\mu\nu}
 +\sum_{l=0}^\infty \left[\lambda_p^{\langle \mu_1 \cdots \mu_l\rangle}
 + \left(g_{\mu\nu}-\frac{p_{\langle\mu\rangle}}{E_p} u_\nu \right) \ms^\nu \eta_p^{\lmur,\langle \mu_1\cdots \mu_l\rangle}\right]
 p_{\langle\mu_1}\cdots p_{\mu_l\rangle}\right\}\;. \label{distrnoneqfin}
\end{equation}

Making use of the local-equilibrium distribution function \eqref{fle}, we can split the components 
\eqref{nphqdef} of the spin tensor into equilibrium and nonequilibrium parts,
\begin{subequations}
\begin{eqnarray}
 {\mathfrak{N}}^{\mu\nu}&\equiv& \mn_0^{\mu\nu} -\frac{1}{2m}u^\mu \mn^\nu \;,\\
 {\mathfrak{P}}^{\mu}&\equiv& \map_0^{\mu}-\frac{1}{6m}\left(m^2 \map^\mu-\mn^\mu\right) \;,\\
{ \mathfrak{H}}^{\lambda\mu}&\equiv& \mh^{\lambda\mu}_0
- \frac{1}{2m} \mathfrak{h}^{\lambda\mu}\;, \\
 {\mathfrak{Q}}^{\lambda\mu\nu}&\equiv& -\frac{1}{2m}\mathfrak{q}^{\lambda\mu\nu}\;,
\end{eqnarray}
\end{subequations}
with the equilibrium quantities
\begin{subequations}
\begin{align}
 {\mathfrak{n}}_0^{\mu\nu}&\equiv -\frac{1}{2m}u^{\mu} 
 \langle E_p^2\, \ms^{\nu}\rangle_\mathrm{eq}\;,\\  {\mathfrak{p}}_0^{\mu}&\equiv -\frac{1}{6m} \langle \Delta^{\rho\sigma} 
 p_\rho p_\sigma\, \ms^\mu\rangle_\mathrm{eq} \;,\\
{ \mathfrak{h}}_0^{\lambda\mu}&\equiv-\frac{1}{2m}\langle  E_p p^{\langle\lambda\rangle}  
\ms^{\mu}\rangle_\mathrm{eq}\;,
\end{align}
\end{subequations}
and the terms
\begin{align} \label{definitions}
\mn^\nu&\equiv\taum_2^\nu, & \map^\mu &\equiv \tau_0^\mu, & 
\mathfrak{h}^{\lambda\mu}&\equiv\taum_1^{\mu,\lambda}, & \mathfrak{q}^{\lambda\mu\nu}
\equiv \tau^{\lambda,\mu\nu}_{0} \; ,
\end{align}
pertaining to nonequilibrium.

It should be noted that not all spin moments are independent, since $\zeta_p^\mu$ has only three independent components because of $p\cdot \zeta_p =0$. Using Eq.\ (\ref{p_decomp}) 
and $\ms \cdot p = 0$, we compute
\begin{equation}
u_\mu \taum_r^{\mu,\mu_1\cdots\mu_n}=\int d\Gamma E_p^r (u\cdot\ms) p^{\langle\mu_1}
\cdots p^{\mu_n\rangle} \delta f_{p \ms} =- \int d\Gamma E_p^{r-1} \ms_\nu p^{\langle\nu\rangle} 
p^{\langle\mu_1}\cdots p^{\mu_n\rangle} \delta f_{p \ms} \; . \label{udottau}
\end{equation}
Rearranging the projection operators, 
the right-hand side can now be expressed in terms of a linear combination
of the spin moments. For specific $n$, this will be shown explicitly below. 
For this reason, in the following we will derive equations of motion only for the components
of $\taum_r^{\mu,\mu_1\cdots\mu_n}$ orthogonal to $u^\mu$, from which also the ones 
parallel to $u^\mu$ can be obtained.

\section{Matching conditions and equations of motion for hydrodynamic variables}
\label{sec:match}

The dynamical degrees of freedom of the local-equilibrium distribution function \eqref{fle} are
the Lagrange multipliers $\alpha_0$, $\beta_0$, $u^\mu$, and $\Omega^{\mu\nu}$. A priori, these 
fields are not specified and constitute additional degrees of freedom. 
By imposing a choice for the hydrodynamic frame, see, e.g., \eq\eqref{landau}, and so-called 
matching conditions for the moments of the distribution function, they can be related
to physical quantities, e.g., the particle number density, the energy density, and the
angular-momentum density of the system, and at the same time one can eliminate some
of the irreducible moments of the nonequilibrium part of the distribution function. 

In order to define $\alpha_0$ and $\beta_0$, we impose the matching conditions
\begin{align}
n= n_0 \equiv \langle E_p \rangle_{\mathrm{eq}}\;,\quad \quad
\epsilon= \epsilon_0 \equiv \langle E_p^2 \rangle_{\mathrm{eq}}\;, \label{epsmatch}
\end{align}
i.e., the particle number and energy densities of the fictitious local-equilibrium state
match those of the actual system.
Furthermore, in order to define the spin potential $\Omega^{\mu\nu}$, we require that
the total angular momentum of the system matches that of the local-equilibrium state,
\begin{equation} 
u_\lambda J^{\lambda,\mu\nu}=u_\lambda J^{\lambda,\mu\nu}_{\mathrm{eq}} \label{matchj}\; ,
\end{equation}
where 
\begin{equation} \label{Jlmn}
J^{\lambda,\mu\nu}\equiv x^\mu T^{\lambda\nu} - x^\nu T^{\lambda\mu} 
+\hbar S^{\lambda,\mu\nu}
\end{equation} 
is the total angular-momentum tensor. This matching condition is chosen 
in analogy to the Landau matching condition for the energy-momentum tensor, such that the total 
angular-momentum density in the fluid rest frame equals its equilibrium value. 
We note that some works \cite{Bhadury:2020cop,Bhadury:2020puc} use the spin tensor 
$S^{\lambda, \mu \nu}$ for the
matching condition \eq\eqref{matchj}. We prefer not to do so, since only the total angular 
momentum is conserved in the presence of nonlocal collisions. Only for
conserved quantities the corresponding global charge transforms as a tensor under 
Lorentz transformations \cite{Speranza:2020ilk}. The latter is not the case for the
generally nonconserved spin tensor $S^{\lambda, \mu \nu}$.

The matching condition (\ref{matchj}) allows to express some of the components of the
spin tensor in terms of the interacting part of the energy-momentum tensor.
Inserting the angular momentum tensor (\ref{Jlmn}) with the spin tensor \eqref{spitendeco} into 
Eq.\ (\ref{matchj}) and using the Landau condition \eqref{landau}  we find
\begin{equation}
u_\lambda  T_\mathrm{i}^{\lambda[\nu}x^{\mu]}+\tilde{\mathfrak{N}}^{\mu\nu}
+ \tilde{\mathfrak{H}}_\lambda^{\ \mu\nu\lambda}=\tilde{\mn}_0^{\mu\nu}
+2 \tilde{\mh}_{0\lambda}^{\ \ \mu\nu\lambda}\;.
\end{equation}
Here, we dropped the terms proportional to the derivatives of $\Pi$ and $\pi^{\mu\nu}$ in 
\eq\eqref{spitendeco}, since these terms would lead to second-order derivatives of dissipative
quantities in the equations of motion, which are generally not considered in second-order 
hydrodynamic theories. Contracting this equation with $\epsilon_{\alpha\beta\mu\nu}$ and then 
either with $u^\alpha$ or $\proj^\alpha_\mu\proj^\beta_\nu$, respectively, results in the following 
relations for the dissipative spin moments,
\begin{subequations}\label{matchcond44}
\begin{align}
\mn^{\langle\mu\rangle}-u_\lambda \mh^{\mu\lambda}
&=2mu_\alpha\epsilon^{\mu\alpha\rho\nu}  u^\lambda x_\rho T_{\mathrm{i}\lambda\nu}\; ,
\label{matchcond44a}\\
{\frac12}\mh^{[\nu\langle\mu\rangle]}&=m \proj^\mu_\alpha \proj^\nu_\beta \epsilon^{\alpha\beta\rho\sigma} 
u^\lambda x_\rho T_{\mathrm{i}\lambda\sigma}\;, \label{matchcond44b}
\end{align}
\end{subequations}
where $A^{[\mu \langle \nu \rangle]} \equiv A^{[\mu}_{\ \ \alpha} \Delta^{\nu] \alpha}$, which lead to
\begin{equation}
{\frac12} \mathfrak{h}^{[\lambda\mu]}=  \frac{1}{2} u^{[\lambda} \taum_2^{\mu]}
 +m \epsilon^{\lambda\mu\alpha\beta} u^\rho x_\alpha T_{\mathrm{i}\rho\beta}\; .
 \label{Hfrommatch}
\end{equation}
In the following, we choose $\hat{t}^\mu=u^\mu$ in Eq.\ (\ref{deltanon}), i.e., we describe collisions 
in the fluid rest frame. In this case, $u_\lambda T^{\lambda\nu}_\mathrm{i}=0$ and the 
right-hand sides of Eqs.\ (\ref{matchcond44}) as well as the last term in 
\eq\eqref{Hfrommatch} vanish. 


From the conservation equations \eqref{eom_n}, \eqref{eom_t} 
we obtain the following comoving derivatives 
[remember that $T_\mathrm{i}^{\mu\nu}$ is neglected in the equation of motion (\ref{eom_t})],
\begin{align}
\dot{\alpha}_0&=\frac{1}{D_{20}}\left\{-I_{30}(n_0 \theta+\partial\cdot n)
+I_{20}[(\epsilon_0+P_0+\Pi)\theta-\pi^{\mu\nu}\sigma_{\mu\nu}] \right\}\;,\label{alphadot}\\
\dot{\beta}_0&=\frac{1}{D_{20}}\left\{-I_{20}(n_0 \theta+\partial\cdot n)
+I_{10}[(\epsilon_0+P_0+\Pi)\theta-\pi^{\mu\nu}\sigma_{\mu\nu}] \right\}\;,\label{betadot}\\
\dot{u}^\mu& = \frac{1}{\epsilon_0+P_0}\left( \nabla^\mu P_0-\Pi \dot{u}^\mu+\nabla^\mu\Pi
-\proj^\mu_\alpha \partial_\beta \pi^{\alpha\beta}\right)\label{udot}\; ,
\end{align}
where we defined 
\begin{equation}
D_{nq}\equiv I_{(n+1)q}I_{(n-1)q}-I_{nq}^2\;.
\end{equation}
Equations (\ref{alphadot}) -- (\ref{udot}) are identical to the ones in standard
second-order hydrodynamics without spin degrees of freedom.  

Analogously, using the decomposition \eqref{spitendeco}, after multiplying the equation of 
motion \eq\eqref{eoms_s} by $\epsilon_{\mu\nu\alpha\beta}u^\beta$ and $u_\nu$, 
respectively, we obtain
\begin{align}
\frac{\hbar}{m^2}\dot{\omega}_0^{\langle\alpha\rangle}=& -\frac{2}{I_{30}- I_{31}} 
\left\{  \left[ \frac{\hbar}{2m^2} ( 2\theta I_{30}-I_{40}\dot{\beta}_0+I_{30}\dot{\alpha}_0)
+\frac{\hbar}{2m^2}( I_{41}\dot{\beta}_0-I_{31}\dot{\alpha}_0)-\frac16\frac{\hbar}{m^2}I_{31}\theta\right]
\omega_0^\alpha\right.\n\\
& +u_\lambda\nabla^\alpha\frac{1}{3m}\left(m^2 \tau_{0}^\lambda-\taum_2^\lambda\right)
-u_\lambda \frac{1}{3m}\left(m^2 \tau_{0}^\lambda-\taum_2^\lambda\right)\dot{u}^\alpha
+\frac{1}{3m} \theta \left(m^2 \tau_{0}^{\langle\alpha\rangle}-\taum_2^{\langle\alpha\rangle}\right) \n\\
&\left.-\frac12\frac{\hbar}{m^2} \epsilon^{\langle\alpha\rangle\lambda\mu\nu}\kappa_{0\nu}
\left( -I_{41}u_\mu\nabla_\lambda\beta_0+I_{31}u_\mu\nabla_\lambda\alpha_0
 +3I_{31} u_\mu \dot{u}_\lambda \right)-\frac12\frac{\hbar}{m^2}\epsilon^{\langle\alpha\rangle\lambda\mu\nu}I_{31} 
u_\mu\nabla_\lambda\kappa_{0\nu}\right.\n\\
&\left.-\frac{\hbar}{m^2} I_{31} (\sigma^{\alpha\lambda}+\omega^{\alpha\lambda})\omega_{0\lambda}-\frac{1}{2m} \proj^\alpha_\beta \nabla_\lambda 
\taum_1^{\rho,(\lambda}\proj^{\beta)}_\rho +\frac{1}{2m} \taum_1^{\langle\alpha\rangle,\nu}\dot{u}_\nu  +\frac1m u_\beta \proj^\alpha_\rho\nabla_\lambda 
\taum_0^{[\beta,\rho]\lambda} \right.\n\\
&\left. -\frac1m u_\beta \taum_0^{\beta,\alpha\lambda}\dot{u}_\lambda
+2 \epsilon^{\alpha\beta\mu\nu} \hbar T_{\mathrm{i}\mu\nu}u_\beta\right\}\;, \label{omegadot}
\end{align}
and
\begin{align}
\frac{\hbar}{m^2}\dot{\kappa}^{\langle\mu\rangle}_0=&  -\frac{1}{I_{31}} \bigg\{  
\frac{\hbar}{2m^2}I_{30} \epsilon^{\mu\nu\alpha\beta} \dot{u}_\alpha \omega_{0\beta} u_\nu+\frac{\hbar}{2m^2}\epsilon^{\alpha\mu\nu\beta} u_\nu 
\left[-I_{31} \nabla_\alpha \omega_{0\beta}+(I_{41}\nabla_\alpha\beta_0
-I_{31}\nabla_\alpha \alpha_0)\omega_{0\beta}\right]\n\\
&-{\frac{1}{3m}}\epsilon^{\alpha\mu\nu\beta}u_\nu\nabla_\alpha \left(m^2\taum_{0\beta}
-\taum_{2\beta}\right)+\frac{1}{3m}\epsilon^{\alpha\mu\nu\beta}u_\nu\dot{u}_\alpha 
\left(m^2\taum_{0\beta}-\taum_{2\beta}\right)-\frac{\hbar}{2m^2} I_{31}(\sigma^{\mu\nu}+\omega^{\mu\nu})\kappa_{0\nu}
\n\\
&+\frac{1}{2m} \epsilon^{\mu\nu\alpha\beta} u_\alpha \taum_{1(\beta,\lambda)}
\left(\sigma^\lambda_{\ \nu}+\omega^\lambda_{\ \nu}\right)
+\frac{\hbar}{m^2}\left( \frac43 I_{31} \theta-I_{41}\dot{\beta}_0+I_{31}\dot{\alpha}_0 \right)
\kappa_0^\mu\n\\
&-\frac{1}{m} \epsilon^{\mu\nu\alpha\beta} u_\nu \left(\nabla^\lambda \taum_{0\beta,\alpha\lambda}
-\dot{u}^\lambda \taum_{0\beta,\alpha\lambda} \right)+2\hbar T_\mathrm{i}^{\mu\nu}u_\nu\bigg\}\; ,
\label{kappadot}
\end{align}
where we used the matching conditions in \eq\eqref{Hfrommatch}.
Using \eqs\eqref{deltanon} and \eqref{ti}, 
the last term in \eq\eqref{omegadot} is given by \begin{align}
2\epsilon^{\alpha\beta\mu\nu}u_\beta \int d\Gamma\, \Delta_\mu p_\nu \mC[f]
=&\frac{\hbar}{m}\int d\Gamma\, \left[(E_p-m) \ms^{\langle\alpha\rangle}
-\frac{E_p}{E_p+m} (u\cdot\ms) p^{\langle\alpha\rangle}\right]\mC[f]\n\\
=& \frac{\hbar}{m}\int d\Gamma\, (E_p-m) \left[ \ms^{\langle\alpha\rangle}-\frac{1}{E_p^2} 
\sum_{j=0}^\infty \left(\frac{m^2}{E_p^2}\right)^j   (u\cdot\ms) p^{\langle\alpha\rangle}\right]\mC[f]\; , 
\label{eudpc}
\end{align}
where we used the geometric series to express $1/(1-m^2/E_p^2)$.
Similarly, we have for the last term in \eq\eqref{kappadot}
\begin{align}
\int d\Gamma E_p \Delta^\mu=-\frac{\hbar}{2m} \epsilon^{\mu\nu\alpha\beta} u_\alpha 
\int d\Gamma\, \frac{1}{E_p} (E_p-m) \sum_{j=0}^\infty \left(\frac{m^2}{E_p^2}\right)^j  
p_{\langle\nu\rangle} \ms_\beta\; . \label{tamunuunu}
\end{align}
Equations \eqref{omegadot} and \eqref{kappadot} thus contain an infinite sum of moments with 
negative $r$, however, we as will show in Sec.\ \ref{1424}, such moments can be expressed in terms of those with positive $r$.

\section{Equations of motion for spin moments}
\label{geomsec}

In this section, we derive the equations of motion for the spin moments 
$\taum_r^{\langle\mu\rangle,\mu_1\ldots\mu_n}$. In our truncation scheme, we
only need these moments up to tensor-rank two in momentum. 
From the definition \eqref{spinmom} we find
\begin{equation}
\dot{\taum}_r^{\mu,\langle \mu_1 \ldots \mu_n\rangle}=\proj^{\mu_1\ldots\mu_n}_{\nu_1\ldots\nu_n} 
\frac{d}{d\tau}\int d\Gamma\, E_p^r\, p^{\langle \nu_1} \cdots p^{\nu_n\rangle} \ms^\mu \delta f_{p\ms}\;,
\label{taueomgen}
\end{equation}
Using \eq\eqref{fsplit1} with \eq\eqref{fle}, up to order $\mathcal{O}(\hbar)$
the Boltzmann equation \eqref{boltz} can be written in the form
\begin{equation}
\delta \dot{f}_{p\ms}=-\dot{f}_{0p}\left( 1+ \frac\hbar4\Omega_{\alpha\beta}\Sigma_\ms^{\alpha\beta}
\right)-\frac\hbar4f_{0p}\dot{\Omega}_{\alpha\beta}\Sigma_\ms^{\alpha\beta}
-E_p^{-1} p\cdot \nabla \left[{f}_{0p}\left( 1+ \frac\hbar4\Omega_{\alpha\beta}\Sigma_\ms^{\alpha\beta}
\right)\right]-E_p^{-1} p\cdot \nabla \delta f_{p\ms} + E_{p}^{-1} \mC[f]\;. \label{Bsplitmom}
\end{equation}
In the following, we define $I^\mu\equiv \nabla^\mu \alpha_0$, the thermodynamic function
\begin{equation}
G_{nm}\equiv I_{n0} I_{m0}-I_{(n-1)0}I_{(m+1)0}\;,
\end{equation}
as well as the collision integrals
\begin{equation} \label{coll_int}
\mC_r^{\mu,\langle \mu_1\cdots \mu_n\rangle}\equiv \int d\Gamma\, E_p^r\, p^{\langle\mu_1} 
\cdots p^{\mu_n\rangle} \ms^\mu \mC[f]\;.
\end{equation}
We also make use of the relation
\begin{equation}
\nabla^\mu P_0= \frac{n_0}{\beta_0}\nabla^\mu \alpha_0
-\frac{\epsilon_0+P_0}{\beta_0}\nabla^\mu \beta_0\;,
\end{equation}
which is unaffected by spin effects up to $\mathcal{O}(\hbar)$.
After a straightforward calculation using the properties of irreducible tensors in 
Appendix \ref{itapp}, we obtain the equation of motion for the spin moment of tensor-rank zero 
in momentum as
\begin{align}
\dot{\tau}^{\langle\mu\rangle}_r-\mC^{\langle\mu\rangle}_{r-1}=& 
\frac{\hbar}{2m}\left[\xi_r^{(0)} \theta+\frac{G_{2(r+1)}}{D_{20}}\Pi \theta
-\frac{G_{2(r+1)}}{D_{20}}\pi^{\lambda\nu}\sigma_{\lambda\nu}
-\frac{G_{3r}}{D_{20}}\partial\cdot n\right]\, \omega_0^\mu 
-\frac{\hbar}{4m} I_{(r+1)1}\proj^\mu_\lambda\nabla_\nu \tilde{\Omega}^{\lambda\nu}\n\\
&-\frac{\hbar}{4m} \tilde{\Omega}^{\langle\mu\rangle\nu}  
\left[ I_{(r+1)1} I_\nu -{I_{(r+2)1}}\frac{\beta_0}{\epsilon_0+P_0} 
\left( -\Pi\dot{u}_\nu+\nabla_\nu \Pi -\proj_{\nu\lambda}\partial_\rho \pi^{\lambda\rho}\right)\right]\n\\
&+ r\, \dot{u}_\nu \taum_{r-1}^{\lmur,\nu}+(r-1)\sigma_{\alpha\beta} \taum_{r-2}^{\lmur,\alpha\beta}
-\proj^\mu_\lambda\nabla_\nu \taum^{\lambda,\nu}_{r-1}-\frac13\left[(r+2)\taum_r^\lmur
-(r-1)m^2\taum_{r-2}^\lmur \right]\theta\n\\
& -\frac{\hbar}{4m} I_{(r+1)0} \epsilon^{\mu\nu\alpha\beta} u_\nu \dot{\Omega}_{\alpha\beta} \;,
\label{eomtaumu}
\end{align}
where we defined 
\begin{equation}
\xi_r^{(0)}\equiv - I_{(r+1)0}-r\, I_{(r+1)1}
-\frac{1}{D_{20}}\left[G_{2(r+1)}(\epsilon_0+P_0) -G_{3(r+1)}n_0\right] \; .
\end{equation}

Furthermore, for the spin moment of tensor-rank one in momentum we find the equation of motion
\begin{align}
\dot{\taum}^{\langle\mu\rangle,\langle\nu\rangle}_r-\mC^{\langle\mu\rangle,\langle\nu\rangle}_{r-1}
=&\frac{\hbar}{4m} \proj^\mu_\rho\proj^\nu_\lambda \tilde{\Omega}^{\rho\lambda}\left[\xi_r^{(1)}\theta+\frac{G_{3(r+2)}}{D_{20}}\partial\cdot n -\frac{G_{2(r+2)}}{D_{20}}\left( \Pi\theta-\pi^{\alpha\beta}\sigma_{\alpha\beta}\right) \right]  +\frac{\hbar}{4m}\proj^\mu_\rho\proj^\nu_\lambda \dot{\tilde{\Omega}}^{\rho\lambda}I_{(r+2)1}\n\\
&+\frac{\hbar}{2m} \omega_0^\mu \left[\frac{\beta_0}{\epsilon_0+P_0}  I_{(r+3)1}
\left(-\Pi\dot{u}^\nu+\nabla^\nu \Pi-\proj^\nu_\lambda \partial_\rho \pi^{\lambda\rho}\right)
-I_{(r+2)1} I^\nu \right]-\frac{\hbar}{2m}\beta_0 I_{(r+3)2}
\tilde{\Omega}^{\langle\mu\rangle}_{\ \lambda} \sigma^{\nu\lambda}\n\\
&-\frac{\hbar}{4m} I_{(r+2)1}\proj^\mu_\rho 
\left(\nabla^\nu\tilde{\Omega}^{\rho\lambda}\right)u_\lambda
+\omega^\nu_{\ \rho}\taum^{\langle\mu\rangle,\rho}_r+\frac13\left[ (r-1)m^2\taum_{r-2}^{\lmur,\nu}
-(r+3)\taum_r^{\lmur,\nu}\right]\theta-\proj^\nu_\lambda  \proj^\mu_\alpha \nabla_\rho  
\taum^{\alpha,\lambda\rho}_{r-1}\n\\
&+r \dot{u}_\rho \taum_{r-1}^{\lmur,\nu\rho}+\frac15\left[2(r-1)m^2\taum^{\lmur,\lambda}_{r-2}
-(2r+3)\taum^{\lmur,\lambda}_r \right] \sigma^\nu_{\ \lambda}
+\frac13 \dot{u}^\nu \left[ m^2 r \taum^\lmur_{r-1}-(r+3)\taum^\lmur_{r+1} \right]\n\\
&-\frac13\proj^\mu_\lambda \nabla^\nu \left( m^2 \taum_{r-1}^\lambda-\taum_{r+1}^\lambda\right)
+(r-1)\sigma_{\lambda\rho}\taum_{r-2}^{\lmur,\nu\lambda\rho}\;,
\label{eomtaumunu}
\end{align}
with
\begin{equation}
    \xi_r^{(1)}\equiv \frac{G_{3(r+2)}}{D_{20}}n_0-\frac{G_{2(r+2)}}{D_{20}}(\epsilon_0+P_0)-\frac53 \beta_0 I_{(r+3)2}\; ,
\end{equation}
and finally for the spin moment of tensor-rank two in momentum the equation of motion
reads
\begin{align}
\dot{\tau}_r^{\lmur,\langle\nu\lambda\rangle}-\mC_{r-1}^{\lmur,\langle\nu\lambda\rangle}
=& \frac{\hbar}{2m} \xi_r^{(2)} \tilde{\Omega}^{\lmur\langle\nu}I^{\lambda\rangle}
+\frac{\hbar}{2m} I_{(r+3)2} \proj^\mu_\rho\proj^{\nu\lambda}_{\alpha\beta} 
\nabla^\alpha \tilde{\Omega}^{\rho\beta}-\frac{\hbar}{2m}\tilde{\Omega}^{\mu\rho}\beta_0 u_\rho 
\sigma^{\nu\lambda} I_{(r+4)2}\n\\
&-\frac{\hbar}{2m} \frac{\beta_0}{\epsilon_0+P_0} I_{(r+4)2} \tilde{\Omega}^{\lmur\langle\nu} 
\left(-\Pi \dot{u}^{\lambda\rangle}+\nabla^{\lambda\rangle}\Pi-\proj^{\lambda\rangle}_\alpha 
\partial_\beta \pi^{\alpha\beta}\right)\n\\
&+r\dot{u}_\rho \taum^{\lmur,\nu\lambda\rho}_{r-1}+\frac25\left[m^2\taum_{r-1}^{\lmur,\langle\nu}
-(r+5) \taum_{r+1}^{\lmur,\langle\nu}\right]\dot{u}^{\lambda\rangle}\n\\
&  -\proj^\mu_\gamma\proj^{\nu\lambda}_{\alpha\beta} \nabla_\rho 
\taum_{r-1}^{\gamma,\alpha\beta\rho}+\proj^\mu_\rho\frac25 \proj^{\nu\lambda}_{\alpha\beta}
\nabla^\beta\left(\taum_{r+1}^{\rho,\alpha}-m^2 \taum_{r-1}^{\rho,\alpha} \right)\n\\
&+ \frac13\left[(r-1)m^2\taum_{r-2}^{\lmur,\nu\lambda}-(r+4)\taum_r^{\lmur,\nu\lambda} \right]\theta
+(r-1)\sigma_{\rho\tau}\taum_{r-2}^{\lmur,\nu\lambda\rho\tau}\n\\
& +\frac27 \left[2(r-1)m^2\taum_{r-2}^{\lmur,\rho\langle\nu}-(2r+5)\taum_r^{\lmur,\rho\langle \nu} \right] 
\sigma^{\lambda\rangle}_{\ \rho}+2\taum_r^{\lmur,\rho\langle\nu}\omega^{\lambda\rangle}_{\ \rho} \n\\
&+\frac{2}{15} \left[ (r-1)m^4 \taum^\lmur_{r-2}-(2r+3)m^2\taum_r^\lmur
+(r+4)\taum^\lmur_{r+2}\right]\sigma^{\nu\lambda}\;,
\label{eomtaumunulambda}
\end{align}
with 
\begin{equation}
\xi_r^{(2)}\equiv I_{(r+3)2}-\frac{n_0}{\epsilon_0+P_0} I_{(r+4)2}\;.
\end{equation}

We note that the equations of motion for the spin-independent irreducible moments
$\rho_r$, $\rho_r^{\mu}$, and $\rho_r^{\mu \nu}$ take the same form as
in Ref.\ \cite{Denicol:2012cn}. The reason is that the terms proportional to $\Sigma_\ms^{\mu \nu}$
and proportional to $\zeta^\mu$ in $\delta f_{p\ms}$
in Eq.\ (\ref{Bsplitmom}) vanish when integrating over spin space. 
This, however, does not mean that these moments do not couple to the spin moments 
$\tau_r^\mu$, $\tau_r^{\mu, \nu}$, and $\tau_r^{\mu, \nu \lambda}$, since such 
a coupling may arise from the collision term in Eq.\ (\ref{Bsplitmom}). We will discuss this further
in the next section.

Apart from the $\hbar$ expansion, which is truncated at first order, the above equations of motion 
are exact. In order to close the system of equations, we need to employ a truncation procedure. 
The Navier-Stokes limit is obtained by taking into account only terms linear in gradients of 
order $\mathcal{O}(L_{\hydro}^{-1})$, see Sec.\ \ref{sec:scheme}. However, in this approximation, the
spin moments are not dynamical. Going beyond the Navier-Stokes limit and keeping terms
which are linear in the product of gradients of order $\mathcal{O}(L_{\hydro}^{-1})$ and a 
dissipative quantity, one arrives at the second-order equations of motion, where
the spin moments are determined dynamically. In principle, one could now
follow the DNMR approach~\cite{Denicol:2012cn} by
considering only the slowest microscopic time scales as dynamical,
approximating the faster time scales by their Navier-Stokes limit, and systematically
resumming higher-order moments in energy. This will be discussed in a forthcoming work. 
In this paper, we will apply a procedure similar to Israel--Stewart theory \cite{Israel:1979wp}, which 
employs an explicit truncation of the moment expansion at tensor-rank two in momentum and in
the lowest order in moments of energy \cite{Denicol:2012es}. 
In conventional hydrodynamics, this is known as the ``14-moment approximation''. 
Since the spin tensor has 24 dynamical degrees of freedom, the analogue of this approximation
in the case of spin hydrodynamics will be referred to as ``14+24-moment approximation".

\section{Collision terms}
\label{sec:lin}

In order to close the system of equations of motion (\ref{eomtaumu}), (\ref{eomtaumunu}), and
(\ref{eomtaumunulambda}), we have to express the collision integrals (\ref{coll_int}) 
in terms of spin-independent irreducible moments and spin moments. 
We will neglect terms of second order in dissipative quantities, which means
that we keep only linear terms in $\phi$ and $\zeta^\mu$ in the collision term \eqref{finalcollisionterm}. 
[This means that terms of second order in inverse Reynolds number are neglected, cf.\
the discussion in Ref.\ \cite{Molnar:2013lta}, where such terms were computed.]
Furthermore, we keep terms of linear order in 
$\hbar$, in gradients of order $\mathcal{O}(L_\mathrm{hydro}^{-1})$, as well as in the product of the
two. Using Eq.\ (\ref{fsplit1}) with Eqs.\ (\ref{fle}) and (\ref{delta_fps}) we obtain
\begin{equation}
\mC[f] \equiv  \bar{\mC}[f] + \widehat{\mC}[f] + \mathcal{O}(\hbar^2, \Delta \partial \delta f)\;,
\label{colcolcol}
\end{equation}
where
\begin{subequations}
\begin{align}
\bar{\mC}[f] \equiv & \int d\Gamma_1 d\Gamma_2 d\Gamma^\prime\,  
{\mathcal{W}}  f_{0p} f_{0p^\prime} 
(\phi_1+\phi_2-\phi-\phi^\prime+\ms_1\cdot\zeta_1+\ms_2\cdot\zeta_2
-\ms\cdot \zeta-\ms^\prime\cdot\zeta^\prime)\;,\\
\widehat{\mC}[f] \equiv &\int d\Gamma_1 d\Gamma_2 d\Gamma^\prime\,  
{\mathcal{W}}  f_{0p} f_{0p^\prime} \bigg\{-(\partial_\mu \beta_0 u_\nu) 
\left[ \Delta_1^\mu p_1^\nu (1+\phi_2)+\Delta_2^\mu p_2^\nu(1+\phi_1)
-\Delta^\mu p^\nu(1+\phi^\prime)+\Delta^{\prime\mu} p^{\prime\nu}(1+\phi)\right]\n\\
&+\frac{\hbar}{4}\Omega_{\mu\nu}\left[ \Sigma_{\ms_1}^{\mu\nu} (1+\phi_2)
+\Sigma_{\ms_2}^{\mu\nu} (1+\phi_1)-\Sigma_{\ms}^{\mu\nu} (1+\phi^\prime)
-\Sigma_{\ms^\prime}^{\mu\nu} (1+\phi) \right]\bigg\}\; , \label{widehatC}
\end{align}
\end{subequations}
where we have abbreviated $\phi \equiv \phi_p$, $\phi' \equiv \phi_{p'}$, $\phi_1 \equiv \phi_{p_1}$,
$\phi_2 \equiv \phi_{p_2}$, $\zeta^\mu \equiv \zeta_p^\mu$, $\zeta^{\prime \mu} \equiv \zeta_{p'}^\mu$, 
$\zeta_1^\mu \equiv \zeta_{p_1}^\mu$, and
$\zeta_2 \equiv \zeta_{p_2}^\mu$, respectively.
Note that $\bar{\mC}[f]$ is the local part of the collision term and, up to the terms
proportional to the spin vectors, formally identical with the collision term in the standard Boltzmann 
equation. On the other hand, $\widehat{\mC}[f]$ corresponds to the nonlocal part of 
the collision term and is responsible for the mutual conversion of orbital angular momentum
and spin. As we shall see below, it is the local part $\bar{\mC}[f]$ which determines the
spin relaxation times, while the nonlocal part $\widehat{\mC}[f]$ enters
the equations of motion for the spin-dependent moments in a similar way as the Navier-Stokes
terms in the equations of motion for the usual dissipative quantities.

Using \eq\eqref{colcolcol}, the spin-dependent collision integrals \eqref{coll_int} 
are split into two parts,
\begin{equation}
{\mC}_{r-1}^{\mu,\langle\mu_1\cdots\mu_n\rangle}
=\bar{\mC}_{r-1}^{\mu,\langle\mu_1\cdots\mu_n\rangle}+ 
\widehat{\mC}_{r-1}^{\mu,\langle\mu_1\cdots\mu_n\rangle}\;,
\end{equation}
where
\begin{subequations}
\begin{align}
\bar{\mC}_{r-1}^{\mu,\langle\mu_1\cdots\mu_n\rangle}
\equiv&  \int d\Gamma\, E_p^{r-1} p^{\langle\mu_1}\cdots p^{\mu_n\rangle} \ms^\mu 
\bar{\mC}[f]\; , \label{cmur-1} \\
\widehat{\mC}_{r-1}^{\mu,\langle\mu_1\cdots\mu_n\rangle}
\equiv&  \int d\Gamma\, E_p^{r-1} p^{\langle\mu_1}\cdots p^{\mu_n\rangle} \ms^\mu 
\widehat{\mC}[f]\; . \label{cmur-2}
\end{align}
\end{subequations}
Inserting the expansion of the distribution function \eqref{distrnoneqfin} we find
\begin{align}
\bar{\mC}_{r-1}^{\mu,\langle\mu_1\cdots\mu_n\rangle}
=& \; 16 \sum_{l=0}^{\infty} \int [dP]\, E_p^{r-1} p^{\langle\mu_1}\cdots p^{\mu_n\rangle}  
f_{0p} f_{0p^\prime} \Big[-\mathcal{W}_0\,  p_{\langle\nu_1}\cdots p_{\nu_l\rangle} 
\eta_p^{\mu,\langle\nu_1\cdots\nu_l\rangle}\n\\
 &+ w^\mu \Big( p_{1\langle\nu_1}\cdots p_{1\nu_l\rangle} 
 \lambda_{p_1}^{\langle\nu_1\cdots\nu_l\rangle}+ p_{2\langle\nu_1}\cdots p_{2\nu_l\rangle}
 \lambda_{p_1}^{\langle\nu_2\cdots\nu_l\rangle}- p_{\langle\nu_1}\cdots p_{\nu_l\rangle}
 \lambda_{p}^{\langle\nu_1\cdots\nu_l\rangle}- p^\prime_{\langle\nu_1}\cdots p^\prime_{\nu_l\rangle}
 \lambda_{p^\prime}^{\langle\nu_1\cdots\nu_l\rangle}\Big)\n\\
&+ w_{1\ \nu}^\mu \eta_{p_1}^{\nu,\langle\nu_1\cdots\nu_l\rangle} 
p_{1\langle\nu_1}\cdots p_{1\nu_l\rangle}+ w_{2\ \nu}^\mu 
\eta_{p_2}^{\nu,\langle\nu_2\cdots\nu_l\rangle} p_{2\langle\nu_1}\cdots p_{2\nu_l\rangle}
- w_{\ \nu}^{\prime\mu} \eta_{p^\prime}^{\nu,\langle\nu_1\cdots\nu_l\rangle} 
p^\prime_{\langle\nu_1}\cdots p^\prime_{\nu_l\rangle}\Big]\; , \label{Cbar}
\end{align}
where we defined
$[d P] \equiv dP dP^\prime dP_1 dP_2$ and 
used $p \cdot \zeta=0$, cf.\ the discussion after \eq\eqref{zetaetaexpand}.
We also defined
\begin{subequations}
\label{differentws}
\begin{align}
\mathcal{W}_0\equiv& \frac{1}{2^4}\int [dS]\, \mathcal{W}\;,\label{wme}\\
w^\mu\equiv &  \frac{1}{2^4}\int [dS]\, \ms^\mu \mathcal{W}\;,\label{wsme}\\
w_i^{\mu\nu}\equiv &  \frac{1}{2^4}\int [dS]\, \ms^\mu \ms_i^\nu \mathcal{W}\;,\label{wssme}
\end{align}
\end{subequations}
with $[dS]\equiv dS(p) dS^\prime(p^\prime) dS_1(p_1) dS_2(p_2)$.

In this work, we focus on parity-conserving interactions, and in particular on scalar and
vector interactions. It is shown in Appendix \ref{meapp} that in the case of a scalar interaction 
\eqs\eqref{wsme} and \eqref{wssme} vanish, respectively. For a vector interaction, \eq\eqref{wsme} 
also vanishes, while \eq\eqref{wssme} is nonzero. However, in the limit
of small momentum transfer,
\eq\eqref{wssme} is zero also for the vector interaction, while the only nonzero contribution comes from 
\eq\eqref{wme}. For this reason, in the following we will consider the situation of either a scalar 
interaction or a vector interaction in the limit of small momentum transfer and drop the terms in
\eqs\eqref{wsme} and \eqref{wssme}. One can then immediately conclude that
the collision integrals for the spin-independent irreducible moments $\rho_r^{\mu_1\cdots \mu_l}$,
\begin{equation}
{\mC}_{r-1}^{\langle\mu_1\cdots\mu_n\rangle}
=\int d\Gamma\, E_p^{r-1} p^{\langle\mu_1}\cdots p^{\mu_n\rangle} \left\{ 
\bar{\mC}[f] +\widehat{\mC}[f]\right\}\;,
\end{equation}
only contain terms proportional to $\mathcal{W}_0$ and do not involve
the spin moments $\tau_r^{\mu,\mu_1\cdots \mu_l}$.
Therefore, the equations of motion for the $\rho_r^{\mu_1\cdots \mu_l}$
are not affected by contributions from spin, at least up to order $\mathcal{O}(\hbar)$, and
decouple from the equations of motion for the spin moments.
Hence, the standard dissipative currents, i.e., the bulk viscous pressure, the particle diffusion current, 
and the shear-stress tensor follow the same equations of motion as derived in 
Ref.~\cite{Denicol:2012cn}.

We remark at this point that \eq\eqref{wsme} is nonzero only for parity-violating interactions and, in this 
case, leads to a coupling between the equations of motion for the spin-independent irreducible
moments $\rho_r^{\mu_1 \cdots \mu_l}$ and the spin moments $\tau_r^{\mu,\mu_1 \cdots \mu_l}$. 
In this case, the time evolution of $\Pi, n^\mu$, and $\pi^{\mu \nu}$ will be influenced by spin effects. 
More detailed studies of this are left for future work.

Keeping only terms proportional to $\mathcal{W}_0$ in Eq.\ (\ref{Cbar}), we obtain
with Eq.\ (\ref{etatauexpand})
\begin{align}
\bar{\mC}_{r-1}^{\mu,\langle\mu_1\cdots\mu_l\rangle}=& -\sum_{m=0}^\infty \sum_{n \in \mathbb{S}_m} 
\left(B_{rn}^{(l)}\right)^{\langle\mu_1\cdots\mu_l\rangle}_{\langle\nu_1\cdots\nu_m\rangle} 
\taum_n^{\mu,\langle\nu_1\cdots\nu_m\rangle}\;, \label{cbartau}
\end{align}
with
\begin{align}
\left(B_{rn}^{(l)}\right)^{\langle\mu_1\cdots\mu_l\rangle}_{\langle\nu_1\cdots\nu_m\rangle}\equiv& 
-16 \int [dP] \mathcal{W}_0 f_{0p} f_{0p^\prime} E_p^{r-1} p^{\langle\mu_1}\cdots p^{\mu_l\rangle}  
\mathcal{H}_{pn}^{(m)}p_{\langle\nu_1}\cdots p_{\nu_m\rangle}\;. \label{bmanymunu}
\end{align}
This tensor can only be nonzero for $l=m$ and it must be traceless and orthogonal to $u^\mu$.
Therefore, cf.\ Appendix \ref{ciapp} and Ref.~\cite{Denicol:2012cn}, we arrive at
\begin{align}
\bar{\mC}_{r-1}^{\mu,\langle\mu_1\cdots\mu_l\rangle}=& -\sum_{n \in \mathbb{S}_l} B_{rn}^{(l)} 
\taum_n^{\mu,\langle\mu_1\cdots\mu_l\rangle}\;,\label{mcbtau}
\end{align}
with
\begin{align}
B_{rn}^{(l)}& \equiv \frac{1}{2l+1} \proj^{\nu_1\cdots\nu_l}_{\mu_1\cdots\mu_l} 
\left(B_{rn}^{(l)}\right)^{\langle\mu_1\cdots\mu_l\rangle}_{\langle\nu_1\cdots\nu_l\rangle}\;. 
\label{BDelB}
\end{align}

Finally, we consider the collision integral \eqref{cmur-2} with \eq\eqref{widehatC}. 
Since we neglect terms proportional to $w_i^{\mu \nu}$, cf.\ Eq.\ (\ref{wssme}),
all terms involving $\ms^\prime$, $\ms_1$, and $\ms_2$ vanish. Using the 
conservation of total angular momentum in binary collisions, we are left with
\begin{align}
\widehat{\mC}_{r-1}^{\mu,\langle\mu_1\cdots\mu_n\rangle}
=& \int [d\Gamma]\, \mathcal{W}\, E_p^{r-1} p^{\langle\mu_1}\cdots p^{\mu_n\rangle} \ms^\mu  
f_{0p} f_{0p^\prime} \left[-  \frac\hbar4 (\Omega_{\alpha\beta}-\varpi_{\alpha\beta}) 
\Sigma_\ms^{\alpha\beta}+\frac12 \partial_{(\alpha}\beta_0 u_{\beta)} \Delta^\alpha p^\beta \right]\; . 
\label{waisdjd}
\end{align}
These terms give corrections to the spin moments which come from the difference between thermal 
vorticity and spin potential and from thermal shear $\partial_{(\alpha}\beta_0 u_{\beta)}/2$. 
Remembering that $\hat{t}^\mu$ was chosen to be equal to $u^\mu$,
we thus obtain for the full collision integrals up to tensor-rank two in momentum 
\begin{subequations} \label{105}
\begin{align}
\mC_{r-1}^\mu&=-\sum_{n \in \mathbb{S}_0} B_{rn}^{(0)} 
\taum_n^{\mu}+g^{(0)}_{r} \left(\tilde{\Omega}^{\mu\nu}
-\tilde{\varpi}^{\mu\nu}\right) u_\nu\;,\\
\mC_{r-1}^{\mu,\langle\nu\rangle}&=-\sum_{n \in \mathbb{S}_1} B_{rn}^{(1)} 
\taum_n^{\mu,\langle\nu\rangle}+ g_r^{(1)} \left(\tilde{\Omega}^{\mu\langle\nu\rangle}
-\tilde{\varpi}^{\mu\langle\nu\rangle} \right)+h_r^{(1)} (\beta_0\dot{u}_\lambda+\nabla_\lambda \beta_0) 
\epsilon^{\mu\nu\alpha\lambda} u_\alpha \;, \label{105b}\\
\mC_{r-1}^{\mu,\langle\nu \lambda\rangle}&=  -\sum_{n \in \mathbb{S}_2} B_{rn}^{(2)} 
\taum_n^{\mu,\langle\nu \lambda\rangle}
+ h_r^{(2)} \beta_0 \sigma_\rho^{\ \langle\nu} \epsilon^{\lambda\rangle\mu\alpha\rho} u_\alpha \;, 
\label{ccbaridontknow}
\end{align}
\end{subequations}
where we used the orthogonality relation \eqref{orthrel} (see also Appendix \ref{ciapp})
and defined
\begin{subequations}
\begin{align}
g_r^{(0)} \equiv & \frac{4\hbar}{m}\int [dP]\, \mathcal{W}_0 E_p^{r} f_{0p} f_{0p^\prime}\;, 
\label{106a}\\
g_r^{(1)} \equiv & \frac{4\hbar}{3m} \int [dP]\, \mathcal{W}_0 E_p^{r-1} f_{0p} f_{0p^\prime} 
(\proj^{\alpha\beta} p_\alpha p_\beta)\;,\\
h_r^{(1)} \equiv & -\frac{4\hbar}{3m}\int [dP]\, \frac{1}{E_p+m} \mathcal{W}_0 E_p^{r-1} 
f_{0p} f_{0p^\prime} (\proj^{\alpha\beta} p_\alpha p_\beta)\;, \\
h_r^{(2)} \equiv & -\frac{16\hbar}{15m}\int [dP]\, \frac{1}{E_p+m} \mathcal{W}_0 E_p^{r-1} 
f_{0p} f_{0p^\prime} (\proj^{\alpha\beta} p_\alpha p_\beta)^2\;. \label{106d}
\end{align}
\end{subequations}

\section{Second-order equations of motion in the 14+24-moment approximation}
\label{1424}

In this section we close the set of equations of motion by a direct truncation of the moment expansion. 
Analogously to the 14-moment approximation, we assume that only the moments which appear in the 
conservation laws contribute to the moment expansion. In this case there are 24 independent 
variables for the spin degrees of freedom, which constitute the minimal number of additional
degrees of freedom in the dissipative case. Together with the 14 moments from the lowest-order
approximation in the spin-independent case we call this truncation ``14+24-moment approximation".

One may wonder what would have happened if we had chosen a different pseudo-gauge. If we had 
used, for example, the canonical currents \cite{Speranza:2020ilk}, we would have had fewer degrees of 
freedom due to the fact that the canonical spin tensor is completely antisymmetric. However, the 
canonical spin tensor is not conserved even in global equilibrium \cite{Speranza:2020ilk}, so its 
equations of motion do not correspond to conservation laws. 
On the other hand, the HW spin tensor is conserved in 
global equilibrium, which is physically more intuitive, since the mutual conversion of
orbital angular momentum into spin should balance to zero in this case. As a consequence, 
(at least some of the) spin dynamics must occur on large, i.e., hydrodynamic scales. 
Hence, it is natural to favor the HW pseudo-gauge over the canonical one. 
  
In order to express the moments which do not appear in the conservation laws in terms of those which 
do appear, we first note that inserting \eq\eqref{distrnoneqfin} into \eq\eqref{spinmom} and using 
the orthogonality relation  \eqref{orthrel} we derive the identity, 
\begin{equation}
\taum_r^{\mu,\mu_1\cdots\mu_l}=  \sum_{n\in \mathbb{S}_l} \taum^{\mu,\mu_1\cdots\mu_l}_n 
\mathfrak{F}_{rn}^{(l)} \;,
\end{equation}
with
\begin{equation}
\mathfrak{F}_{rn}^{(l)}\equiv \frac{2l!}{(2l+1)!!} \int dP\, E_p^r\, 
\mathcal{H}_{pn}^{(l)} \left(\proj^{\alpha\beta}p_\alpha p_\beta\right)^{l} f_{0p}\;,
\end{equation}
cf.\ Ref.~\cite{Denicol:2012es}. This relation is exact for $ r \in \mathbb{S}_l$ and
approximately valid for all other values of $r$.
Keeping only the moments which appear in \eqs\eqref{nphqdef} we obtain
$\mathbb{S}_0 = \{ 0,2\}$, $\mathbb{S}_1 = \{ 1 \}$, $\mathbb{S}_2 = \{0 \}$, 
while $\mathbb{S}_l$ is an empty set for $l \geq 3$. We thus arrive at
the following approximate relations,
\begin{subequations} \label{38momapprox}
\begin{align}
\taum_r^\lmur&\simeq \mathfrak{F}_{r0}^{(0)} \mft^\lmur+ \mathfrak{F}_{r2}^{(0)}\mfu^\lmur\;,\\
\taum_r^{\lmur,\nu}&\simeq \mathfrak{F}_{r1}^{(1)} \mathfrak{h}^{\lmur \nu}
=\frac12  \mathfrak{F}_{r1}^{(1)} \mfv^{\mu\nu}\;,\\
\taum_r^{\lmur,\nu\lambda}&\simeq  \mathfrak{F}_{r0}^{(2)} \mfw^{\lmur \nu\lambda}\;,\\
\taum_r^{\mu,\nu\lambda\rho\cdots}&\simeq 0 \;,
\end{align}
\end{subequations}
where we defined $\mfv^{\mu\nu}\equiv \mathfrak{h}^{(\langle\mu\rangle\nu)}$
and used Eq.\ (\ref{matchcond44b}). 
The components of the spin moments parallel to the fluid velocity are then obtained from 
\eq\eqref{udottau} as
\begin{subequations}\label{tauparallel24}
\begin{align}
u_\mu \taum_r^\mu =& - \taum_{r-1,\mu}^{\mu}\n\\
\simeq& -\frac12   \mathfrak{F}_{(r-1)1}^{(1)} \mfv^{\mu}_{\ \mu}\;,\\
u_\mu \taum_r^{\mu,\nu}=& -\taum_{r-1\,\mu}^{\mu,\nu}
-\frac13 \left(m^2 \taum_{r-1}^{\langle\nu\rangle}-\taum_{r+1}^{\langle\nu\rangle}\right)\n\\
\simeq& - \mathfrak{F}_{(r-1)0}^{(2)} \mfw^{\mu\nu}_{\ \ \mu}
-\frac13\left(m^2 \mathfrak{F}_{(r-1)0}^{(0)}-\mathfrak{F}_{(r+1)0}^{(0)}\right) 
\mft^{\langle\nu\rangle}-\frac13\left(m^2 \mathfrak{F}_{(r-1)2}^{(0)}-\mathfrak{F}_{(r+1)2}^{(0)}\right)
\mfu^{\langle\nu\rangle}\;, \label{tauparallel24b}\\
u_\mu \taum_r^{\mu,\nu\lambda}=& -\taum_{r-1\ \mu}^{\mu,\nu\lambda}
+\frac{2}{15} \left(m^2 \taum^\mu_{r-1,\mu}-\taum^\mu_{r+1,\mu} \right)\proj^{\nu\lambda}
-\frac15 \left( m^2 \taum_{r-1}^{(\langle\nu\rangle,\lambda)}
-\taum_{r+1}^{(\langle\nu\rangle,\lambda)}\right)\n \\
\simeq& \frac{1}{15} \left(m^2 \mathfrak{F}_{(r-1)1}^{(1)}-\mathfrak{F}_{(r+1)1}^{(1)}\right) 
\mfv^{\mu}_{\ \mu} \proj^{\nu\lambda}-\frac15\left(m^2 \mathfrak{F}_{(r-1)1}^{(1)}
-\mathfrak{F}_{(r+1)1}^{(1)}\right) \mfv^{\nu\lambda}\;. 
\end{align}
\end{subequations}
From Eqs.\ \eqref{matchcond44a} and the first line of Eq.\ (\ref{tauparallel24b})
for $r=1$ we conclude with the definitions (\ref{definitions}) that
\begin{equation}
\frac23  \mfu^{\langle\nu\rangle}= - \mfw^{\mu\nu}_{\ \ \mu}-\frac13 m^2 \mft^{\langle\nu\rangle}\; . \label{fixn}
\end{equation}
Hence, $\mfu^{\langle\nu\rangle}$ can be expressed through the other dynamical moments and 
we do not need to consider an additional equation of motion for this quantity. 

We note that the matrices whose components $B_{rn}^{(l)}$ appear in \eq\eqref{mcbtau} are invertible.
For $l\neq1$, we can thus straightforwardly obtain
\begin{align}
\taum^{\mu,\mu_1\cdots\mu_l}_n&=-\sum_{r \in \mathbb{S}_l} 
\mathfrak{T}_{nr}^{(l)}\, \bar{\mathfrak{C}}_{r-1}^{\mu,\langle\mu_1\cdots\mu_l\rangle}\;,
\label{invertB}
\end{align}
where we defined the matrix
\begin{equation} \label{TBinv}
\mathfrak{T}^{(l)}\equiv \left( B^{(l)}\right)^{-1}\;.
\end{equation}
Since $\tau_2^{\langle\nu\rangle}$ is fixed by the matching conditions, cf.\ \eq\eqref{fixn}, $r=2$ has 
to be excluded from the set $\mathbb{S}_0$ when performing the sum in \eq\eqref{invertB} for $l=0$. 
The reason for this is that an equation analogous to \eq\eqref{udottau} relates the components of 
$u_\mu \mC^{\mu,\langle\mu_1\cdots\mu_n\rangle}$ to the components of
$\mC^{\langle \mu \rangle,\langle\mu_1\cdots\mu_n\rangle}$,
implying that more than six components of the collision integrals can be related to 
collisional invariants. 
On the other hand, for $l=1$ \eq\eqref{mcbtau} reads with the approximation \eqref{38momapprox}
\begin{equation}
\bar{\mC}_{r-1}^{\mu,\langle\nu\rangle}=- \sum_{n \in \mathbb{S}_1} B_{rn}^{(1)} \taum_n^{\mu,\nu}
=-\frac12  \sum_{n \in \mathbb{S}_1} B_{rn}^{(1)}  
\mathfrak{F}_{n1}^{(1)} \left(\taum_1^{(\langle\mu\rangle,\nu)}
+u^{\mu} \taum_2^{\langle\nu\rangle} \right)\;, \label{cmunu47}
\end{equation}
where we have used Eqs.\ (\ref{definitions}) and (\ref{matchcond44a}).
This implies for the symmetric part orthogonal to the fluid velocity
\begin{equation}
\taum_1^{(\langle\mu\rangle,\nu)}=-\sum_{r \in \mathbb{S}_1}\mathfrak{T}_{1r}^{(1)} 
 \bar{\mC}_{r-1}^{(\langle\mu\rangle,\langle\nu\rangle)}\;. 
\end{equation}


We now multiply the equations of motion for the spin moments \eqref{eomtaumu}, \eqref{eomtaumunu},
and \eqref{eomtaumunulambda} with $\mathfrak{T}_{nr}^{(l)}$, sum over $r$ in each equation, and use 
\eq\eqref{38momapprox}. We then obtain with $\mathbb{S}_0=\{0\}$, $\mathbb{S}_1=\{1\}$, 
and $\mathbb{S}_2=\{0\}$ up to linear order in the product of gradients with dissipative 
quantities (which includes gradients of dissipative quantities) the following equation of motion
for $\mft^\lmur$,
\begin{align}
 \tau_{\mft}  \proj^\mu_\nu \frac{d}{d\tau}  {\mft}^{\langle\nu\rangle}+\mft^\lmur=&\mathfrak{e}^{(0)}\left(\tilde{\Omega}^{\mu\nu}
 -\tilde{\varpi}^{\mu\nu}\right) u_\nu+ \left(\tc^{(0)}_{\theta\omega} \theta
 + \tc^{(0)}_{\theta\omega\Pi} \Pi\theta+ \tc^{(0)}_{\pi\sigma\omega}\pi^{\lambda\nu}
 \sigma_{\lambda\nu}+ \tc^{(0)}_{n\omega}\partial\cdot n\right) \omega_0^\mu\n\\
&+\left[  \tc^{(0)}_{I\Omega}I_\nu+ \tc^{(0)}_{\Pi\Omega}\left( -\Pi\dot{u}_\nu+\nabla_\nu \Pi 
-\proj_{\nu\lambda}\partial_\rho \pi^{\lambda\rho}\right)\right]\tilde{\Omega}^{\lmur\nu}
+ \tc^{(0)}_{\nabla\Omega} \proj^\mu_\lambda\nabla_\nu \tilde{\Omega}^{\lambda\nu}+\tc^{(0)}_{\dot{\Omega}} \left( \dot{\omega}_0^\lmur-\tilde{\Omega}^{\lmur\nu}\dot{u}_\nu\right)\n\\
&+ \tct^{(0)}_1 \mfv^{\mu\nu}F_\nu+\tct_2^{(0)}\sigma_{\alpha\beta}\mfw^{\lmur\alpha\beta}
-\tct_3^{(0)} \proj^\mu_\lambda\nabla_\nu \mfv^{\lambda\nu}+\tct_4^{(0)} \theta \mft^\lmur-\tct_5^{(0)}\theta\mfw^{\nu\mu}_{\ \ \nu}+\tct_6^{(0)} \mfv^{\mu\nu} I_\nu\n\\
&+\left(\tct_7^{(0)} \mft_\nu
+\tct_8^{(0)} \mfw^{\lambda}_{\ \nu\lambda}\right)\left(\sigma^{\nu\mu}+\omega^{\nu\mu}\right)+\tct^{(0)}_9 \mfv^\nu_{\ \nu} F^\mu\;. \label{eomqqq}
\end{align}
Here we defined $F^\mu\equiv \nabla^\mu P_0$. We also
converted derivatives of the thermodynamic integrals $I_{nq}$ by the chain rule into
derivatives of $\alpha_0$ and $\beta_0$. In principle, also $\dot{u}^\mu$ could be replaced by \eq\eqref{udot}, keeping only terms 
up to linear order in the product of gradients with dissipative quantities. 
The transport coefficients appearing in front of the various terms are listed in Appendix \ref{tcapp}.

The equation of motion for $\mfv^{\mu\nu}$ is obtained from the symmetric part of
\eq\eqref{eomtaumunu} following similar steps as in \eq\eqref{eomqqq},
\begin{align}
\tau_{\mfv}\proj^\mu_\lambda \proj^\nu_\rho \frac{d}{d\tau}{\mfv}^{\lambda\rho}+\mfv^{\mu\nu}
+\tau_{\mfv}\omega^{\ (\nu}_{\rho} \mfv^{\mu)\rho} =
& \left[\tc^{(1)}_{\omega\Pi}\left(-\Pi\dot{u}^{(\nu}+\nabla^{(\nu} \Pi-\proj^{(\nu}_\lambda \partial_\rho 
\pi^{\lambda\rho}\right)-\tc^{(1)}_{\omega I} I^{(\nu} \right]\omega_0^{\mu)}+\tc_{\Omega\sigma}^{(1)}\tilde{\Omega}^{\ (\langle\mu\rangle}_{\lambda} \sigma^{\nu)\lambda}\n\\
&
-\tc^{(1)}_{\nabla\Omega}\proj^{(\mu}_\rho \left(\nabla^{\nu)}\tilde{\Omega}^{\rho\lambda}\right)
u_\lambda +\tct_1^{(1)} \mfv^{\mu\nu} \theta+\tct_2^{(1)} \proj^{(\nu}_\lambda \proj^{\mu)}_\tau \nabla_\rho 
(\proj^\tau_\alpha\mfw^{\alpha,\lambda\rho})\n\\
&+\tct_{3}^{(1)} (\nabla_\rho u^{(\mu}) \mfv^{\langle\nu)\rho\rangle}+\tct_4^{(1)} \mfw^{(\langle\mu\rangle,\nu)\lambda}F_\lambda+\tct_5^{(1)} 
\sigma^{\ (\nu}_{\lambda}\mfv^{\mu)\lambda} +\tct_6^{(1)} F^{(\nu} \mft^{\langle\mu\rangle)}+\tct_{7}^{(1)}  \mfw^{\rho(\nu}_{\ \ \rho}F^{\mu)}\n\\
&
+\tct_8^{(1)}\proj^{(\mu}_\lambda \nabla^{\nu)} (\proj^\lambda_\rho \mft^\rho)+\tct_9^{(1)}
\proj^{(\mu}_\lambda \nabla^{\nu)} \mfw^{\rho\lambda}_{\ \ \rho}
+\tct_{10}^{(1)}(\nabla^{(\nu} u^{\mu)}) 
\mfv^\lambda_{\ \lambda}\;,
\label{mfveom2nd}
\end{align}
with the transport coefficients again given in Appendix \ref{tcapp}.

Finally, the equation of motion for $\mfw^{\langle\mu\rangle\nu\lambda}$ is given by
\begin{align}
\tau_\mfw\proj^\mu_\rho \proj^{\nu\lambda}_{\alpha\beta}
\frac{d}{d\tau}{\mfw}^{\langle\rho\rangle\alpha\beta}+\mfw^{\lmur\nu\lambda}
-2\tau_\mfw \mfw^{\lmur,\rho\langle\nu}
\omega^{\lambda\rangle}_{\ \rho}=& - \mathfrak{d}^{(2)} 
\beta_0 \sigma_\rho^{\ \langle\nu} \epsilon^{\lambda\rangle\mu\alpha\rho} u_\alpha
+\tc_{\Omega I}^{(2)} \tilde{\Omega}^{\lmur\langle\nu}I^{\lambda\rangle}+\tc_{\nabla\Omega}^{(2)} \proj^{\mu}_\rho\proj^{\nu\lambda}_{\alpha\beta} \nabla^\alpha \tilde{\Omega}^{\rho\beta} \n\\
&- \tc_{\omega\sigma}^{(2)}
\sigma^{\nu\lambda} \omega_0^\mu-\tc_{\Omega\Pi}^{(2)} \tilde{\Omega}^{\lmur\langle\nu} \left(-\Pi \dot{u}^{\lambda\rangle}
+\nabla^{\lambda\rangle}\Pi-\proj^{\lambda\rangle}_\alpha \partial_\beta \pi^{\alpha\beta}\right)
\n \\
& +\tct_1^{(2)}\mfv^{\mu\langle\nu}F^{\lambda\rangle}+\tct_2^{(2)}\mfv^{\mu\langle\nu}
I^{\lambda\rangle}+\tct_3^{(2)} \proj^\mu_\rho \proj^{\nu\lambda}_{\alpha\beta}\nabla^\beta\mfv^{\rho\alpha}
+\tct_4^{(2)} \mfw^{\lmur,\nu\lambda} \theta\n \\
& +\tct_5^{(2)} \mfw^{\lmur,\rho\langle\nu} 
\sigma^{\lambda\rangle}_{\ \rho} +\tct_6^{(2)} \mft ^\lmur\sigma^{\nu\lambda}
-6\tct_7^{(2)} \mfw^{\rho\mu}_{\ \ \rho}
\sigma^{\nu\lambda}+\tct_8^{(2)} F^\mu \mfv^{\langle\nu\lambda\rangle}\n\\
&+ \tct_9^{(2)}  \mft^{\langle\nu}
\nabla^{\lambda\rangle} u^\mu+\tct_{10}^{(2)} \mfw^{\rho\langle\nu}_{\ \ \rho}\nabla^{\lambda\rangle} 
u^\mu\;, \label{eommfw2nd}
\end{align}
with the transport coefficients listed in Appendix \ref{tcapp}.

In \eq\eqref{eomqqq}, $\dot{\omega}_0^{\lmur}$ should in principle be replaced by \eq\eqref{omegadot}.
However, we refrain from doing so at this point in order to keep \eq\eqref{eomqqq} more compact. 
The equation of motion for $\omega_0^\mu$ contains the antisymmetric part of the energy-momentum 
tensor, which has to be expressed in terms of the dynamical spin moments.
In the 14+24-moment approximation, we restrict the sums in \eqs\eqref{eudpc} and \eqref{tamunuunu}, 
respectively,  to $\mathbb{S}_1$ and obtain
\begin{subequations}
\begin{align}
2\epsilon^{\alpha\beta\mu\nu}u_\beta T_{\mu\nu}=&
 \frac{\hbar}{m}\left[ \mC_1^{\langle\alpha\rangle}-m\mC_0^{\langle\alpha\rangle}
 -\sum_{j \in \mathbb{S}_1} {m^{2j}} u_\lambda \left(\mC^{\lambda,\alpha}_{-2j-1}
 -m\mC^{\lambda,\alpha}_{-2j-2}\right)\right]\;,\\
T^{[\mu\nu]} u_\nu=& -\frac{\hbar}{2m} \epsilon^{\mu\nu\alpha\beta} u_\alpha  
\sum_{j \in \mathbb{S}_1} m^{2j} \left( \mC_{-2j\beta,\nu}-m\mC_{(-2j-1)\beta,\nu}\right)\;  ,
\end{align}
\end{subequations}
where the collision terms can be expressed in terms of the dynamical spin moments using 
\eqs\eqref{cbartau}, \eqref{105}, \eqref{38momapprox}, and \eqref{tauparallel24}. 

We note that the spin relaxation times\footnote{\bbb{In the literature, the term ``spin relaxation time'' has sometimes a different meaning than the one used here. In this work, it is the time scale on which a dissipative spin moment approaches its Navier-Stokes limit, see Sec.\ \ref{NSLspin}.}} $ \tau_{\mft}$, $\tau_{\mfv}$, and $\tau_\mfw$ arise
from the inversion of the matrices $B^{(l)}$, cf.\ Eq.\ (\ref{TBinv}), see also Appendix \ref{tcapp}.
Thus, as claimed above, they originate from the local part of the collision term.
On the other hand, the first terms on the right-hand sides of Eqs.\ (\ref{eomqqq}) and 
(\ref{eommfw2nd}) arise from the nonlocal part of the collision term, cf.\ Eqs.\ (\ref{waisdjd}) and
(\ref{105}), and as mentioned above, appear in a similar way as the Navier-Stokes terms
in ordinary dissipative hydrodynamics. Note that there is no such term in Eq.\ (\ref{mfveom2nd}),
as $\mfv^{\mu \nu}$ is a symmetric rank-2 tensor, while the corresponding terms
in Eq.\ (\ref{105b}) are antisymmetric.

\begin{figure}[t]
        \includegraphics[width=0.7\textwidth]{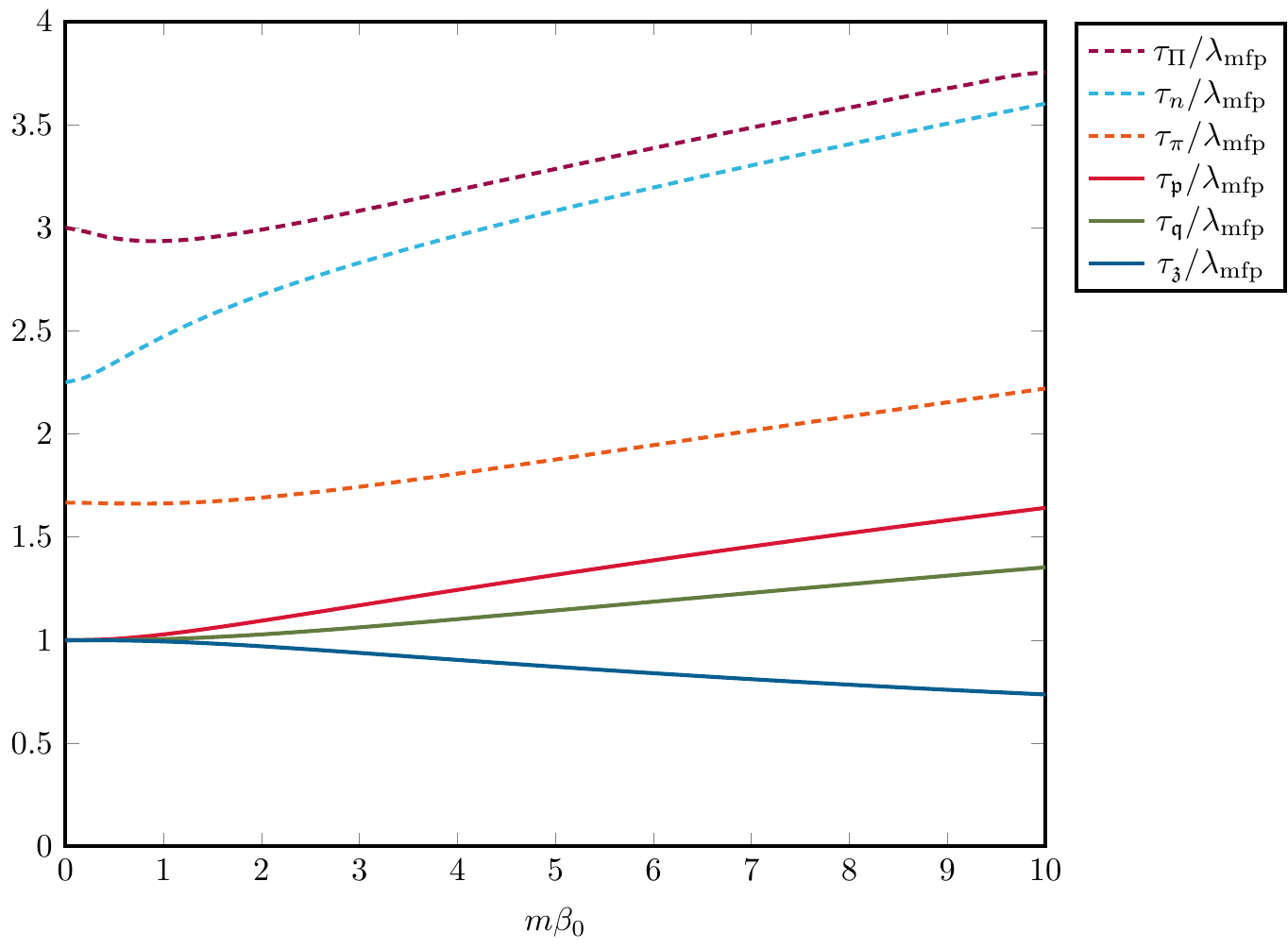}
		\caption{Relaxation times for the dissipative components of the spin tensor 
		$\mathfrak{p}^\mu$,  $\mathfrak{z}^{\mu\nu}$, $\mathfrak{q}^{\lambda\mu\nu}$
		(full lines), in comparison to those for the 
		standard dissipative quantities $\Pi$, $n^\mu$, $\pi^{\mu\nu}$ 
		(dashed lines) in the 14+24-moment 
		approximation. Note that the values for $\tau_n$ and $\tau_\pi$ in the limit 
		$m\beta_0\to 0$ agree with those obtained in Ref.\ \cite{Denicol:2012cn}.}
		\label{fig:tau_spin}
	\end{figure}

The calculation of the relaxation times requires the evaluation of certain collision integrals,
which is delegated to Appendix \ref{ciapp}.
In Fig.\ \ref{fig:tau_spin} we show the spin relaxation times  
$ \tau_{\mft}$, $\tau_{\mfv}$, and $\tau_\mfw$ (solid lines) 
in comparison to the relaxation times $\tau_\Pi$, $\tau_n$, and $\tau_\pi$ (dashed lines) of
the usual dissipative quantities as a function of $m\beta_0$.
We choose a range for $m\beta_0$ from $0$ (for \bbb{very high temperature}) to $m\beta_0=10$
(corresponding to a typical hadronic particle of mass $m \sim$ 1 GeV at a temperature
$T \sim$ 100 MeV). \bbb{It should be noted that the particle mass cannot be zero due to conditions \eqref{coco} and \eqref{cococo}, therefore, the limit $m\beta_0\rightarrow0$ corresponds to the limit $T\rightarrow\infty$.} One observes that the spin relaxation times are smaller than the 
usual relaxation times by a factor of at least 1.6 for all values of $m\beta_0$.
This means that spin dissipates on a slightly faster time scale than particle number or 
energy-momentum. Nevertheless, the order of magnitude of spin relaxation is the same as
for the usual dissipative quantities, such that it makes sense to treat spin as a dynamical
degree of freedom in second-order dissipative hydrodynamics.
By the same argument it would also be justified to consider higher-order moments
for the usual dissipative quantities as dynamical degrees of freedom, cf.\ Ref.\ \cite{Denicol:2012vq},
as the corresponding relaxation times are of a similar order as the spin relaxation times, but
here we refrain from doing so in order to keep the discussion as simple as possible.
\bbb{We observe that} all spin relaxation times converge to the same value
of $\lambda_{\mfp}$ when the \bbb{high-temperature} limit is approached. \bbb{This feature is most likely an artifact of assuming a constant cross section.}

We note that spin relaxation time has been studied using perturbative QCD techniques 
\cite{Li:2019qkf,Kapusta:2019sad,Kapusta:2020npk,Hongo:2022izs}, the Nambu--Jona-Lasinio 
model \cite{Kapusta:2019ktm}, and an effective vertex for the interaction with the thermal vorticity 
\cite{Ayala:2019iin,Ayala:2020ndx}. In Ref.\ \cite{Kapusta:2019sad}, the spin relaxation time was 
estimated based on the probability to change helicity in a collision, with a result
which is orders of magnitude larger than our results. The reason for this is that, 
in our case, the spin relaxation times include also other processes where spin is dissipated,
not only particular helicity-changing processes.

\section{Pauli-Lubanski vector}
\label{PLsec}

Comparing the polarization of hadrons measured in heavy-ion collisions with
theoretical calculations requires knowledge of the so-called 
Pauli-Lubanski vector \cite{Becattini:2013fla,Becattini:2020sww,Speranza:2020ilk,Tinti:2020gyh}. 
The latter quantity can be expressed in terms of the axial-vector component of the Wigner function 
\begin{equation}
\mathcal{A}^\mu(x,p)=\int  dS(p)\, \ms^\mu f(x,p,\ms)
\end{equation}
as \cite{Becattini:2020sww,Speranza:2020ilk,Tinti:2020gyh}
\begin{equation}
\Pi^\mu(p) = \frac{1}{2\mathcal{N}} \int d\Sigma_{\lambda} p^\lambda  \mathcal{A}^\mu(x,p)\;, \label{pi}
\end{equation}
where $d\Sigma_\lambda$ denotes the integration over the freeze-out hypersurface and we defined
\begin{equation}
\mathcal{N}\equiv   \int d\Sigma_{\lambda} p^\lambda \int  dS(p) f(x,p,\ms)\; .
\end{equation}
Inserting the distribution function \eqref{distrnoneqfin} and using the 14+24-moment approximation, 
we obtain
\begin{align}
  \Pi^\mu(p) = & \frac{1}{2\mathcal{N}} 
\int d\Sigma_{\lambda} p^\lambda f_{0p} \bigg[ -\frac{\hbar}{2m} \tilde{\Omega}^{\mu\rho}
\left(E_p u_\rho +p_{\langle\rho\rangle}\right)+ \left(g^\mu_{\nu}-\frac{p_{\langle\nu\rangle}}{E_p} u^\mu \right)\Big( \chi_\mft \mft^{\langle\nu\rangle}
-6\chi_\mfu \mfw^{\rho\nu}_{\ \ \rho}+\chi_\mfv\mfv^{\nu\alpha}p_{\langle\alpha\rangle}\n\\
&+\chi_\mfw\mfw^{\langle\nu\rangle\alpha\beta} p_{\langle\alpha} p_{\beta\rangle}\Big)\bigg]  \label{Pilocal}
\end{align}
where we defined 
\begin{align}
\chi_\mft& \equiv -2\sum_{n\in\mathbb{S}_0} \mathcal{H}_{pn}^{(0)}\left( \mathfrak{F}_{n0}^{(0)} 
-\frac12m^2 \mathfrak{F}_{n2}^{(0)} \right)\; ,\n\\
\chi_\mfu&\equiv -\frac12\sum_{n\in\mathbb{S}_0} \mathcal{H}_{pn}^{(0)} \mathfrak{F}_{n2}^{(0)}\; , \n\\
\chi_\mfv&\equiv -\sum_{n\in\mathbb{S}_1} \mathcal{H}_{pn}^{(1)} \mathfrak{F}_{n1}^{(1)}\; , \n\\
\chi_\mfw&\equiv -2\sum_{n\in\mathbb{S}_2} \mathcal{H}_{pn}^{(2)} \mathfrak{F}_{n0}^{(2)}\; .
\end{align}
In the theory of second-order dissipative spin hydrodynamics, the spin moments are treated 
dynamically and follow the equations of motion derived in Sec.\ \ref{1424}. On long time scales, they 
approach their Navier-Stokes values, namely the first-order terms on the right-hand sides of
\eqs\eqref{eomqqq}, \eqref{mfveom2nd}, and \eqref{eommfw2nd}, respectively, cf.\
Sec.\ \ref{NSLspin}. 

The global polarization is given as
\begin{equation}
\bar{\Pi}^\mu = \frac{1}{2{N}} \int d\Gamma
\int d\Sigma_{\lambda} p^\lambda  \, \ms^\mu f(x,p,\ms)\;, \label{pibar}
\end{equation}
with $N\equiv \int dP\, \mathcal{N}$.
We obtain
\begin{align}
\bar{\Pi}^\mu 
=& \frac{1}{2N} \int d\Sigma_\lambda \left[ -u^\lambda \frac{\hbar}{2m} \omega_0^\mu I_{20}
+\frac{\hbar}{2m}\proj^\lambda_\nu  \tilde{\Omega}^{\mu\nu}I_{21}+u^\lambda 
\left(\mathfrak{y}_\mft \mft^{\langle\mu\rangle} +\mathfrak{y}_\mfw \mfw^{\rho\mu}_{\ \ \rho} 
- \mathfrak{y}_\mfv u^\mu \mfv^\rho_{\ \rho} \right)+\mathfrak{y}_\mfv \mfv^{\mu\lambda} 
+ u^\mu \left( \mathfrak{w}_\mft \mft^{\langle\lambda\rangle}
+\mathfrak{w}_\mfw \mfw^{\rho\lambda}_{\ \ \rho}\right) \right]\; ,
\end{align}
with 
\begin{align}
\mathfrak{y}_\mft&\equiv \mathfrak{F}_{10}^{(0)}-\frac12m^2\mathfrak{F}_{12}^{(0)}\;, & \mathfrak{y}_\mfw&\equiv-\frac32 \mathfrak{F}_{12}^{(0)}\; , 
& \mathfrak{y}_\mfv&\equiv \frac12 \mathfrak{F}_{01}^{(1)}\;,\n\\
\mathfrak{w}_\mft&\equiv -\frac13\left(m^2 \mathfrak{F}_{-10}^{(0)}-\mathfrak{F}_{10}^{(0)}\right) 
+ \frac{m^2}{6}\left(m^2 \mathfrak{F}_{-12}^{(0)}-\mathfrak{F}_{12}^{(0)}\right)\; , 
& \mathfrak{w}_\mfw&\equiv -\mathfrak{F}_{-10}^{(2)}+\frac12 \left(m^2\mathfrak{F}^{(0)}_{-12}
-\mathfrak{F}^{(0)}_{12} \right)\; .
\end{align}
\bbb{We remark that, although the form of the Pauli-Lubanski vector given by \eq\eqref{pi} is independent of the choice of pseudo-gauge, the truncation scheme used in a hydrodynamic framework can implicitly induce a pseudo-gauge dependence on the polarization, see also Refs.~\cite{Becattini:2018duy,Becattini:2020sww,Speranza:2020ilk,Buzzegoli:2021wlg} for related discussions. In our case, the pseudo-gauge dependence enters through the choice of dynamical spin moments in the expansion of the distribution function in \eq\eqref{Pilocal}.}

\section{Navier-Stokes limit} \label{NSLspin}

The Navier-Stokes limit is obtained by considering only terms up to first order in gradients in the 
equations of motion in Sec.\ \ref{1424}.
In this context it is important to note that, in our power-counting scheme, the 
vorticity is considered to be different  than standard gradients, cf.\ 
Sec.\ \ref{sec:scheme}, which allows to account for a global-equilibrium state with arbitrary
rotation. Thus, we neglect terms
of linear order in the product of gradients and dissipative quantities, e.g., terms like
$\nabla^\lambda \tau_r^{\nu, \mu_1 \cdots \mu_n}$. However, this does not pertain to
terms of linear order in the product of vorticity and dissipative quantities as, for instance,
appear in expressions 
$\sim \nabla^\lambda \Delta^\lambda_\alpha \tau_r^{\alpha, \mu_1 \cdots \mu_n}$, 
where the space-like gradient also acts on the three-space projector.
Therefore we have in the Navier-Stokes limit
\begin{subequations}\label{NSV1}
\begin{align}
\mft^{\lmur}_\ns&=\mft^\lmur_\nr\; ,\\
\mfv^{\mu\nu}_\ns&=\mfv^{\mu\nu}_\nr+\tau_{\mfv}\omega^{\ (\nu}_{\rho} \mfv^{\mu)\rho}_\ns\; ,\\
\mfw^{\lmur \nu\lambda}_\ns&=\mfw^{\lmur \nu\lambda}_\nr-2\tau_\mfw \mfw_\ns^{\lmur \rho\langle\nu}
\omega^{\lambda\rangle}_{\ \rho}\; , 
\end{align}
\end{subequations}
with 
\begin{subequations}\label{NSnonrot}
\begin{align}
\mft^\lmur_\nr&= \mathfrak{e}^{(0)}\left(\tilde{\Omega}^{\mu\nu}-\tilde{\varpi}^{\mu\nu}\right) u_\nu
+ \tc^{(0)}_{\theta\omega} \theta\omega_0^\mu+  \tc^{(0)}_{I\Omega}I_\nu\tilde{\Omega}^{\lmur\nu}+ \tc^{(0)}_{\nabla\Omega} \proj^\mu_\lambda\nabla_\nu \tilde{\Omega}^{\lambda\nu}+\tc^{(0)}_{\dot{\Omega}}  \dot{\omega}_0^\lmur\; , \label{129a}\\
\mfv^{\mu\nu}_\nr&= -\tc^{(1)}_{\nabla\Omega}\proj^{(\mu}_\rho \left(\nabla^{\nu)}
\tilde{\Omega}^{\rho\lambda}\right)u_\lambda- \tc^{(1)}_{\omega I} I^{(\nu} \omega_0^{\mu)}
+\tc_{\Omega\sigma}^{(1)}\tilde{\Omega}^{\ (\langle\mu\rangle}_{\lambda} \sigma^{\nu)\lambda}\; ,\\
\mfw^{\lmur\nu\lambda}_\nr&=-\mathfrak{d}^{(2)} \beta_0 \sigma_\rho^{\ \langle\nu} 
\epsilon^{\lambda\rangle\mu\alpha\rho} u_\alpha+\tc_{\Omega I}^{(2)} \tilde{\Omega}^{\lmur\langle\nu}
I^{\lambda\rangle}+\tc_{\nabla\Omega}^{(2)} \proj^\mu_\rho \proj^{\nu\lambda}_{\alpha\beta} \nabla^\alpha 
\tilde{\Omega}^{\rho\beta}- \tc_{\omega\sigma}^{(2)}\sigma^{\nu\lambda} \omega_0^\mu\; , 
\label{qmunulambda}
\end{align}
\end{subequations}
being the Navier-Stokes values for relaxation to a nonrotating equilibrium state.
The first terms on the right-hand sides of Eqs.\ \eqref{129a} and \eqref{qmunulambda}
arise from the nonlocal collision term \eqref{waisdjd}.
This is also apparent from the fact that the coefficient $\mathfrak{e}^{(0)}$ 
is of order $\sim \Delta/\beta_0$ times the dimension of $\mft^\mu$,
while $\mathfrak{d}^{(2)}$ is of order $\sim \Delta/\beta_0$
times the dimension of $\mfw^{\mu \nu \lambda}$. This follows
from the definition of these quantities in Eqs.\ \eqref{e0} and \eqref{d2} and the
fact that $\tau_\mft$ and $\tau_\mfw$ are of order $\sim \lambda_{\text{mfp}}$, while
$g^{(0)}_0$ and $h_0^{(2)}$, cf.\ Eqs.\ \eqref{106a} and \eqref{106d}, are
$\sim \Delta/\lambda_{\text{mfp}}$ times appropriate powers of temperature to
give the correct dimensions. Thus, the first terms on the right-hand sides
of Eqs.\ \eqref{129a} and \eqref{qmunulambda} are of order 
$\sim \Delta/L_{\text{hydro}}$ times the dimensions of $\mft^\mu$ and
$\mfw^{\mu \nu \lambda}$, respectively. On the other hand, the other terms
$\sim \tc^{(l)}_i$ in Eqs.\ \eqref{129a} -- \eqref{qmunulambda} are
of order $\sim (\Delta/L_{\text{hydro}}) (\lambda_{\text{mfp}}/\ell_{\text{vort}})$ times 
the dimensions of $\mft^\mu$ and $\mfw^{\mu \nu \lambda}$, respectively.
While this is formally of order $\text{Kn}^2$, cf.\ Eq.\ \eqref{samepc}, we cannot
simply neglect this term, as one would usually do for the Navier-Stokes limit.
The reason is that in our power-counting scheme we were not forced to specify the ratio 
$\lambda_{\text{mfp}}/\ell_{\text{vort}}$, such that in principle it can be of order unity.
Then, the terms $\sim \tc^{(l)}_i$ are of the same order as the first
terms $\sim \mathfrak{e}^{(0)},\, \mathfrak{d}^{(2)}$. 
Only if $\lambda_{\text{mfp}}/\ell_{\text{vort}} \ll 1$, we may drop these terms.

We note that, when inserting \eq\eqref{qmunulambda} into \eq\eqref{Pilocal}, we
obtain a term 
\begin{equation}
\Pi^\mu(p) \ni \int d\Sigma_{\lambda} p^\lambda f_{0p}  
\chi_\mfw \mathfrak{d}^{(2)} \beta_0  \epsilon^{\mu \beta \sigma\rho} u_\sigma \sigma_{\rho}^{\ \alpha}  
p_{\langle \alpha} p_{\beta \rangle}\;. \label{shearpol}
\end{equation}
This term has a similar structure as the coupling term
between spin and thermal shear obtained in 
Refs.\ \cite{Liu:2021uhn,Fu:2021pok,Becattini:2021suc,Becattini:2021iol}.
As we have just argued, this term arises from the nonlocal
collision term and thus is of order $\sim \Delta/L_{\text{hydro}}$
($\chi_\mfw$ is just a combination of thermodynamic integrals).
However, due to the fact that it arises from collisions we are hesitant to call such a term
nondissipative. Therefore, it may have a different origin in the approach of Refs.\ \cite{Liu:2021uhn,Fu:2021pok,Becattini:2021suc,Becattini:2021iol}. \bbb{We remark that the contribution \eqref{shearpol} to the local polarization  vanishes after integration over four-momentum, therefore it does not affect the global polarization \eqref{pibar}.}

The solution of \eqs\eqref{NSV1} is obtained analogously to the calculation outlined in 
Ref.~\cite{Denicol:2018rbw} as 
\begin{subequations} \label{solutionNS}
\begin{align}
\mfv^{\mu\nu}_\ns=& \left[2 \lambda_0 \proj^{\mu\nu\alpha\beta}+\lambda_1\left(\proj^{\mu\nu} 
-\frac32 \Xi^{\mu\nu}\right)\left(\proj^{\alpha\beta} -\frac32 \Xi^{\alpha\beta}\right)
-2\lambda_2\Xi^{\alpha(\mu}\hat{\omega}^{\nu)}\hat{\omega}^\beta
-2\lambda_3\Xi^{\alpha(\mu} \hat{\omega}^{\nu)\beta}
+2\lambda_4 \hat{\omega}^{\alpha(\mu} \hat{\omega}^{\nu)}\hat{\omega}^\beta\right.\n\\
&\left.+ \frac13\proj^{\mu\nu}\proj^{\alpha\beta}  \right] \mfv_{\nr\alpha\beta}\; ,\\
\mfw^{\lmur\nu\lambda}_\ns=&\left[2 \eta_0 \proj^{\nu\lambda\alpha\beta}
+\eta_1\left(\proj^{\nu\lambda} -\frac32 \Xi^{\nu\lambda}\right)\left(\proj^{\alpha\beta} 
-\frac32 \Xi^{\alpha\beta}\right)-2\eta_2\Xi^{\alpha(\nu}\hat{\omega}^{\lambda)}
\hat{\omega}^\beta-2\eta_3\Xi^{\alpha(\nu} \hat{\omega}^{\lambda)\beta}
+2\eta_4 \hat{\omega}^{\alpha(\nu} \hat{\omega}^{\lambda)}\hat{\omega}^\beta \right]\n\\
&\times \mfw^{\lmur}_{\nr\ \alpha\beta}\; ,
\end{align}
\end{subequations}
where we defined $\omega\equiv \sqrt{\omega^{\mu\nu}\omega_{\mu\nu}/2}$ and 
$\hat{\omega}^{\mu\nu}\equiv -\omega^{\mu\nu}/\omega$. Furthermore, we defined the unit vector 
along the vorticity direction $\hat{\omega}^\mu\equiv \omega^\mu/\omega$ and the projector 
orthogonal to both the fluid velocity and vorticity vector,
\begin{equation}
\Xi^{\mu\nu}\equiv g^{\mu\nu}-u^\mu u^\nu+\hat{\omega}^\mu \hat{\omega}^\nu\;.
\end{equation}
The coefficients in Eqs.\ (\ref{solutionNS}) are given in Appendix \ref{tcapp}.

\section{Conclusions}

In this paper, we derived the second-order dissipative equations of motion for relativistic spin 
hydrodynamics using the method of moments. The starting point was the quantum kinetic theory 
for massive spin-1/2 particles developed in Refs.\
\cite{Weickgenannt:2019dks,Weickgenannt:2020aaf,Weickgenannt:2021cuo}, which takes 
into account nonlocal collisions. We constructed relativistic spin hydrodynamics using the 
HW pseudo-gauge for the energy-momentum and spin tensors. In our framework, we treated the 
components of the HW spin tensor as the dynamical variables of the theory. Furthermore, we 
argued that the choice of the pseudo-gauge affects the evolution of the system, since in 
different pseudo-gauges different moments are treated dynamically. The equations of motion 
of the HW spin tensor correspond to conservation laws in global equilibrium, unlike in the
case of the canonical pseudo-gauge \cite{Speranza:2020ilk}. As a consequence, (at least some) spin dynamics 
must occur on large, i.e., hydrodynamic scales. 

In order to define our expansion, we proposed a novel power-counting scheme, which 
generalizes the concept of local equilibrium in the presence of spin dynamics with 
nonlocal collisions. We then extended the method of moments presented in 
Ref.\ \cite{Denicol:2012cn} to include spin dynamics. In particular, we expanded the distribution 
function around a local-equilibrium state in terms of the usual irreducible moments in 
momentum space \cite{Denicol:2012cn} and, in addition, of so-called spin moments 
containing the phase-space variable $\ms^\mu$. For the truncation we chose the 
``14+24-moment approximation", where ``14" corresponds to the usual components of the 
charge current and the energy-momentum tensor and ``24" to the components of the spin 
tensor. Our result is a closed set of equations of motion for the dynamical spin moments, 
where the latter approach their Navier-Stokes limits on time scales corresponding to 
characteristic relaxation times. Remarkably, the spin relaxation times are determined
by the local part of the collision term in the Boltzmann equation.
On the other hand, the nonlocal part, which is
responsible for the mutual conversion of orbital angular momentum and spin, 
gives rise to terms which appear
in the same way as the conventional Navier-Stokes terms, albeit their
power counting is different: they are of order $\Delta/L_{\text{hydro}}$, not of order Kn.
Moreover, we find that the spin relaxation times are 
comparable in magnitude to the relaxation times of the conventional dissipative quantities 
such as bulk, shear stress, and particle diffusion. This implies that it is reasonable to treat  
spin as dynamical variable in relativistic second-order spin hydrodynamics. Finally, we  
gave an expression for the Pauli-Lubanski vector which takes into account dissipative spin 
effects. In the Navier-Stokes limit, we obtained a
coupling term between spin and the shear-stress tensor, similar as in
Refs.\ \cite{Liu:2021uhn,Fu:2021pok,Becattini:2021suc,Becattini:2021iol}, although
the origin of this term in our approach may be different. 

Our work establishes a theory of relativistic second-order dissipative spin hydrodynamics for
applications in heavy-ion collisions as well as astrophysics. In particular, it 
can be used to solve and understand the puzzle related to the longitudinal 
polarization of Lambda particles. 
The equations of motion derived in this work provide the starting point for a numerical 
implementation of relativistic spin hydrodynamics. 
In this respect, a crucial future task is to analyze the conditions under which
our theory of spin hydrodynamics is causal and stable. This is challenging because of the 
larger number of variables and equations of motion of relativistic spin hydrodynamics 
compared to the conventional case, and due to the presence of vorticity fields 
\cite{Speranza:2021bxf}. It will also be interesting to investigate how the choice of different 
hydrodynamic frames (i.e., different matching conditions) \cite{Bemfica:2017wps,Bemfica:2019knx,Bemfica:2020zjp,Kovtun:2019hdm,Hoult:2020eho,Speranza:2021bxf,Noronha:2021syv} 
affect the theory presented here as far as its causality and stability is concerned.
For a study regarding conventional relativistic hydrodynamics using the method of moments 
with general matching conditions see Ref.~\cite{Rocha:2021lze}.

\section*{Acknowledgments}

The work of D.H.R., D.W., and N.W.\ is supported by the
Deutsche Forschungsgemeinschaft (DFG, German Research Foundation)
through the Collaborative Research Center TransRegio
CRC-TR 211 ``Strong- interaction matter under extreme conditions'' -- project number
315477589 -- TRR 211 and by the State of Hesse within the Research Cluster
ELEMENTS (Project ID 500/10.006).
D.W.\ acknowledges support by the Studienstiftung des deutschen Volkes 
(German Academic Scholarship Foundation).

\begin{appendix}

\section{Properties of irreducible tensors}
\label{itapp}

In the calculations in Sec.\ \ref{geomsec} we make use of the following relations 
\cite{coope1970irreducible}:
\begin{align}
p^\mu&= E_p u^\mu+p^{\langle \mu \rangle}\;,\\
\proj^{\alpha\beta}p_\alpha p_\beta &=m^2-E_p^2\;,\\
p^{\langle \mu\rangle} p^{\langle \nu \rangle} &= p^{\langle \mu} p^{\nu\rangle} 
+\frac13(m^2-E_p^2) \Delta^{\mu\nu}\;,\\
p^{\langle \mu\rangle} p^{\langle \nu \rangle} p^{\langle \lambda\rangle}&=p^{\langle \mu} p^{\nu} p^{\lambda\rangle} +\frac15(m^2-E_p^2)\left(p^{\langle\mu\rangle}\proj^{\nu\lambda}+p^{\langle \nu\rangle}\proj^{\lambda\mu}+p^{\langle \lambda\rangle}\proj^{\mu\nu}\right) \; , \\
p^{\langle \mu\rangle} p^{\langle \nu \rangle} p^{\langle \lambda\rangle}p^{\langle\rho\rangle}&=p^{\langle \mu} p^{\nu} p^{\lambda} p^{\rho\rangle} +\frac17 \left(  p^{\langle\mu\rangle} p^{\langle\nu\rangle}\proj^{\lambda\rho} + p^{\langle\mu\rangle} p^{\langle\lambda\rangle}\proj^{\nu\rho}+p^{\langle\mu\rangle} p^{\langle\rho\rangle}\proj^{\lambda\nu}+p^{\langle\lambda\rangle} p^{\langle\nu\rangle}\proj^{\mu\rho}+p^{\langle\rho\rangle} p^{\langle\nu\rangle}\proj^{\lambda\mu}\right.\n\\
&\left.+p^{\langle\lambda\rangle} p^{\langle\rho\rangle}\proj^{\mu\nu} \right)(m^2-E_p^2)-\frac{1}{35}\left(\proj^{\mu\nu}\proj^{\lambda\rho}+\proj^{\mu\lambda}\proj^{\nu\rho}+\proj^{\mu\rho}\proj^{\nu\lambda} \right) (m^2-E_p^2)^2 \n\\
 &=p^{\langle \mu} p^{\nu} p^{\lambda} p^{\rho\rangle} +\frac17 \left(  p^{\langle\mu} p^{\nu\rangle}\proj^{\lambda\rho} + p^{\langle\mu} p^{\lambda\rangle}\proj^{\nu\rho}+p^{\langle\mu} p^{\rho\rangle}\proj^{\lambda\nu}+p^{\langle\lambda} p^{\nu\rangle}\proj^{\mu\rho}+p^{\langle\rho} p^{\nu\rangle}\proj^{\lambda\mu}\right.\n\\
&\left.+p^{\langle\lambda} p^{\rho\rangle}\proj^{\mu\nu} \right)(m^2-E_p^2)+\frac{1}{15}\left(\proj^{\mu\nu}\proj^{\lambda\rho}+\proj^{\mu\lambda}\proj^{\nu\rho}+\proj^{\mu\rho}\proj^{\nu\lambda} \right) (m^2-E_p^2)^2 \; ,  \\
p^{\langle \mu\rangle} p^{\langle \nu \rangle} p^{\langle \lambda\rangle}p^{\langle\rho\rangle} p^{\langle\sigma\rangle}&=p^{\langle \mu} p^{\nu} p^{\lambda} p^{\rho} p^{\sigma\rangle}+\frac19 (m^2-E_p^2)\left(p^{\langle\mu\rangle}p^{\langle\nu\rangle} p^{\langle\lambda\rangle} \proj^{\rho\sigma}+\text{perm.}\right)\n\\
&-\frac{1}{63}(m^2-E_p^2)^2\left(p^{\langle\mu\rangle}\proj^{\nu\lambda}\proj^{\rho\sigma}+\text{perm}\right)\n\\
&=p^{\langle \mu} p^{\nu} p^{\lambda} p^{\rho} p^{\sigma\rangle}+\frac19 (m^2-E_p^2)\left(p^{\langle\mu}p^{\nu} p^{\lambda\rangle} \proj^{\rho\sigma}+\text{perm.}\right)\n\\
&+\frac{1}{35}(m^2-E_p^2)^2\left(p^{\langle\mu\rangle}\proj^{\nu\lambda}\proj^{\rho\sigma}+\text{perm.}\right)\; ,\\
\end{align}
where "perm." denotes all distinct permutations of indices. We also utilize the orthogonality relation 
\begin{equation}
\int dP\, p^{\langle\mu_1}\cdots p^{\mu_m\rangle} p_{\langle \nu_1}\cdots p_{\nu_n\rangle} F(E_p) = \frac{m!\delta_{mn}}{(2m+1)!!} \proj^{\mu_1\cdots \mu_m}_{\nu_1\cdots\nu_m}\int dP\, \left( \proj^{\alpha\beta} p_\alpha p_\beta\right)^m  F(E_p) \label{orthrel}
\end{equation}
for an arbitrary function $F(E_p)$.

\section{Scattering matrix elements}
\label{meapp}

In this appendix we show some details of the calculation of the scattering matrix elements in the 
collision term. The vacuum scattering matrix elements are defined as 
\cite{siskens1977transition,DeGroot:1980dk}
\begin{equation}
\langle{p,p^\prime;r,r^\prime|t|p_1,p_2;s_1,s_2}\rangle= \langle{p,p^\prime;r,r^\prime|: {H}_I(0):
|p_1,p_2;s_1,s_2}\rangle\;,
\end{equation}
where $H_I$ is the effective interaction Hamiltonian, which is here taken to be of an NJL-type form 
\cite{Nambu:1961tp,Nambu:1961fr}
\begin{equation}
H_I(x)= G\, \bar{\psi}(x) \Gamma^a \psi(x)\, \bar{\psi}(x) \Gamma_a \psi(x)\;,
\end{equation}
where $G$ is a coupling constant and $\Gamma_a$ is in general a linear combination of
elements of the Clifford algebra in a given representation of the Lorentz group, e.g., for
scalar interactions $\Gamma^a = 1$, for vector interactions $\Gamma^a  = \gamma^\mu$, 
and for parity-violating interactions $\Gamma^a = (1- \gamma_5) \gamma^\mu$.
Inserting the free-field expansion of the spinors, 
\begin{equation}
\label{psin}
\psi(x)=\sqrt{\frac{2}{(2\pi\hbar)^3}}\sum_{r} \int dP \,
e^{-\frac i\hbar p\cdot x} u_r(p) a_{r}(p)  \; ,
\end{equation}
and making use of the anticommutation relation of the creation and annihilation operators
\begin{equation}
\{ a_r(p), a^\dagger_s(p^\prime)\}= p^0 \delta^{(3)}(\mathbf{p}-\mathbf{p}^\prime) \delta_{rs}\;,
\end{equation}
we find
\begin{align}
\langle{p,p^\prime;r,r^\prime|t|p_1,p_2;r_1,r_2}\rangle=& \bar{G} \left[ \bar{u}_r(p) 
\Gamma^a u_{r_2}(p_2) \bar{u}_{r^\prime}(p^\prime)\Gamma_a u_{r_1}(p_1)
- \bar{u}_r(p) \Gamma^a u_{r_1}(p_1) \bar{u}_{r^\prime}(p^\prime)\Gamma_a u_{r_2}(p_2)\right]
\;,
\end{align}
with $\bar{G}\equiv 8/(2\pi\hbar)^6 G $. For \eq\eqref{wme} we then obtain using 
\eq\eqref{local_col_GLW_after}
\begin{align}
\int [dS]\, \mathcal{W}
=& |\bar{G}|^2 \delta^{(4)}(p+p^\prime-p_1-p_2)\,  \left[ \bar{u}_r(p) \Gamma^a u_{r_2}(p_2) 
\bar{u}_{r^\prime}(p^\prime)\Gamma_a u_{r_1}(p_1)- \bar{u}_r(p) \Gamma^a u_{r_1}(p_1) 
\bar{u}_{r^\prime}(p^\prime)\Gamma_a u_{r_2}(p_2)\right]\n\\
&\times  \left[ \bar{u}_r(p) \Gamma^b u_{r_2}(p_2) \bar{u}_{r^\prime}(p^\prime)\Gamma_b 
u_{r_1}(p_1)- \bar{u}_r(p) \Gamma^b u_{r_1}(p_1) \bar{u}_{r^\prime}(p^\prime)\Gamma_b 
u_{r_2}(p_2)\right]^\dagger\n\\
=& |\bar{G}|^2 \delta^{(4)}(p+p^\prime-p_1-p_2)\, \bigg\{ \Tr\left[\left(\slashed{p}+m\right)
\Gamma^a\left(\slashed{p}_2+m\right)\Gamma^b \right] \Tr\left[\left(\slashed{p}^\prime+m\right)
\Gamma_a\left(\slashed{p}_1+m\right)\Gamma_b \right]\n\\
& -\Tr \left[\left(\slashed{p}+m\right)\Gamma^a \left(\slashed{p}_2+m\right)\Gamma_b 
\left(\slashed{p}^\prime+m\right)\Gamma_a \left(\slashed{p}_1+m\right)\Gamma^b\right] \n\\
& - \Tr \left[\left(\slashed{p}+m\right)\Gamma^a \left(\slashed{p}_1+m\right)\Gamma_b 
\left(\slashed{p}^\prime+m\right)\Gamma_a \left(\slashed{p}_2+m\right)\Gamma^b\right]  \n\\
&+\Tr\left[\left(\slashed{p}+m\right)\Gamma^a\left(\slashed{p}_1+m\right)\Gamma^b \right] 
\Tr\left[\left(\slashed{p}^\prime+m\right)\Gamma_a\left(\slashed{p}_2+m\right)\Gamma_b \right] 
\bigg\} \;. \label{wmeap}
\end{align}
Furthermore we have for \eq\eqref{wsme}
\begin{align}
\int [dS]\, \ms^\alpha \mathcal{W}
=&- |\bar{G}|^2 \delta^{(4)}(p+p^\prime-p_1-p_2)\,  \left[ \bar{u}_r(p) \Gamma^a u_{r_2}(p_2) 
\bar{u}_{r^\prime}(p^\prime)\Gamma_a u_{r_1}(p_1)- \bar{u}_r(p) \Gamma^a u_{r_1}(p_1) 
\bar{u}_{r^\prime}(p^\prime)\Gamma_a u_{r_2}(p_2)\right]\n\\
&\times  \left[ \bar{u}_s(p) \Gamma^b u_{r_2}(p_2) \bar{u}_{r^\prime}(p^\prime)\Gamma_b
u_{r_1}(p_1)- \bar{u}_s(p) \Gamma^b u_{r_1}(p_1) \bar{u}_{r^\prime}(p^\prime)\Gamma_b 
u_{r_2}(p_2)\right]^\dagger  \frac{1}{2m} \bar{u}_s(p) \gamma^5 \gamma^\alpha u_r(p) \n\\
=& -\frac{|\bar{G}|^2}{2m} \delta^{(4)}(p+p^\prime-p_1-p_2)\, \bigg\{ \Tr\left[\left(\slashed{p}+m\right)
\gamma^5\gamma^\alpha \left(\slashed{p}+m\right)\Gamma^a\left(\slashed{p}_2+m\right)
\Gamma^b \right]\n\\
&\times\Tr\left[\left(\slashed{p}^\prime+m\right)\Gamma_a\left(\slashed{p}_1+m\right)
\Gamma_b \right]\n\\
& -\Tr \left[\left(\slashed{p}+m\right)\gamma^5\gamma^\alpha\left(\slashed{p}+m\right)\Gamma^a 
\left(\slashed{p}_2+m\right)\Gamma_b \left(\slashed{p}^\prime+m\right)\Gamma_a 
\left(\slashed{p}_1+m\right)\Gamma^b\right]  \n\\
& - \Tr \left[\left(\slashed{p}+m\right)\gamma^5\gamma^\alpha\left(\slashed{p}+m\right)\Gamma^a 
\left(\slashed{p}_1+m\right)\Gamma_b \left(\slashed{p}^\prime+m\right)\Gamma_a 
\left(\slashed{p}_2+m\right)\Gamma^b\right]  \n\\
&+\Tr\left[\left(\slashed{p}+m\right)\gamma^5\gamma^\alpha \left(\slashed{p}+m\right)
\Gamma^a\left(\slashed{p}_1+m\right)\Gamma^b \right] \Tr\left[\left(\slashed{p}^\prime+m\right)
\Gamma_a\left(\slashed{p}_2+m\right)\Gamma_b \right] \bigg\}\;,\label{wsmeap}
\end{align}
and for \eq\eqref{wssme}
\begin{align}
\int [dS]\, \ms^\alpha \ms^{\prime\beta} \mathcal{W}
=& \frac{|\bar{G}|^2}{(2m)^2} \delta^{(4)}(p+p^\prime-p_1-p_2)\, \bigg\{ \Tr\left[\left(\slashed{p}+m\right)
\gamma^5\gamma^\alpha \left(\slashed{p}+m\right)\Gamma^a\left(\slashed{p}_2+m\right)
\Gamma^b \right] \n\\
&\times\Tr\left[\left(\slashed{p}^\prime+m\right)\gamma^5\gamma^\beta 
\left(\slashed{p}^\prime+m\right)\Gamma_a\left(\slashed{p}_1+m\right)\Gamma_b \right]\n\\
& -\Tr \left[\left(\slashed{p}+m\right)\gamma^5\gamma^\alpha\left(\slashed{p}+m\right)\Gamma^a 
\left(\slashed{p}_2+m\right)\Gamma_b \left(\slashed{p}^\prime+m\right) \gamma^5\gamma^\beta 
\left(\slashed{p}^\prime+m\right)\Gamma_a \left(\slashed{p}_1+m\right)\Gamma^b\right]  \n\\
& - \Tr \left[\left(\slashed{p}+m\right)\gamma^5\gamma^\alpha\left(\slashed{p}+m\right)\Gamma^a 
\left(\slashed{p}_1+m\right)\Gamma_b \left(\slashed{p}^\prime+m\right)
\gamma^5\gamma^\beta\left(\slashed{p}^\prime+m\right)\Gamma_a \left(\slashed{p}_2+m\right)
\Gamma^b\right]  \n\\
&+\Tr\left[\left(\slashed{p}+m\right)\gamma^5\gamma^\alpha \left(\slashed{p}+m\right)
\Gamma^a\left(\slashed{p}_1+m\right)\Gamma^b \right] \Tr\left[\left(\slashed{p}^\prime+m\right)
\gamma^5\gamma^\beta\left(\slashed{p}^\prime+m\right)\Gamma_a\left(\slashed{p}_2+m\right)
\Gamma_b \right] \bigg\}\;.\label{wssmeap}
\end{align}

In the case of a scalar intercation, $\Gamma^a=1$, evaluating the traces of Dirac matrices yields that 
\eqs\eqref{wsmeap} and \eqref{wssmeap} vanish, respectively. For a vector interaction, 
$\Gamma^a=\gamma^\mu$,  \eq\eqref{wsmeap} also vanishes, while \eq\eqref{wssmeap} is nonzero. 
However, in the limit of small momentum transfer 
($s = 4 m^2$, $t=u=0$) \eq\eqref{wssmeap} is zero also for 
the vector interaction, while the only nonzero contribution comes from \eq\eqref{wmeap}.

\section{Transport coefficients}
\label{tcapp}

The relaxation times and transport coefficients in \eq\eqref{eomqqq} are given by 
\begin{subequations}
\begin{align}
\tau_{\mft}&= \mathfrak{T}_{00}^{(0)}\;,\\
\mathfrak{e}^{(0)}&= \tau_{\mft}\,g_0^{(0)}\;, \label{e0}\\
\tc^{(0)}_{\theta\omega}&= \frac{\hbar}{2m}\tau_{\mft}\,\xi_0^{(0)}\;,\\
\tc^{(0)}_{\theta\omega\Pi}&= \frac{\hbar}{2m}\, \frac{G_{21}}{D_{20}}\tau_{\mft}
=-\tc^{(0)}_{\pi\sigma\omega}\;,\\
\tc^{(0)}_{n\omega}&= -\frac{\hbar}{2m}\, \frac{G_{30}}{D_{20}}\tau_{\mft}\;,\\
\tc^{(0)}_{I\Omega}&= -\frac{\hbar}{4m}I_{11}\tau_{\mft}
=\tc^{(0)}_{\nabla\Omega}\;,\\
\tc^{(0)}_{\Pi\Omega}&= \frac{\hbar}{4m} 
\frac{\beta_0I_{21}}{\epsilon_0+P_0}\tau_{\mft}\;,\\
\tc^{(0)}_{\dot{\Omega}}&= -\frac{\hbar}{4m}I_{10}\tau_{\mft}\;,\\
\tct^{(0)}_1&=   \frac12   \tau_{\mft}
\frac{\partial \mathfrak{F}_{-11}^{(1)}}{\partial\beta_0} \frac{\beta_0}{\epsilon_0+P_0}\;,\\
\tct^{(0)}_2&=   -  \tau_{\mft}  \mathfrak{F}_{-20}^{(2)}\;,\\
\tct^{(0)}_3&=  \frac12 \tau_{\mft} \mathfrak{F}_{-11}^{(1)}\;,\\
\tct^{(0)}_4&= -\frac19 \tau_{\mft} \left(7-4m^2\mathfrak{F}_{-20}^{(0)}
+2m^4\mathfrak{F}_{-22}^{(0)} \right)\;,\\
\tct^{(0)}_5&= \frac13 \tau_{\mft} \left(2 m^2\mathfrak{F}_{-22}^{(0)} 
-\mathfrak{F}_{-20}^{(2)} \right)\;,\\
\tct^{(0)}_6&=-  \frac12\tau_{\mft}\left(  \frac{\partial 
\mathfrak{F}_{-11}^{(1)}}{\partial\alpha_0}+  \frac{\partial 
\mathfrak{F}_{-11}^{(1)}}{\partial\beta_0}\frac{n_0}{\epsilon_0+P_0} \right)\;,\\
\tct_7^{(0)} &= \frac13  \tau_{\mft} \left(m^2 \mathfrak{F}_{-20}^{(0)}
-1-\frac12m^4 \mathfrak{F}_{-22}^{(0)} \right)\;,\\
\tct_8^{(0)} &=   \tau_{\mft} \left( \mathfrak{F}_{-20}^{(2)}
-\frac12m^2 \mathfrak{F}^{(0)}_{-22} \right)\;,\\
\tct_9^{(0)}&= \frac{1}{2(\epsilon_0+P_0)}   \tau_{\mft} \mathfrak{F}^{(1)}_{-11}\;.
\end{align}
\end{subequations}
Furthermore, the relaxation time and transport coefficients of \eq\eqref{mfveom2nd} read
\begin{subequations}
\begin{align}
\tau_\mfv&=   \mathfrak{T}_{11}^{(1)}\; , \\
\tc_{\omega\Pi}^{(1)}&={\frac{\hbar}{2m}} \frac{\beta_0 I_{41}}{\epsilon_0+P_0} 
 \tau_\mfv\; , \\
\tc_{\omega I}^{(1)}&={\frac{\hbar}{2m}} \frac{\beta_0 I_{31}}{\epsilon_0+P_0} 
\tau_\mfv \; , \\
\tc_{\Omega\sigma}^{(1)}&=-{\frac{\hbar}{2m}} \beta_0 I_{42} \tau_\mfv \; , \\
\tc_{\nabla\Omega}^{(1)}&={\frac{\hbar}{4m}}  I_{31} \tau_\mfv\; ,\\
\tct_1^{(1)}&=-\frac43  \tau_\mfv \; , \\
\tct_2^{(1)}&=-  \tau_\mfv\; , \\
\tct_{3}^{(1)}&= \frac15  \tau_\mfv\, \left( m^2\mathfrak{F}_{-11}^{(1)}-1\right)\; ,\\
\tct_4^{(1)}&=   \tau_\mfv\,\frac{1}{\epsilon_0+P_0}\; , \\
\tct_5^{(1)}&= -\frac{1}{2}  \tau_\mfv\, \; , \\
\tct_6^{(1)}&= \frac32 \frac{m^2}{(\epsilon_0+P_0)}  \tau_\mfv \; ,\\
\tct_7^{(1)}&=\frac72 \frac{1}{\epsilon_0+P_0}\tau_\mfv\; ,\\
\tct_8^{(1)}&=-\frac12m^2 \tau_\mfv \; , \\
\tct_9^{(1)}&=-\frac12  \tau_\mfv\; , \\
\tct_{10}^{(1)}&= \frac16  \tau_\mfv
\left(m^2\mathfrak{F}_{-11}^{(1)}-1\right)\; .
\end{align}
\end{subequations}
Finally, the relaxation time and transport coefficients in \eq\eqref{eommfw2nd} are
\begin{subequations}
\begin{align}
\tau_\mfw&= \mathfrak{T}_{00}^{(2)}\; ,\\
\mathfrak{d}^{(2)}&= \tau_\mfw\,h_0^{(2)}\; , \label{d2}\\
\tc_{\Omega I}^{(2)}&=\frac{\hbar}{2 m} \tau_\mfw\,\xi_0^{(2)}\; ,\\
\tc_{\nabla\Omega}^{(2)}&={\frac{\hbar}{2 m}}  I_{32}\tau_\mfw\; ,\\
\tc_{\omega\sigma}^{(2)}&= \frac{\hbar}{2 m}  \beta_0 I_{42}\tau_\mfw\; ,\\
\tc_{\Omega\Pi}^{(2)}&= \frac{\hbar}{2 m} 
\frac{\beta_0I_{42}}{\epsilon_0+P_0} \tau_\mfw\; ,\\
\tct_1^{(2)}&= \frac15\tau_\mfw \frac{1}{\epsilon_0+P_0} 
\left(m^2\mathfrak{F}^{(1)}_{-11}-5\mathfrak{F}^{(1)}_{11}+m^2 
\frac{\partial\mathfrak{F}^{(1)}_{-11}}{\partial\beta_0} \beta_0\right)\; ,\\
\tct_2^{(2)}&=  -\frac{m^2}{5} \tau_\mfw
\left( \frac{\partial\mathfrak{F}^{(1)}_{-11}}{\partial\alpha_0}+\frac{n_0}{\epsilon_0+P_0} 
  \frac{\partial\mathfrak{F}^{(1)}_{-11}}{\partial\beta_0} \right)\; ,\\
\tct_3^{(2)}&= \frac15 \tau_\mfw \left( \mathfrak{F}^{(1)}_{11}
-m^2 \mathfrak{F}^{(1)}_{-11} \right)\; ,\\
\tct_4^{(2)}&= \frac13\tau_\mfw\left(-m^2\mathfrak{F}^{(2)}_{-20}-4 \right)\; ,\\
\tct_5^{(2)}&= \frac27  \tau_\mfw \left(-2m^2 \mathfrak{F}^{(2)}_{-20}-5\right)\; ,\\
\tct_6^{(2)}&= \frac{2m^2}{15}  \tau_\mfw
\left(\frac12m^4 \mathfrak{F}^{(0)}_{-22}-m^2 \mathfrak{F}^{(0)}_{-20}- 5 \right)\; ,\\
\tct_7^{(2)}&=  \frac{2}{15} \tau_\mfw 
\left(1-\frac{m^4}{4} \mathfrak{F}^{(0)}_{-22}\right)\; ,\\
\tct_8^{(2)}&= \frac{1}{5(\epsilon_0+P_0)} \tau_\mfw
\left(m^2\mathfrak{F}^{(1)}_{-11}-1 \right)\; ,\\
\tct_9^{(2)}&= -\frac{2}{15}\tau_\mfw \left(\frac52 m^2
-m^4\mathfrak{F}^{(2)}_{-20}+\frac12m^6\mathfrak{F}^{(0)}_{-22} \right)\; ,\\
\tct_{10}^{(2)}&= \frac25 \tau_\mfw\left(-\frac32
+m^2\mathfrak{F}^{(2)}_{-20}
-\frac12m^4 \mathfrak{F}^{(0)}_{-22} \right) \; .
\end{align}
\end{subequations}

Furthermore, the coefficients of the Navier-Stokes values in Sec.\ \ref{NSLspin} read
\begin{subequations}
\begin{align}
\lambda_0& = \frac{1}{1+4(\tau_\mfv \omega)^2}\; ,\\
\lambda_1&= \frac{16}{3} (\tau_\mfv\omega)^2\, \lambda_0 \; ,\\
\lambda_2&= (\tau_\mfv\omega)^2 \left( \lambda_0 + \lambda_4 \right)\; ,\\
\lambda_3&=  \tau_\mfv\omega\, \lambda_0\; ,\\
\lambda_4&= -\frac{\tau_\mfv\omega}{1+(\tau_\mfv \omega)^2}\; ,\\
\eta_0& = \frac{1}{1+4(\tau_\mfw \omega)^2} \; ,\\
\eta_1&= \frac{16}{3} (\tau_\mfw\omega)^2\, \eta_0 \; ,\\
\eta_2&= (\tau_\mfw\omega)^2 \left( \eta_0 + \eta_4\right)\; ,\\
\eta_3&=   \tau_\mfw\omega\, \eta_0\; ,\\
\eta_4&= -\frac{\tau_\mfw\omega}{1+(\tau_\mfw \omega)^2} \;.
\end{align}
\end{subequations}

\section{Collision integrals}
\label{ciapp}

\subsection{Reducing tensor structures through orthogonality relations}

Here we explain how to obtain \eqs\eqref{mcbtau} and \eqref{105}. 
Consider an integral of the form \eqref{bmanymunu}
\begin{equation}
 \int [dP] \mathcal{W} f_{0p} f_{0p^\prime} E_p^{r-1} p^{\langle\mu_1}\cdots p^{\mu_l\rangle}  
 \mathcal{H}_{pn}^{(m)}p_{\langle\nu_1}\cdots p_{\nu_m\rangle}\; ,
\end{equation}
where $\mathcal{W}$ is a function of $p$, $p^\prime$, $p_1$, and $p_2$ and $\mathcal{H}_{pn}^{(m)}$ 
can be expressed as a polynomial of $E_p$. After integration over $p^\prime$, $p_1$, and $p_2$ the 
result must assume the form 
\begin{equation}
\int dP\, p^{\langle\mu_1} \cdots p^{\mu_m\rangle} p_{\langle \nu_1} \cdots p_{\nu_n\rangle} F(E_p) \;,
\end{equation}
and one can apply \eq\eqref{orthrel}. An analogous argument can be used 
to simplify \eq\eqref{waisdjd}.

\subsection{Collision integrals for constant cross section}

Here we show the calculation of the collision integrals used to compute the relaxation times in Sec.\
\ref{1424}. The procedure is similar to the one presented in Ref.~\cite{Denicol:2012cn}, however, we 
do not consider the ultrarelativistic limit, since we intend to describe massive particles. 
The matrix $B$ in \eq\eqref{BDelB} is given as
\begin{equation}
B_{rn}^{(l)}= \frac{16}{2l+1} \int [dP] \mathcal{W}_0 f_{0p} f_{0p^\prime} E_p^{r-1} p^{\langle\mu_1}\cdots 
p^{\mu_l\rangle}  \mathcal{H}_{pn}^{(l)}p_{\langle\mu_1}\cdots p_{\mu_l\rangle}\; . \label{Bapp}
\end{equation}
The transition amplitude $\mathcal{W}_0$ is taken to be of the from
\begin{equation}
\mathcal{W}_0= \delta^{(4)}(p+p^\prime-p_1-p_2)\, s\, \sigma(s,\Theta)\;,
\end{equation}
where $s\equiv (p+p^\prime)^2$ is a Mandelstam variable and $\cos\Theta\equiv(p-p^\prime)\cdot
(p_1-p_2)/(p-p^\prime)^2$. Furthermore, we introduced the differential cross section 
$\sigma(s,\Theta)$. We also define the total cross section
\begin{equation}
\sigma_T\equiv 2\pi \int d\Theta\, \sin\Theta\, \sigma(s,\Theta)\;,
\end{equation}
which is here assumed to be constant.
First performing the $p_1$ and $p_2$ integrations in \eq\eqref{Bapp} in the center-of-momentum 
frame yields
\begin{align}
\int dP_1\, dP_2\, \mathcal{W}_0=\frac{1}{8}\sqrt{s} \sqrt{s-4m^2}\, \sigma_T \; .
\end{align}
We then insert \eqs\eqref{Hcoeffdef} and \eqref{polynomialE} to obtain
\begin{subequations} \label{Bapp2}
\begin{align}
B_{rn}^{(0)}={}& 2\sigma_T {w^{(0)}}\sum_{m \in \mathbb{S}_0, m \geq n} 
a_{mn}^{(0)} \sum_{q=0}^m 
a_{mq}^{(0)}  \int dP dP^\prime\, f_{0p} f_{0p^\prime} E_p^{r-1+q}    \sqrt{s} \sqrt{s-4m^2}\; ,\\
B_{rn}^{(1)}={}& 2 \sigma_T \frac{w^{(1)}}{3}\sum_{m \in \mathbb{S}_1, m \geq n} a_{mn}^{(1)} 
\sum_{q=0}^m a_{mq}^{(1)}  \int dP dP^\prime\, f_{0p} f_{0p^\prime} E_p^{r-1+q} 
p_{\langle\mu\rangle} p^\mu   \sqrt{s} \sqrt{s-4m^2}\; , \\
 B_{rn}^{(2)}={}&  \sigma_T \frac{w^{(2)}}{5}\sum_{m \in \mathbb{S}_2, m \geq n} a_{mn}^{(2)} 
 \sum_{q=0}^m  a_{mq}^{(2)}  \int dP dP^\prime\, f_{0p} f_{0p^\prime} E_p^{r-1+q} 
 p_{\langle\mu} p_{\nu \rangle} p^\mu p^\nu \sqrt{s} \sqrt{s-4m^2}\; .
\end{align}
\end{subequations}
The remaining integrals are then solved numerically.

\end{appendix}

\bibliography{biblio_paper_long}{}

\begin{thebibliography}{104}
\expandafter\ifx\csname natexlab\endcsname\relax\def\natexlab#1{#1}\fi
\expandafter\ifx\csname bibnamefont\endcsname\relax
  \def\bibnamefont#1{#1}\fi
\expandafter\ifx\csname bibfnamefont\endcsname\relax
  \def\bibfnamefont#1{#1}\fi
\expandafter\ifx\csname citenamefont\endcsname\relax
  \def\citenamefont#1{#1}\fi
\expandafter\ifx\csname url\endcsname\relax
  \def\url#1{\texttt{#1}}\fi
\expandafter\ifx\csname urlprefix\endcsname\relax\def\urlprefix{URL }\fi
\providecommand{\bibinfo}[2]{#2}
\providecommand{\eprint}[2][]{\url{#2}}

\bibitem[{\citenamefont{Florkowski
  et~al.}(2018{\natexlab{a}})\citenamefont{Florkowski, Friman, Jaiswal, and
  Speranza}}]{Florkowski:2017ruc}
\bibinfo{author}{\bibfnamefont{W.}~\bibnamefont{Florkowski}},
  \bibinfo{author}{\bibfnamefont{B.}~\bibnamefont{Friman}},
  \bibinfo{author}{\bibfnamefont{A.}~\bibnamefont{Jaiswal}}, \bibnamefont{and}
  \bibinfo{author}{\bibfnamefont{E.}~\bibnamefont{Speranza}},
  \bibinfo{journal}{Phys. Rev.} \textbf{\bibinfo{volume}{C97}},
  \bibinfo{pages}{041901} (\bibinfo{year}{2018}{\natexlab{a}}),
  \eprint{1705.00587}.

\bibitem[{\citenamefont{Florkowski
  et~al.}(2018{\natexlab{b}})\citenamefont{Florkowski, Friman, Jaiswal,
  Ryblewski, and Speranza}}]{Florkowski:2017dyn}
\bibinfo{author}{\bibfnamefont{W.}~\bibnamefont{Florkowski}},
  \bibinfo{author}{\bibfnamefont{B.}~\bibnamefont{Friman}},
  \bibinfo{author}{\bibfnamefont{A.}~\bibnamefont{Jaiswal}},
  \bibinfo{author}{\bibfnamefont{R.}~\bibnamefont{Ryblewski}},
  \bibnamefont{and} \bibinfo{author}{\bibfnamefont{E.}~\bibnamefont{Speranza}},
  \bibinfo{journal}{Phys. Rev.} \textbf{\bibinfo{volume}{D97}},
  \bibinfo{pages}{116017} (\bibinfo{year}{2018}{\natexlab{b}}),
  \eprint{1712.07676}.

\bibitem[{\citenamefont{Hidaka et~al.}(2018)\citenamefont{Hidaka, Pu, and
  Yang}}]{Hidaka:2017auj}
\bibinfo{author}{\bibfnamefont{Y.}~\bibnamefont{Hidaka}},
  \bibinfo{author}{\bibfnamefont{S.}~\bibnamefont{Pu}}, \bibnamefont{and}
  \bibinfo{author}{\bibfnamefont{D.-L.} \bibnamefont{Yang}},
  \bibinfo{journal}{Phys. Rev.} \textbf{\bibinfo{volume}{D97}},
  \bibinfo{pages}{016004} (\bibinfo{year}{2018}), \eprint{1710.00278}.

\bibitem[{\citenamefont{Florkowski
  et~al.}(2018{\natexlab{c}})\citenamefont{Florkowski, Speranza, and
  Becattini}}]{Florkowski:2018myy}
\bibinfo{author}{\bibfnamefont{W.}~\bibnamefont{Florkowski}},
  \bibinfo{author}{\bibfnamefont{E.}~\bibnamefont{Speranza}}, \bibnamefont{and}
  \bibinfo{author}{\bibfnamefont{F.}~\bibnamefont{Becattini}},
  \bibinfo{journal}{Acta Phys. Polon.} \textbf{\bibinfo{volume}{B49}},
  \bibinfo{pages}{1409} (\bibinfo{year}{2018}{\natexlab{c}}),
  \eprint{1803.11098}.

\bibitem[{\citenamefont{Weickgenannt et~al.}(2019)\citenamefont{Weickgenannt,
  Sheng, Speranza, Wang, and Rischke}}]{Weickgenannt:2019dks}
\bibinfo{author}{\bibfnamefont{N.}~\bibnamefont{Weickgenannt}},
  \bibinfo{author}{\bibfnamefont{X.-L.} \bibnamefont{Sheng}},
  \bibinfo{author}{\bibfnamefont{E.}~\bibnamefont{Speranza}},
  \bibinfo{author}{\bibfnamefont{Q.}~\bibnamefont{Wang}}, \bibnamefont{and}
  \bibinfo{author}{\bibfnamefont{D.~H.} \bibnamefont{Rischke}},
  \bibinfo{journal}{Phys. Rev.} \textbf{\bibinfo{volume}{D100}},
  \bibinfo{pages}{056018} (\bibinfo{year}{2019}), \eprint{1902.06513}.

\bibitem[{\citenamefont{Bhadury
  et~al.}(2021{\natexlab{a}})\citenamefont{Bhadury, Florkowski, Jaiswal, Kumar,
  and Ryblewski}}]{Bhadury:2020puc}
\bibinfo{author}{\bibfnamefont{S.}~\bibnamefont{Bhadury}},
  \bibinfo{author}{\bibfnamefont{W.}~\bibnamefont{Florkowski}},
  \bibinfo{author}{\bibfnamefont{A.}~\bibnamefont{Jaiswal}},
  \bibinfo{author}{\bibfnamefont{A.}~\bibnamefont{Kumar}}, \bibnamefont{and}
  \bibinfo{author}{\bibfnamefont{R.}~\bibnamefont{Ryblewski}},
  \bibinfo{journal}{Phys. Lett. B} \textbf{\bibinfo{volume}{814}},
  \bibinfo{pages}{136096} (\bibinfo{year}{2021}{\natexlab{a}}),
  \eprint{2002.03937}.

\bibitem[{\citenamefont{Weickgenannt
  et~al.}(2021{\natexlab{a}})\citenamefont{Weickgenannt, Speranza, Sheng, Wang,
  and Rischke}}]{Weickgenannt:2020aaf}
\bibinfo{author}{\bibfnamefont{N.}~\bibnamefont{Weickgenannt}},
  \bibinfo{author}{\bibfnamefont{E.}~\bibnamefont{Speranza}},
  \bibinfo{author}{\bibfnamefont{X.-l.} \bibnamefont{Sheng}},
  \bibinfo{author}{\bibfnamefont{Q.}~\bibnamefont{Wang}}, \bibnamefont{and}
  \bibinfo{author}{\bibfnamefont{D.~H.} \bibnamefont{Rischke}},
  \bibinfo{journal}{Phys. Rev. Lett.} \textbf{\bibinfo{volume}{127}},
  \bibinfo{pages}{052301} (\bibinfo{year}{2021}{\natexlab{a}}),
  \eprint{2005.01506}.

\bibitem[{\citenamefont{Shi et~al.}(2021)\citenamefont{Shi, Gale, and
  Jeon}}]{Shi:2020htn}
\bibinfo{author}{\bibfnamefont{S.}~\bibnamefont{Shi}},
  \bibinfo{author}{\bibfnamefont{C.}~\bibnamefont{Gale}}, \bibnamefont{and}
  \bibinfo{author}{\bibfnamefont{S.}~\bibnamefont{Jeon}},
  \bibinfo{journal}{Phys. Rev. C} \textbf{\bibinfo{volume}{103}},
  \bibinfo{pages}{044906} (\bibinfo{year}{2021}), \eprint{2008.08618}.

\bibitem[{\citenamefont{Speranza and Weickgenannt}(2021)}]{Speranza:2020ilk}
\bibinfo{author}{\bibfnamefont{E.}~\bibnamefont{Speranza}} \bibnamefont{and}
  \bibinfo{author}{\bibfnamefont{N.}~\bibnamefont{Weickgenannt}},
  \bibinfo{journal}{Eur. Phys. J. A} \textbf{\bibinfo{volume}{57}},
  \bibinfo{pages}{155} (\bibinfo{year}{2021}), \eprint{2007.00138}.

\bibitem[{\citenamefont{Bhadury
  et~al.}(2021{\natexlab{b}})\citenamefont{Bhadury, Florkowski, Jaiswal, Kumar,
  and Ryblewski}}]{Bhadury:2020cop}
\bibinfo{author}{\bibfnamefont{S.}~\bibnamefont{Bhadury}},
  \bibinfo{author}{\bibfnamefont{W.}~\bibnamefont{Florkowski}},
  \bibinfo{author}{\bibfnamefont{A.}~\bibnamefont{Jaiswal}},
  \bibinfo{author}{\bibfnamefont{A.}~\bibnamefont{Kumar}}, \bibnamefont{and}
  \bibinfo{author}{\bibfnamefont{R.}~\bibnamefont{Ryblewski}},
  \bibinfo{journal}{Phys. Rev. D} \textbf{\bibinfo{volume}{103}},
  \bibinfo{pages}{014030} (\bibinfo{year}{2021}{\natexlab{b}}),
  \eprint{2008.10976}.

\bibitem[{\citenamefont{Singh et~al.}(2021)\citenamefont{Singh, Sophys, and
  Ryblewski}}]{Singh:2020rht}
\bibinfo{author}{\bibfnamefont{R.}~\bibnamefont{Singh}},
  \bibinfo{author}{\bibfnamefont{G.}~\bibnamefont{Sophys}}, \bibnamefont{and}
  \bibinfo{author}{\bibfnamefont{R.}~\bibnamefont{Ryblewski}},
  \bibinfo{journal}{Phys. Rev. D} \textbf{\bibinfo{volume}{103}},
  \bibinfo{pages}{074024} (\bibinfo{year}{2021}), \eprint{2011.14907}.

\bibitem[{\citenamefont{Bhadury
  et~al.}(2021{\natexlab{c}})\citenamefont{Bhadury, Bhatt, Jaiswal, and
  Kumar}}]{Bhadury:2021oat}
\bibinfo{author}{\bibfnamefont{S.}~\bibnamefont{Bhadury}},
  \bibinfo{author}{\bibfnamefont{J.}~\bibnamefont{Bhatt}},
  \bibinfo{author}{\bibfnamefont{A.}~\bibnamefont{Jaiswal}}, \bibnamefont{and}
  \bibinfo{author}{\bibfnamefont{A.}~\bibnamefont{Kumar}},
  \bibinfo{journal}{Eur. Phys. J. ST} \textbf{\bibinfo{volume}{230}},
  \bibinfo{pages}{655} (\bibinfo{year}{2021}{\natexlab{c}}),
  \eprint{2101.11964}.

\bibitem[{\citenamefont{Peng et~al.}(2021)\citenamefont{Peng, Zhang, Sheng, and
  Wang}}]{Peng:2021ago}
\bibinfo{author}{\bibfnamefont{H.-H.} \bibnamefont{Peng}},
  \bibinfo{author}{\bibfnamefont{J.-J.} \bibnamefont{Zhang}},
  \bibinfo{author}{\bibfnamefont{X.-L.} \bibnamefont{Sheng}}, \bibnamefont{and}
  \bibinfo{author}{\bibfnamefont{Q.}~\bibnamefont{Wang}},
  \bibinfo{journal}{Chin. Phys. Lett.} \textbf{\bibinfo{volume}{38}},
  \bibinfo{pages}{116701} (\bibinfo{year}{2021}), \eprint{2107.00448}.

\bibitem[{\citenamefont{Sheng et~al.}(2021)\citenamefont{Sheng, Weickgenannt,
  Speranza, Rischke, and Wang}}]{Sheng:2021kfc}
\bibinfo{author}{\bibfnamefont{X.-L.} \bibnamefont{Sheng}},
  \bibinfo{author}{\bibfnamefont{N.}~\bibnamefont{Weickgenannt}},
  \bibinfo{author}{\bibfnamefont{E.}~\bibnamefont{Speranza}},
  \bibinfo{author}{\bibfnamefont{D.~H.} \bibnamefont{Rischke}},
  \bibnamefont{and} \bibinfo{author}{\bibfnamefont{Q.}~\bibnamefont{Wang}},
  \bibinfo{journal}{Phys. Rev. D} \textbf{\bibinfo{volume}{104}},
  \bibinfo{pages}{016029} (\bibinfo{year}{2021}), \eprint{2103.10636}.

\bibitem[{\citenamefont{Sheng et~al.}(2022)\citenamefont{Sheng, Wang, and
  Rischke}}]{Sheng:2022ssd}
\bibinfo{author}{\bibfnamefont{X.-L.} \bibnamefont{Sheng}},
  \bibinfo{author}{\bibfnamefont{Q.}~\bibnamefont{Wang}}, \bibnamefont{and}
  \bibinfo{author}{\bibfnamefont{D.~H.} \bibnamefont{Rischke}}
  (\bibinfo{year}{2022}), \eprint{2202.10160}.

\bibitem[{\citenamefont{Hu}(2021)}]{Hu:2021pwh}
\bibinfo{author}{\bibfnamefont{J.}~\bibnamefont{Hu}} (\bibinfo{year}{2021}),
  \eprint{2111.03571}.

\bibitem[{\citenamefont{Hu}(2022)}]{Hu:2022lpi}
\bibinfo{author}{\bibfnamefont{J.}~\bibnamefont{Hu}} (\bibinfo{year}{2022}),
  \eprint{2202.07373}.

\bibitem[{\citenamefont{Fang et~al.}(2022)\citenamefont{Fang, Pu, and
  Yang}}]{Fang:2022ttm}
\bibinfo{author}{\bibfnamefont{S.}~\bibnamefont{Fang}},
  \bibinfo{author}{\bibfnamefont{S.}~\bibnamefont{Pu}}, \bibnamefont{and}
  \bibinfo{author}{\bibfnamefont{D.-L.} \bibnamefont{Yang}},
  \bibinfo{journal}{Phys. Rev. D} \textbf{\bibinfo{volume}{106}},
  \bibinfo{pages}{016002} (\bibinfo{year}{2022}), \eprint{2204.11519}.

\bibitem[{\citenamefont{Wang}(2022)}]{Wang:2022yli}
\bibinfo{author}{\bibfnamefont{Z.}~\bibnamefont{Wang}} (\bibinfo{year}{2022}),
  \eprint{2205.09334}.

\bibitem[{\citenamefont{Montenegro and Torrieri}(2019)}]{Montenegro:2018bcf}
\bibinfo{author}{\bibfnamefont{D.}~\bibnamefont{Montenegro}} \bibnamefont{and}
  \bibinfo{author}{\bibfnamefont{G.}~\bibnamefont{Torrieri}},
  \bibinfo{journal}{Phys. Rev. D} \textbf{\bibinfo{volume}{100}},
  \bibinfo{pages}{056011} (\bibinfo{year}{2019}), \eprint{1807.02796}.

\bibitem[{\citenamefont{Montenegro and Torrieri}(2020)}]{Montenegro:2020paq}
\bibinfo{author}{\bibfnamefont{D.}~\bibnamefont{Montenegro}} \bibnamefont{and}
  \bibinfo{author}{\bibfnamefont{G.}~\bibnamefont{Torrieri}},
  \bibinfo{journal}{Phys. Rev. D} \textbf{\bibinfo{volume}{102}},
  \bibinfo{pages}{036007} (\bibinfo{year}{2020}), \eprint{2004.10195}.

\bibitem[{\citenamefont{Gallegos et~al.}(2021)\citenamefont{Gallegos, G\"ursoy,
  and Yarom}}]{Gallegos:2021bzp}
\bibinfo{author}{\bibfnamefont{A.~D.} \bibnamefont{Gallegos}},
  \bibinfo{author}{\bibfnamefont{U.}~\bibnamefont{G\"ursoy}}, \bibnamefont{and}
  \bibinfo{author}{\bibfnamefont{A.}~\bibnamefont{Yarom}},
  \bibinfo{journal}{SciPost Phys.} \textbf{\bibinfo{volume}{11}},
  \bibinfo{pages}{041} (\bibinfo{year}{2021}), \eprint{2101.04759}.

\bibitem[{\citenamefont{Hattori et~al.}(2019)\citenamefont{Hattori, Hongo,
  Huang, Matsuo, and Taya}}]{Hattori:2019lfp}
\bibinfo{author}{\bibfnamefont{K.}~\bibnamefont{Hattori}},
  \bibinfo{author}{\bibfnamefont{M.}~\bibnamefont{Hongo}},
  \bibinfo{author}{\bibfnamefont{X.-G.} \bibnamefont{Huang}},
  \bibinfo{author}{\bibfnamefont{M.}~\bibnamefont{Matsuo}}, \bibnamefont{and}
  \bibinfo{author}{\bibfnamefont{H.}~\bibnamefont{Taya}},
  \bibinfo{journal}{Phys. Lett.} \textbf{\bibinfo{volume}{B795}},
  \bibinfo{pages}{100} (\bibinfo{year}{2019}), \eprint{1901.06615}.

\bibitem[{\citenamefont{Fukushima and Pu}(2021)}]{Fukushima:2020ucl}
\bibinfo{author}{\bibfnamefont{K.}~\bibnamefont{Fukushima}} \bibnamefont{and}
  \bibinfo{author}{\bibfnamefont{S.}~\bibnamefont{Pu}}, \bibinfo{journal}{Phys.
  Lett. B} \textbf{\bibinfo{volume}{817}}, \bibinfo{pages}{136346}
  (\bibinfo{year}{2021}), \eprint{2010.01608}.

\bibitem[{\citenamefont{Li et~al.}(2021)\citenamefont{Li, Stephanov, and
  Yee}}]{Li:2020eon}
\bibinfo{author}{\bibfnamefont{S.}~\bibnamefont{Li}},
  \bibinfo{author}{\bibfnamefont{M.~A.} \bibnamefont{Stephanov}},
  \bibnamefont{and} \bibinfo{author}{\bibfnamefont{H.-U.} \bibnamefont{Yee}},
  \bibinfo{journal}{Phys. Rev. Lett.} \textbf{\bibinfo{volume}{127}},
  \bibinfo{pages}{082302} (\bibinfo{year}{2021}), \eprint{2011.12318}.

\bibitem[{\citenamefont{She et~al.}(2021)\citenamefont{She, Huang, Hou, and
  Liao}}]{She:2021lhe}
\bibinfo{author}{\bibfnamefont{D.}~\bibnamefont{She}},
  \bibinfo{author}{\bibfnamefont{A.}~\bibnamefont{Huang}},
  \bibinfo{author}{\bibfnamefont{D.}~\bibnamefont{Hou}}, \bibnamefont{and}
  \bibinfo{author}{\bibfnamefont{J.}~\bibnamefont{Liao}}
  (\bibinfo{year}{2021}), \eprint{2105.04060}.

\bibitem[{\citenamefont{Wang et~al.}(2021{\natexlab{a}})\citenamefont{Wang,
  Fang, and Pu}}]{Wang:2021ngp}
\bibinfo{author}{\bibfnamefont{D.-L.} \bibnamefont{Wang}},
  \bibinfo{author}{\bibfnamefont{S.}~\bibnamefont{Fang}}, \bibnamefont{and}
  \bibinfo{author}{\bibfnamefont{S.}~\bibnamefont{Pu}}, \bibinfo{journal}{Phys.
  Rev. D} \textbf{\bibinfo{volume}{104}}, \bibinfo{pages}{114043}
  (\bibinfo{year}{2021}{\natexlab{a}}), \eprint{2107.11726}.

\bibitem[{\citenamefont{Wang et~al.}(2021{\natexlab{b}})\citenamefont{Wang,
  Xie, Fang, and Pu}}]{Wang:2021wqq}
\bibinfo{author}{\bibfnamefont{D.-L.} \bibnamefont{Wang}},
  \bibinfo{author}{\bibfnamefont{X.-Q.} \bibnamefont{Xie}},
  \bibinfo{author}{\bibfnamefont{S.}~\bibnamefont{Fang}}, \bibnamefont{and}
  \bibinfo{author}{\bibfnamefont{S.}~\bibnamefont{Pu}}
  (\bibinfo{year}{2021}{\natexlab{b}}), \eprint{2112.15535}.

\bibitem[{\citenamefont{Hongo et~al.}(2021)\citenamefont{Hongo, Huang,
  Kaminski, Stephanov, and Yee}}]{Hongo:2021ona}
\bibinfo{author}{\bibfnamefont{M.}~\bibnamefont{Hongo}},
  \bibinfo{author}{\bibfnamefont{X.-G.} \bibnamefont{Huang}},
  \bibinfo{author}{\bibfnamefont{M.}~\bibnamefont{Kaminski}},
  \bibinfo{author}{\bibfnamefont{M.}~\bibnamefont{Stephanov}},
  \bibnamefont{and} \bibinfo{author}{\bibfnamefont{H.-U.} \bibnamefont{Yee}},
  \bibinfo{journal}{JHEP} \textbf{\bibinfo{volume}{11}}, \bibinfo{pages}{150}
  (\bibinfo{year}{2021}), \eprint{2107.14231}.

\bibitem[{\citenamefont{Liang and Wang}(2005)}]{Liang:2004ph}
\bibinfo{author}{\bibfnamefont{Z.-T.} \bibnamefont{Liang}} \bibnamefont{and}
  \bibinfo{author}{\bibfnamefont{X.-N.} \bibnamefont{Wang}},
  \bibinfo{journal}{Phys. Rev. Lett.} \textbf{\bibinfo{volume}{94}},
  \bibinfo{pages}{102301} (\bibinfo{year}{2005}), \bibinfo{note}{[Erratum:
  Phys. Rev. Lett.96,039901(2006)]}, \eprint{nucl-th/0410079}.

\bibitem[{\citenamefont{Voloshin}(2004)}]{Voloshin:2004ha}
\bibinfo{author}{\bibfnamefont{S.~A.} \bibnamefont{Voloshin}}
  (\bibinfo{year}{2004}), \eprint{nucl-th/0410089}.

\bibitem[{\citenamefont{Betz et~al.}(2007)\citenamefont{Betz, Gyulassy, and
  Torrieri}}]{Betz:2007kg}
\bibinfo{author}{\bibfnamefont{B.}~\bibnamefont{Betz}},
  \bibinfo{author}{\bibfnamefont{M.}~\bibnamefont{Gyulassy}}, \bibnamefont{and}
  \bibinfo{author}{\bibfnamefont{G.}~\bibnamefont{Torrieri}},
  \bibinfo{journal}{Phys. Rev. C} \textbf{\bibinfo{volume}{76}},
  \bibinfo{pages}{044901} (\bibinfo{year}{2007}), \eprint{0708.0035}.

\bibitem[{\citenamefont{Becattini et~al.}(2008)\citenamefont{Becattini,
  Piccinini, and Rizzo}}]{Becattini:2007sr}
\bibinfo{author}{\bibfnamefont{F.}~\bibnamefont{Becattini}},
  \bibinfo{author}{\bibfnamefont{F.}~\bibnamefont{Piccinini}},
  \bibnamefont{and} \bibinfo{author}{\bibfnamefont{J.}~\bibnamefont{Rizzo}},
  \bibinfo{journal}{Phys. Rev.} \textbf{\bibinfo{volume}{C77}},
  \bibinfo{pages}{024906} (\bibinfo{year}{2008}), \eprint{0711.1253}.

\bibitem[{\citenamefont{Barnett}(1935)}]{Barnett:1935}
\bibinfo{author}{\bibfnamefont{S.~J.} \bibnamefont{Barnett}},
  \bibinfo{journal}{Rev. Mod. Phys.} \textbf{\bibinfo{volume}{7}},
  \bibinfo{pages}{129} (\bibinfo{year}{1935}).

\bibitem[{\citenamefont{Adamczyk et~al.}(2017)}]{STAR:2017ckg}
\bibinfo{author}{\bibfnamefont{L.}~\bibnamefont{Adamczyk}} \bibnamefont{et~al.}
  (\bibinfo{collaboration}{STAR}), \bibinfo{journal}{Nature}
  \textbf{\bibinfo{volume}{548}}, \bibinfo{pages}{62} (\bibinfo{year}{2017}),
  \eprint{1701.06657}.

\bibitem[{\citenamefont{Adam et~al.}(2018)}]{Adam:2018ivw}
\bibinfo{author}{\bibfnamefont{J.}~\bibnamefont{Adam}} \bibnamefont{et~al.}
  (\bibinfo{collaboration}{STAR}), \bibinfo{journal}{Phys. Rev.}
  \textbf{\bibinfo{volume}{C98}}, \bibinfo{pages}{014910}
  (\bibinfo{year}{2018}), \eprint{1805.04400}.

\bibitem[{\citenamefont{Acharya et~al.}(2020)}]{ALICE:2019aid}
\bibinfo{author}{\bibfnamefont{S.}~\bibnamefont{Acharya}} \bibnamefont{et~al.}
  (\bibinfo{collaboration}{ALICE}), \bibinfo{journal}{Phys. Rev. Lett.}
  \textbf{\bibinfo{volume}{125}}, \bibinfo{pages}{012301}
  (\bibinfo{year}{2020}), \eprint{1910.14408}.

\bibitem[{\citenamefont{Mohanty et~al.}(2021)\citenamefont{Mohanty, Kundu,
  Singha, and Singh}}]{Mohanty:2021vbt}
\bibinfo{author}{\bibfnamefont{B.}~\bibnamefont{Mohanty}},
  \bibinfo{author}{\bibfnamefont{S.}~\bibnamefont{Kundu}},
  \bibinfo{author}{\bibfnamefont{S.}~\bibnamefont{Singha}}, \bibnamefont{and}
  \bibinfo{author}{\bibfnamefont{R.}~\bibnamefont{Singh}},
  \bibinfo{journal}{Mod. Phys. Lett. A} \textbf{\bibinfo{volume}{36}},
  \bibinfo{pages}{2130026} (\bibinfo{year}{2021}), \eprint{2112.04816}.

\bibitem[{\citenamefont{Becattini
  et~al.}(2013{\natexlab{a}})\citenamefont{Becattini, Csernai, and
  Wang}}]{Becattini:2013vja}
\bibinfo{author}{\bibfnamefont{F.}~\bibnamefont{Becattini}},
  \bibinfo{author}{\bibfnamefont{L.}~\bibnamefont{Csernai}}, \bibnamefont{and}
  \bibinfo{author}{\bibfnamefont{D.~J.} \bibnamefont{Wang}},
  \bibinfo{journal}{Phys. Rev.} \textbf{\bibinfo{volume}{C88}},
  \bibinfo{pages}{034905} (\bibinfo{year}{2013}{\natexlab{a}}),
  \bibinfo{note}{[Erratum: Phys. Rev.C93,no.6,069901(2016)]},
  \eprint{1304.4427}.

\bibitem[{\citenamefont{Becattini
  et~al.}(2013{\natexlab{b}})\citenamefont{Becattini, Chandra, Del~Zanna, and
  Grossi}}]{Becattini:2013fla}
\bibinfo{author}{\bibfnamefont{F.}~\bibnamefont{Becattini}},
  \bibinfo{author}{\bibfnamefont{V.}~\bibnamefont{Chandra}},
  \bibinfo{author}{\bibfnamefont{L.}~\bibnamefont{Del~Zanna}},
  \bibnamefont{and} \bibinfo{author}{\bibfnamefont{E.}~\bibnamefont{Grossi}},
  \bibinfo{journal}{Annals Phys.} \textbf{\bibinfo{volume}{338}},
  \bibinfo{pages}{32} (\bibinfo{year}{2013}{\natexlab{b}}), \eprint{1303.3431}.

\bibitem[{\citenamefont{Becattini et~al.}(2015)\citenamefont{Becattini,
  Inghirami, Rolando, Beraudo, Del~Zanna, De~Pace, Nardi, Pagliara, and
  Chandra}}]{Becattini:2015ska}
\bibinfo{author}{\bibfnamefont{F.}~\bibnamefont{Becattini}},
  \bibinfo{author}{\bibfnamefont{G.}~\bibnamefont{Inghirami}},
  \bibinfo{author}{\bibfnamefont{V.}~\bibnamefont{Rolando}},
  \bibinfo{author}{\bibfnamefont{A.}~\bibnamefont{Beraudo}},
  \bibinfo{author}{\bibfnamefont{L.}~\bibnamefont{Del~Zanna}},
  \bibinfo{author}{\bibfnamefont{A.}~\bibnamefont{De~Pace}},
  \bibinfo{author}{\bibfnamefont{M.}~\bibnamefont{Nardi}},
  \bibinfo{author}{\bibfnamefont{G.}~\bibnamefont{Pagliara}}, \bibnamefont{and}
  \bibinfo{author}{\bibfnamefont{V.}~\bibnamefont{Chandra}},
  \bibinfo{journal}{Eur. Phys. J.} \textbf{\bibinfo{volume}{C75}},
  \bibinfo{pages}{406} (\bibinfo{year}{2015}), \bibinfo{note}{[Erratum: Eur.
  Phys. J.C78,no.5,354(2018)]}, \eprint{1501.04468}.

\bibitem[{\citenamefont{Becattini et~al.}(2017)\citenamefont{Becattini,
  Karpenko, Lisa, Upsal, and Voloshin}}]{Becattini:2016gvu}
\bibinfo{author}{\bibfnamefont{F.}~\bibnamefont{Becattini}},
  \bibinfo{author}{\bibfnamefont{I.}~\bibnamefont{Karpenko}},
  \bibinfo{author}{\bibfnamefont{M.}~\bibnamefont{Lisa}},
  \bibinfo{author}{\bibfnamefont{I.}~\bibnamefont{Upsal}}, \bibnamefont{and}
  \bibinfo{author}{\bibfnamefont{S.}~\bibnamefont{Voloshin}},
  \bibinfo{journal}{Phys. Rev.} \textbf{\bibinfo{volume}{C95}},
  \bibinfo{pages}{054902} (\bibinfo{year}{2017}), \eprint{1610.02506}.

\bibitem[{\citenamefont{Karpenko and Becattini}(2017)}]{Karpenko:2016jyx}
\bibinfo{author}{\bibfnamefont{I.}~\bibnamefont{Karpenko}} \bibnamefont{and}
  \bibinfo{author}{\bibfnamefont{F.}~\bibnamefont{Becattini}},
  \bibinfo{journal}{Eur. Phys. J.} \textbf{\bibinfo{volume}{C77}},
  \bibinfo{pages}{213} (\bibinfo{year}{2017}), \eprint{1610.04717}.

\bibitem[{\citenamefont{Pang et~al.}(2016)\citenamefont{Pang, Petersen, Wang,
  and Wang}}]{Pang:2016igs}
\bibinfo{author}{\bibfnamefont{L.-G.} \bibnamefont{Pang}},
  \bibinfo{author}{\bibfnamefont{H.}~\bibnamefont{Petersen}},
  \bibinfo{author}{\bibfnamefont{Q.}~\bibnamefont{Wang}}, \bibnamefont{and}
  \bibinfo{author}{\bibfnamefont{X.-N.} \bibnamefont{Wang}},
  \bibinfo{journal}{Phys. Rev. Lett.} \textbf{\bibinfo{volume}{117}},
  \bibinfo{pages}{192301} (\bibinfo{year}{2016}), \eprint{1605.04024}.

\bibitem[{\citenamefont{Xie et~al.}(2017)\citenamefont{Xie, Wang, and
  Csernai}}]{Xie:2017upb}
\bibinfo{author}{\bibfnamefont{Y.}~\bibnamefont{Xie}},
  \bibinfo{author}{\bibfnamefont{D.}~\bibnamefont{Wang}}, \bibnamefont{and}
  \bibinfo{author}{\bibfnamefont{L.~P.} \bibnamefont{Csernai}},
  \bibinfo{journal}{Phys. Rev.} \textbf{\bibinfo{volume}{C95}},
  \bibinfo{pages}{031901} (\bibinfo{year}{2017}), \eprint{1703.03770}.

\bibitem[{\citenamefont{Becattini and Karpenko}(2018)}]{Becattini:2017gcx}
\bibinfo{author}{\bibfnamefont{F.}~\bibnamefont{Becattini}} \bibnamefont{and}
  \bibinfo{author}{\bibfnamefont{I.}~\bibnamefont{Karpenko}},
  \bibinfo{journal}{Phys. Rev. Lett.} \textbf{\bibinfo{volume}{120}},
  \bibinfo{pages}{012302} (\bibinfo{year}{2018}), \eprint{1707.07984}.

\bibitem[{\citenamefont{Becattini and Lisa}(2020)}]{Becattini:2020ngo}
\bibinfo{author}{\bibfnamefont{F.}~\bibnamefont{Becattini}} \bibnamefont{and}
  \bibinfo{author}{\bibfnamefont{M.~A.} \bibnamefont{Lisa}},
  \bibinfo{journal}{Ann. Rev. Nucl. Part. Sci.} \textbf{\bibinfo{volume}{70}},
  \bibinfo{pages}{395} (\bibinfo{year}{2020}), \eprint{2003.03640}.

\bibitem[{\citenamefont{Florkowski
  et~al.}(2019{\natexlab{a}})\citenamefont{Florkowski, Kumar, Ryblewski, and
  Singh}}]{Florkowski:2019qdp}
\bibinfo{author}{\bibfnamefont{W.}~\bibnamefont{Florkowski}},
  \bibinfo{author}{\bibfnamefont{A.}~\bibnamefont{Kumar}},
  \bibinfo{author}{\bibfnamefont{R.}~\bibnamefont{Ryblewski}},
  \bibnamefont{and} \bibinfo{author}{\bibfnamefont{R.}~\bibnamefont{Singh}},
  \bibinfo{journal}{Phys. Rev.} \textbf{\bibinfo{volume}{C99}},
  \bibinfo{pages}{044910} (\bibinfo{year}{2019}{\natexlab{a}}),
  \eprint{1901.09655}.

\bibitem[{\citenamefont{Florkowski
  et~al.}(2019{\natexlab{b}})\citenamefont{Florkowski, Kumar, Ryblewski, and
  Mazeliauskas}}]{Florkowski:2019voj}
\bibinfo{author}{\bibfnamefont{W.}~\bibnamefont{Florkowski}},
  \bibinfo{author}{\bibfnamefont{A.}~\bibnamefont{Kumar}},
  \bibinfo{author}{\bibfnamefont{R.}~\bibnamefont{Ryblewski}},
  \bibnamefont{and}
  \bibinfo{author}{\bibfnamefont{A.}~\bibnamefont{Mazeliauskas}},
  \bibinfo{journal}{Phys. Rev.} \textbf{\bibinfo{volume}{C100}},
  \bibinfo{pages}{054907} (\bibinfo{year}{2019}{\natexlab{b}}),
  \eprint{1904.00002}.

\bibitem[{\citenamefont{Zhang et~al.}(2019)\citenamefont{Zhang, Fang, Wang, and
  Wang}}]{Zhang:2019xya}
\bibinfo{author}{\bibfnamefont{J.-j.} \bibnamefont{Zhang}},
  \bibinfo{author}{\bibfnamefont{R.-h.} \bibnamefont{Fang}},
  \bibinfo{author}{\bibfnamefont{Q.}~\bibnamefont{Wang}}, \bibnamefont{and}
  \bibinfo{author}{\bibfnamefont{X.-N.} \bibnamefont{Wang}},
  \bibinfo{journal}{Phys. Rev.} \textbf{\bibinfo{volume}{C100}},
  \bibinfo{pages}{064904} (\bibinfo{year}{2019}), \eprint{1904.09152}.

\bibitem[{\citenamefont{Becattini
  et~al.}(2019{\natexlab{a}})\citenamefont{Becattini, Cao, and
  Speranza}}]{Becattini:2019ntv}
\bibinfo{author}{\bibfnamefont{F.}~\bibnamefont{Becattini}},
  \bibinfo{author}{\bibfnamefont{G.}~\bibnamefont{Cao}}, \bibnamefont{and}
  \bibinfo{author}{\bibfnamefont{E.}~\bibnamefont{Speranza}},
  \bibinfo{journal}{Eur. Phys. J.} \textbf{\bibinfo{volume}{C79}},
  \bibinfo{pages}{741} (\bibinfo{year}{2019}{\natexlab{a}}),
  \eprint{1905.03123}.

\bibitem[{\citenamefont{Xia et~al.}(2019)\citenamefont{Xia, Li, Huang, and
  Huang}}]{Xia:2019fjf}
\bibinfo{author}{\bibfnamefont{X.-L.} \bibnamefont{Xia}},
  \bibinfo{author}{\bibfnamefont{H.}~\bibnamefont{Li}},
  \bibinfo{author}{\bibfnamefont{X.-G.} \bibnamefont{Huang}}, \bibnamefont{and}
  \bibinfo{author}{\bibfnamefont{H.~Z.} \bibnamefont{Huang}},
  \bibinfo{journal}{Phys. Rev.} \textbf{\bibinfo{volume}{C100}},
  \bibinfo{pages}{014913} (\bibinfo{year}{2019}), \eprint{1905.03120}.

\bibitem[{\citenamefont{Wu et~al.}(2019)\citenamefont{Wu, Pang, Huang, and
  Wang}}]{Wu:2019eyi}
\bibinfo{author}{\bibfnamefont{H.-Z.} \bibnamefont{Wu}},
  \bibinfo{author}{\bibfnamefont{L.-G.} \bibnamefont{Pang}},
  \bibinfo{author}{\bibfnamefont{X.-G.} \bibnamefont{Huang}}, \bibnamefont{and}
  \bibinfo{author}{\bibfnamefont{Q.}~\bibnamefont{Wang}},
  \bibinfo{journal}{Phys. Rev. Research.} \textbf{\bibinfo{volume}{1}},
  \bibinfo{pages}{033058} (\bibinfo{year}{2019}), \eprint{1906.09385}.

\bibitem[{\citenamefont{Sun and Ko}(2019)}]{Sun:2018bjl}
\bibinfo{author}{\bibfnamefont{Y.}~\bibnamefont{Sun}} \bibnamefont{and}
  \bibinfo{author}{\bibfnamefont{C.~M.} \bibnamefont{Ko}},
  \bibinfo{journal}{Phys. Rev.} \textbf{\bibinfo{volume}{C99}},
  \bibinfo{pages}{011903} (\bibinfo{year}{2019}), \eprint{1810.10359}.

\bibitem[{\citenamefont{Liu et~al.}(2020)\citenamefont{Liu, Sun, and
  Ko}}]{Liu:2019krs}
\bibinfo{author}{\bibfnamefont{S.~Y.~F.} \bibnamefont{Liu}},
  \bibinfo{author}{\bibfnamefont{Y.}~\bibnamefont{Sun}}, \bibnamefont{and}
  \bibinfo{author}{\bibfnamefont{C.~M.} \bibnamefont{Ko}},
  \bibinfo{journal}{Phys. Rev. Lett.} \textbf{\bibinfo{volume}{125}},
  \bibinfo{pages}{062301} (\bibinfo{year}{2020}), \eprint{1910.06774}.

\bibitem[{\citenamefont{Florkowski et~al.}(2022)\citenamefont{Florkowski,
  Ryblewski, Singh, and Sophys}}]{Florkowski:2021wvk}
\bibinfo{author}{\bibfnamefont{W.}~\bibnamefont{Florkowski}},
  \bibinfo{author}{\bibfnamefont{R.}~\bibnamefont{Ryblewski}},
  \bibinfo{author}{\bibfnamefont{R.}~\bibnamefont{Singh}}, \bibnamefont{and}
  \bibinfo{author}{\bibfnamefont{G.}~\bibnamefont{Sophys}},
  \bibinfo{journal}{Phys. Rev. D} \textbf{\bibinfo{volume}{105}},
  \bibinfo{pages}{054007} (\bibinfo{year}{2022}), \eprint{2112.01856}.

\bibitem[{\citenamefont{Liu and Yin}(2021)}]{Liu:2021uhn}
\bibinfo{author}{\bibfnamefont{S.~Y.~F.} \bibnamefont{Liu}} \bibnamefont{and}
  \bibinfo{author}{\bibfnamefont{Y.}~\bibnamefont{Yin}},
  \bibinfo{journal}{JHEP} \textbf{\bibinfo{volume}{07}}, \bibinfo{pages}{188}
  (\bibinfo{year}{2021}), \eprint{2103.09200}.

\bibitem[{\citenamefont{Fu et~al.}(2021)\citenamefont{Fu, Liu, Pang, Song, and
  Yin}}]{Fu:2021pok}
\bibinfo{author}{\bibfnamefont{B.}~\bibnamefont{Fu}},
  \bibinfo{author}{\bibfnamefont{S.~Y.~F.} \bibnamefont{Liu}},
  \bibinfo{author}{\bibfnamefont{L.}~\bibnamefont{Pang}},
  \bibinfo{author}{\bibfnamefont{H.}~\bibnamefont{Song}}, \bibnamefont{and}
  \bibinfo{author}{\bibfnamefont{Y.}~\bibnamefont{Yin}},
  \bibinfo{journal}{Phys. Rev. Lett.} \textbf{\bibinfo{volume}{127}},
  \bibinfo{pages}{142301} (\bibinfo{year}{2021}), \eprint{2103.10403}.

\bibitem[{\citenamefont{Becattini
  et~al.}(2021{\natexlab{a}})\citenamefont{Becattini, Buzzegoli, and
  Palermo}}]{Becattini:2021suc}
\bibinfo{author}{\bibfnamefont{F.}~\bibnamefont{Becattini}},
  \bibinfo{author}{\bibfnamefont{M.}~\bibnamefont{Buzzegoli}},
  \bibnamefont{and} \bibinfo{author}{\bibfnamefont{A.}~\bibnamefont{Palermo}},
  \bibinfo{journal}{Phys. Lett. B} \textbf{\bibinfo{volume}{820}},
  \bibinfo{pages}{136519} (\bibinfo{year}{2021}{\natexlab{a}}),
  \eprint{2103.10917}.

\bibitem[{\citenamefont{Becattini
  et~al.}(2021{\natexlab{b}})\citenamefont{Becattini, Buzzegoli, Inghirami,
  Karpenko, and Palermo}}]{Becattini:2021iol}
\bibinfo{author}{\bibfnamefont{F.}~\bibnamefont{Becattini}},
  \bibinfo{author}{\bibfnamefont{M.}~\bibnamefont{Buzzegoli}},
  \bibinfo{author}{\bibfnamefont{G.}~\bibnamefont{Inghirami}},
  \bibinfo{author}{\bibfnamefont{I.}~\bibnamefont{Karpenko}}, \bibnamefont{and}
  \bibinfo{author}{\bibfnamefont{A.}~\bibnamefont{Palermo}},
  \bibinfo{journal}{Phys. Rev. Lett.} \textbf{\bibinfo{volume}{127}},
  \bibinfo{pages}{272302} (\bibinfo{year}{2021}{\natexlab{b}}),
  \eprint{2103.14621}.

\bibitem[{\citenamefont{Heinz and Snellings}(2013)}]{Heinz:2013th}
\bibinfo{author}{\bibfnamefont{U.}~\bibnamefont{Heinz}} \bibnamefont{and}
  \bibinfo{author}{\bibfnamefont{R.}~\bibnamefont{Snellings}},
  \bibinfo{journal}{Ann. Rev. Nucl. Part. Sci.} \textbf{\bibinfo{volume}{63}},
  \bibinfo{pages}{123} (\bibinfo{year}{2013}), \eprint{1301.2826}.

\bibitem[{\citenamefont{Florkowski
  et~al.}(2018{\natexlab{d}})\citenamefont{Florkowski, Heller, and
  Spalinski}}]{Florkowski:2017olj}
\bibinfo{author}{\bibfnamefont{W.}~\bibnamefont{Florkowski}},
  \bibinfo{author}{\bibfnamefont{M.~P.} \bibnamefont{Heller}},
  \bibnamefont{and}
  \bibinfo{author}{\bibfnamefont{M.}~\bibnamefont{Spalinski}},
  \bibinfo{journal}{Rept. Prog. Phys.} \textbf{\bibinfo{volume}{81}},
  \bibinfo{pages}{046001} (\bibinfo{year}{2018}{\natexlab{d}}),
  \eprint{1707.02282}.

\bibitem[{\citenamefont{Gallegos and G\"ursoy}(2020)}]{Gallegos:2020otk}
\bibinfo{author}{\bibfnamefont{A.~D.} \bibnamefont{Gallegos}} \bibnamefont{and}
  \bibinfo{author}{\bibfnamefont{U.}~\bibnamefont{G\"ursoy}},
  \bibinfo{journal}{JHEP} \textbf{\bibinfo{volume}{11}}, \bibinfo{pages}{151}
  (\bibinfo{year}{2020}), \eprint{2004.05148}.

\bibitem[{\citenamefont{Garbiso and Kaminski}(2020)}]{Garbiso:2020puw}
\bibinfo{author}{\bibfnamefont{M.}~\bibnamefont{Garbiso}} \bibnamefont{and}
  \bibinfo{author}{\bibfnamefont{M.}~\bibnamefont{Kaminski}},
  \bibinfo{journal}{JHEP} \textbf{\bibinfo{volume}{12}}, \bibinfo{pages}{112}
  (\bibinfo{year}{2020}), \eprint{2007.04345}.

\bibitem[{\citenamefont{Cartwright et~al.}(2021)\citenamefont{Cartwright,
  Amano, Kaminski, Noronha, and Speranza}}]{Cartwright:2021qpp}
\bibinfo{author}{\bibfnamefont{C.}~\bibnamefont{Cartwright}},
  \bibinfo{author}{\bibfnamefont{M.~G.} \bibnamefont{Amano}},
  \bibinfo{author}{\bibfnamefont{M.}~\bibnamefont{Kaminski}},
  \bibinfo{author}{\bibfnamefont{J.}~\bibnamefont{Noronha}}, \bibnamefont{and}
  \bibinfo{author}{\bibfnamefont{E.}~\bibnamefont{Speranza}}
  (\bibinfo{year}{2021}), \eprint{2112.10781}.

\bibitem[{\citenamefont{Hehl}(1976)}]{Hehl:1976vr}
\bibinfo{author}{\bibfnamefont{F.~W.} \bibnamefont{Hehl}},
  \bibinfo{journal}{Rept. Math. Phys.} \textbf{\bibinfo{volume}{9}},
  \bibinfo{pages}{55} (\bibinfo{year}{1976}).

\bibitem[{\citenamefont{Becattini
  et~al.}(2019{\natexlab{b}})\citenamefont{Becattini, Florkowski, and
  Speranza}}]{Becattini:2018duy}
\bibinfo{author}{\bibfnamefont{F.}~\bibnamefont{Becattini}},
  \bibinfo{author}{\bibfnamefont{W.}~\bibnamefont{Florkowski}},
  \bibnamefont{and} \bibinfo{author}{\bibfnamefont{E.}~\bibnamefont{Speranza}},
  \bibinfo{journal}{Phys. Lett.} \textbf{\bibinfo{volume}{B789}},
  \bibinfo{pages}{419} (\bibinfo{year}{2019}{\natexlab{b}}),
  \eprint{1807.10994}.

\bibitem[{\citenamefont{Buzzegoli}(2022)}]{Buzzegoli:2021wlg}
\bibinfo{author}{\bibfnamefont{M.}~\bibnamefont{Buzzegoli}},
  \bibinfo{journal}{Phys. Rev. C} \textbf{\bibinfo{volume}{105}},
  \bibinfo{pages}{044907} (\bibinfo{year}{2022}), \eprint{2109.12084}.

\bibitem[{\citenamefont{Das et~al.}(2021)\citenamefont{Das, Florkowski,
  Ryblewski, and Singh}}]{Das:2021aar}
\bibinfo{author}{\bibfnamefont{A.}~\bibnamefont{Das}},
  \bibinfo{author}{\bibfnamefont{W.}~\bibnamefont{Florkowski}},
  \bibinfo{author}{\bibfnamefont{R.}~\bibnamefont{Ryblewski}},
  \bibnamefont{and} \bibinfo{author}{\bibfnamefont{R.}~\bibnamefont{Singh}},
  \bibinfo{journal}{Phys. Rev. D} \textbf{\bibinfo{volume}{103}},
  \bibinfo{pages}{L091502} (\bibinfo{year}{2021}), \eprint{2103.01013}.

\bibitem[{\citenamefont{Daher et~al.}(2022)\citenamefont{Daher, Das,
  Florkowski, and Ryblewski}}]{Daher:2022xon}
\bibinfo{author}{\bibfnamefont{A.}~\bibnamefont{Daher}},
  \bibinfo{author}{\bibfnamefont{A.}~\bibnamefont{Das}},
  \bibinfo{author}{\bibfnamefont{W.}~\bibnamefont{Florkowski}},
  \bibnamefont{and} \bibinfo{author}{\bibfnamefont{R.}~\bibnamefont{Ryblewski}}
  (\bibinfo{year}{2022}), \eprint{2202.12609}.

\bibitem[{\citenamefont{Hilgevoord and Wouthuysen}(1963)}]{HILGEVOORD19631}
\bibinfo{author}{\bibfnamefont{J.}~\bibnamefont{Hilgevoord}} \bibnamefont{and}
  \bibinfo{author}{\bibfnamefont{S.}~\bibnamefont{Wouthuysen}},
  \bibinfo{journal}{Nuclear Physics} \textbf{\bibinfo{volume}{40}},
  \bibinfo{pages}{1 } (\bibinfo{year}{1963}), ISSN \bibinfo{issn}{0029-5582}.

\bibitem[{\citenamefont{Denicol
  et~al.}(2012{\natexlab{a}})\citenamefont{Denicol, Moln\'ar, Niemi, and
  Rischke}}]{Denicol:2012es}
\bibinfo{author}{\bibfnamefont{G.~S.} \bibnamefont{Denicol}},
  \bibinfo{author}{\bibfnamefont{E.}~\bibnamefont{Moln\'ar}},
  \bibinfo{author}{\bibfnamefont{H.}~\bibnamefont{Niemi}}, \bibnamefont{and}
  \bibinfo{author}{\bibfnamefont{D.~H.} \bibnamefont{Rischke}},
  \bibinfo{journal}{Eur. Phys. J. A} \textbf{\bibinfo{volume}{48}},
  \bibinfo{pages}{170} (\bibinfo{year}{2012}{\natexlab{a}}),
  \eprint{1206.1554}.

\bibitem[{\citenamefont{Denicol
  et~al.}(2012{\natexlab{b}})\citenamefont{Denicol, Niemi, Molnar, and
  Rischke}}]{Denicol:2012cn}
\bibinfo{author}{\bibfnamefont{G.~S.} \bibnamefont{Denicol}},
  \bibinfo{author}{\bibfnamefont{H.}~\bibnamefont{Niemi}},
  \bibinfo{author}{\bibfnamefont{E.}~\bibnamefont{Molnar}}, \bibnamefont{and}
  \bibinfo{author}{\bibfnamefont{D.}~\bibnamefont{Rischke}},
  \bibinfo{journal}{Phys.\ Rev.\ D} \textbf{\bibinfo{volume}{85}},
  \bibinfo{pages}{114047} (\bibinfo{year}{2012}{\natexlab{b}}),
  \bibinfo{note}{[Erratum: Phys.Rev.D 91, 039902 (2015)]}, \eprint{1202.4551}.

\bibitem[{\citenamefont{Weickgenannt
  et~al.}(2021{\natexlab{b}})\citenamefont{Weickgenannt, Speranza, Sheng, Wang,
  and Rischke}}]{Weickgenannt:2021cuo}
\bibinfo{author}{\bibfnamefont{N.}~\bibnamefont{Weickgenannt}},
  \bibinfo{author}{\bibfnamefont{E.}~\bibnamefont{Speranza}},
  \bibinfo{author}{\bibfnamefont{X.-l.} \bibnamefont{Sheng}},
  \bibinfo{author}{\bibfnamefont{Q.}~\bibnamefont{Wang}}, \bibnamefont{and}
  \bibinfo{author}{\bibfnamefont{D.~H.} \bibnamefont{Rischke}},
  \bibinfo{journal}{Phys. Rev. D} \textbf{\bibinfo{volume}{104}},
  \bibinfo{pages}{016022} (\bibinfo{year}{2021}{\natexlab{b}}),
  \eprint{2103.04896}.

\bibitem[{\citenamefont{Yang et~al.}(2020)\citenamefont{Yang, Hattori, and
  Hidaka}}]{Yang:2020hri}
\bibinfo{author}{\bibfnamefont{D.-L.} \bibnamefont{Yang}},
  \bibinfo{author}{\bibfnamefont{K.}~\bibnamefont{Hattori}}, \bibnamefont{and}
  \bibinfo{author}{\bibfnamefont{Y.}~\bibnamefont{Hidaka}},
  \bibinfo{journal}{JHEP} \textbf{\bibinfo{volume}{20}}, \bibinfo{pages}{070}
  (\bibinfo{year}{2020}), \eprint{2002.02612}.

\bibitem[{\citenamefont{Wang et~al.}(2020)\citenamefont{Wang, Guo, and
  Zhuang}}]{Wang:2020pej}
\bibinfo{author}{\bibfnamefont{Z.}~\bibnamefont{Wang}},
  \bibinfo{author}{\bibfnamefont{X.}~\bibnamefont{Guo}}, \bibnamefont{and}
  \bibinfo{author}{\bibfnamefont{P.}~\bibnamefont{Zhuang}}
  (\bibinfo{year}{2020}), \eprint{2009.10930}.

\bibitem[{\citenamefont{De~Groot et~al.}(1980)\citenamefont{De~Groot,
  Van~Leeuwen, and Van~Weert}}]{DeGroot:1980dk}
\bibinfo{author}{\bibfnamefont{S.~R.} \bibnamefont{De~Groot}},
  \bibinfo{author}{\bibfnamefont{W.~A.} \bibnamefont{Van~Leeuwen}},
  \bibnamefont{and} \bibinfo{author}{\bibfnamefont{C.~G.}
  \bibnamefont{Van~Weert}}, \emph{\bibinfo{title}{Relativistic Kinetic Theory.
  Principles and Applications}} (\bibinfo{publisher}{North-Holland},
  \bibinfo{year}{1980}).

\bibitem[{\citenamefont{Weickgenannt et~al.}(2022)\citenamefont{Weickgenannt,
  Wagner, and Speranza}}]{Weickgenannt:2022jes}
\bibinfo{author}{\bibfnamefont{N.}~\bibnamefont{Weickgenannt}},
  \bibinfo{author}{\bibfnamefont{D.}~\bibnamefont{Wagner}}, \bibnamefont{and}
  \bibinfo{author}{\bibfnamefont{E.}~\bibnamefont{Speranza}},
  \bibinfo{journal}{Phys. Rev. D} \textbf{\bibinfo{volume}{105}},
  \bibinfo{pages}{116026} (\bibinfo{year}{2022}), \eprint{2204.01797}.

\bibitem[{\citenamefont{Becattini}(2012)}]{Becattini:2012tc}
\bibinfo{author}{\bibfnamefont{F.}~\bibnamefont{Becattini}},
  \bibinfo{journal}{Phys. Rev. Lett.} \textbf{\bibinfo{volume}{108}},
  \bibinfo{pages}{244502} (\bibinfo{year}{2012}), \eprint{1201.5278}.

\bibitem[{\citenamefont{Israel and Stewart}(1979)}]{Israel:1979wp}
\bibinfo{author}{\bibfnamefont{W.}~\bibnamefont{Israel}} \bibnamefont{and}
  \bibinfo{author}{\bibfnamefont{J.~M.} \bibnamefont{Stewart}},
  \bibinfo{journal}{Annals Phys.} \textbf{\bibinfo{volume}{118}},
  \bibinfo{pages}{341} (\bibinfo{year}{1979}).

\bibitem[{\citenamefont{Moln\'ar et~al.}(2014)\citenamefont{Moln\'ar, Niemi,
  Denicol, and Rischke}}]{Molnar:2013lta}
\bibinfo{author}{\bibfnamefont{E.}~\bibnamefont{Moln\'ar}},
  \bibinfo{author}{\bibfnamefont{H.}~\bibnamefont{Niemi}},
  \bibinfo{author}{\bibfnamefont{G.~S.} \bibnamefont{Denicol}},
  \bibnamefont{and} \bibinfo{author}{\bibfnamefont{D.~H.}
  \bibnamefont{Rischke}}, \bibinfo{journal}{Phys. Rev. D}
  \textbf{\bibinfo{volume}{89}}, \bibinfo{pages}{074010}
  (\bibinfo{year}{2014}), \eprint{1308.0785}.

\bibitem[{\citenamefont{Denicol et~al.}(2014)\citenamefont{Denicol, Niemi,
  Bouras, Molnar, Xu, Rischke, and Greiner}}]{Denicol:2012vq}
\bibinfo{author}{\bibfnamefont{G.~S.} \bibnamefont{Denicol}},
  \bibinfo{author}{\bibfnamefont{H.}~\bibnamefont{Niemi}},
  \bibinfo{author}{\bibfnamefont{I.}~\bibnamefont{Bouras}},
  \bibinfo{author}{\bibfnamefont{E.}~\bibnamefont{Molnar}},
  \bibinfo{author}{\bibfnamefont{Z.}~\bibnamefont{Xu}},
  \bibinfo{author}{\bibfnamefont{D.~H.} \bibnamefont{Rischke}},
  \bibnamefont{and} \bibinfo{author}{\bibfnamefont{C.}~\bibnamefont{Greiner}},
  \bibinfo{journal}{Phys. Rev. D} \textbf{\bibinfo{volume}{89}},
  \bibinfo{pages}{074005} (\bibinfo{year}{2014}), \eprint{1207.6811}.

\bibitem[{\citenamefont{Li and Yee}(2019)}]{Li:2019qkf}
\bibinfo{author}{\bibfnamefont{S.}~\bibnamefont{Li}} \bibnamefont{and}
  \bibinfo{author}{\bibfnamefont{H.-U.} \bibnamefont{Yee}},
  \bibinfo{journal}{Phys.\ Rev.\ D} \textbf{\bibinfo{volume}{100}},
  \bibinfo{pages}{056022} (\bibinfo{year}{2019}), \eprint{1905.10463}.

\bibitem[{\citenamefont{Kapusta
  et~al.}(2020{\natexlab{a}})\citenamefont{Kapusta, Rrapaj, and
  Rudaz}}]{Kapusta:2019sad}
\bibinfo{author}{\bibfnamefont{J.~I.} \bibnamefont{Kapusta}},
  \bibinfo{author}{\bibfnamefont{E.}~\bibnamefont{Rrapaj}}, \bibnamefont{and}
  \bibinfo{author}{\bibfnamefont{S.}~\bibnamefont{Rudaz}},
  \bibinfo{journal}{Phys. Rev. C} \textbf{\bibinfo{volume}{101}},
  \bibinfo{pages}{024907} (\bibinfo{year}{2020}{\natexlab{a}}),
  \eprint{1907.10750}.

\bibitem[{\citenamefont{Kapusta
  et~al.}(2020{\natexlab{b}})\citenamefont{Kapusta, Rrapaj, and
  Rudaz}}]{Kapusta:2020npk}
\bibinfo{author}{\bibfnamefont{J.~I.} \bibnamefont{Kapusta}},
  \bibinfo{author}{\bibfnamefont{E.}~\bibnamefont{Rrapaj}}, \bibnamefont{and}
  \bibinfo{author}{\bibfnamefont{S.}~\bibnamefont{Rudaz}},
  \bibinfo{journal}{Phys. Rev. C} \textbf{\bibinfo{volume}{102}},
  \bibinfo{pages}{064911} (\bibinfo{year}{2020}{\natexlab{b}}),
  \eprint{2004.14807}.

\bibitem[{\citenamefont{Hongo et~al.}(2022)\citenamefont{Hongo, Huang,
  Kaminski, Stephanov, and Yee}}]{Hongo:2022izs}
\bibinfo{author}{\bibfnamefont{M.}~\bibnamefont{Hongo}},
  \bibinfo{author}{\bibfnamefont{X.-G.} \bibnamefont{Huang}},
  \bibinfo{author}{\bibfnamefont{M.}~\bibnamefont{Kaminski}},
  \bibinfo{author}{\bibfnamefont{M.}~\bibnamefont{Stephanov}},
  \bibnamefont{and} \bibinfo{author}{\bibfnamefont{H.-U.} \bibnamefont{Yee}}
  (\bibinfo{year}{2022}), \eprint{2201.12390}.

\bibitem[{\citenamefont{Kapusta
  et~al.}(2020{\natexlab{c}})\citenamefont{Kapusta, Rrapaj, and
  Rudaz}}]{Kapusta:2019ktm}
\bibinfo{author}{\bibfnamefont{J.~I.} \bibnamefont{Kapusta}},
  \bibinfo{author}{\bibfnamefont{E.}~\bibnamefont{Rrapaj}}, \bibnamefont{and}
  \bibinfo{author}{\bibfnamefont{S.}~\bibnamefont{Rudaz}},
  \bibinfo{journal}{Phys. Rev. C} \textbf{\bibinfo{volume}{101}},
  \bibinfo{pages}{031901} (\bibinfo{year}{2020}{\natexlab{c}}),
  \eprint{1910.12759}.

\bibitem[{\citenamefont{Ayala et~al.}(2020{\natexlab{a}})\citenamefont{Ayala,
  De~La~Cruz, Hernández-Ortíz, Hernández, and Salinas}}]{Ayala:2019iin}
\bibinfo{author}{\bibfnamefont{A.}~\bibnamefont{Ayala}},
  \bibinfo{author}{\bibfnamefont{D.}~\bibnamefont{De~La~Cruz}},
  \bibinfo{author}{\bibfnamefont{S.}~\bibnamefont{Hernández-Ortíz}},
  \bibinfo{author}{\bibfnamefont{L.}~\bibnamefont{Hernández}},
  \bibnamefont{and} \bibinfo{author}{\bibfnamefont{J.}~\bibnamefont{Salinas}},
  \bibinfo{journal}{Phys. Lett. B} \textbf{\bibinfo{volume}{801}},
  \bibinfo{pages}{135169} (\bibinfo{year}{2020}{\natexlab{a}}),
  \eprint{1909.00274}.

\bibitem[{\citenamefont{Ayala et~al.}(2020{\natexlab{b}})\citenamefont{Ayala,
  de~la Cruz, Hern\'andez, and Salinas}}]{Ayala:2020ndx}
\bibinfo{author}{\bibfnamefont{A.}~\bibnamefont{Ayala}},
  \bibinfo{author}{\bibfnamefont{D.}~\bibnamefont{de~la Cruz}},
  \bibinfo{author}{\bibfnamefont{L.~A.} \bibnamefont{Hern\'andez}},
  \bibnamefont{and} \bibinfo{author}{\bibfnamefont{J.}~\bibnamefont{Salinas}},
  \bibinfo{journal}{Phys. Rev. D} \textbf{\bibinfo{volume}{102}},
  \bibinfo{pages}{056019} (\bibinfo{year}{2020}{\natexlab{b}}),
  \eprint{2003.06545}.

\bibitem[{\citenamefont{Becattini}(2021)}]{Becattini:2020sww}
\bibinfo{author}{\bibfnamefont{F.}~\bibnamefont{Becattini}},
  \bibinfo{journal}{Lect. Notes Phys.} \textbf{\bibinfo{volume}{987}},
  \bibinfo{pages}{15} (\bibinfo{year}{2021}), \eprint{2004.04050}.

\bibitem[{\citenamefont{Tinti and Florkowski}(2020)}]{Tinti:2020gyh}
\bibinfo{author}{\bibfnamefont{L.}~\bibnamefont{Tinti}} \bibnamefont{and}
  \bibinfo{author}{\bibfnamefont{W.}~\bibnamefont{Florkowski}}
  (\bibinfo{year}{2020}), \eprint{2007.04029}.

\bibitem[{\citenamefont{Denicol et~al.}(2018)\citenamefont{Denicol, Huang,
  Moln\'{a}r, Monteiro, Niemi, Noronha, Rischke, and Wang}}]{Denicol:2018rbw}
\bibinfo{author}{\bibfnamefont{G.~S.} \bibnamefont{Denicol}},
  \bibinfo{author}{\bibfnamefont{X.-G.} \bibnamefont{Huang}},
  \bibinfo{author}{\bibfnamefont{E.}~\bibnamefont{Moln\'{a}r}},
  \bibinfo{author}{\bibfnamefont{G.~M.} \bibnamefont{Monteiro}},
  \bibinfo{author}{\bibfnamefont{H.}~\bibnamefont{Niemi}},
  \bibinfo{author}{\bibfnamefont{J.}~\bibnamefont{Noronha}},
  \bibinfo{author}{\bibfnamefont{D.~H.} \bibnamefont{Rischke}},
  \bibnamefont{and} \bibinfo{author}{\bibfnamefont{Q.}~\bibnamefont{Wang}},
  \bibinfo{journal}{Phys. Rev.} \textbf{\bibinfo{volume}{D98}},
  \bibinfo{pages}{076009} (\bibinfo{year}{2018}), \eprint{1804.05210}.

\bibitem[{\citenamefont{Speranza et~al.}(2021)\citenamefont{Speranza, Bemfica,
  Disconzi, and Noronha}}]{Speranza:2021bxf}
\bibinfo{author}{\bibfnamefont{E.}~\bibnamefont{Speranza}},
  \bibinfo{author}{\bibfnamefont{F.~S.} \bibnamefont{Bemfica}},
  \bibinfo{author}{\bibfnamefont{M.~M.} \bibnamefont{Disconzi}},
  \bibnamefont{and} \bibinfo{author}{\bibfnamefont{J.}~\bibnamefont{Noronha}}
  (\bibinfo{year}{2021}), \eprint{2104.02110}.

\bibitem[{\citenamefont{Bemfica et~al.}(2018)\citenamefont{Bemfica, Disconzi,
  and Noronha}}]{Bemfica:2017wps}
\bibinfo{author}{\bibfnamefont{F.~S.} \bibnamefont{Bemfica}},
  \bibinfo{author}{\bibfnamefont{M.~M.} \bibnamefont{Disconzi}},
  \bibnamefont{and} \bibinfo{author}{\bibfnamefont{J.}~\bibnamefont{Noronha}},
  \bibinfo{journal}{Phys. Rev. D} \textbf{\bibinfo{volume}{98}},
  \bibinfo{pages}{104064} (\bibinfo{year}{2018}), \eprint{1708.06255}.

\bibitem[{\citenamefont{Bemfica et~al.}(2019)\citenamefont{Bemfica, Disconzi,
  and Noronha}}]{Bemfica:2019knx}
\bibinfo{author}{\bibfnamefont{F.~S.} \bibnamefont{Bemfica}},
  \bibinfo{author}{\bibfnamefont{M.~M.} \bibnamefont{Disconzi}},
  \bibnamefont{and} \bibinfo{author}{\bibfnamefont{J.}~\bibnamefont{Noronha}},
  \bibinfo{journal}{Phys. Rev. D} \textbf{\bibinfo{volume}{100}},
  \bibinfo{pages}{104020} (\bibinfo{year}{2019}), \eprint{1907.12695}.

\bibitem[{\citenamefont{Bemfica et~al.}(2020)\citenamefont{Bemfica, Disconzi,
  and Noronha}}]{Bemfica:2020zjp}
\bibinfo{author}{\bibfnamefont{F.~S.} \bibnamefont{Bemfica}},
  \bibinfo{author}{\bibfnamefont{M.~M.} \bibnamefont{Disconzi}},
  \bibnamefont{and} \bibinfo{author}{\bibfnamefont{J.}~\bibnamefont{Noronha}}
  (\bibinfo{year}{2020}), \eprint{2009.11388}.

\bibitem[{\citenamefont{Kovtun}(2019)}]{Kovtun:2019hdm}
\bibinfo{author}{\bibfnamefont{P.}~\bibnamefont{Kovtun}},
  \bibinfo{journal}{JHEP} \textbf{\bibinfo{volume}{10}}, \bibinfo{pages}{034}
  (\bibinfo{year}{2019}), \eprint{1907.08191}.

\bibitem[{\citenamefont{Hoult and Kovtun}(2020)}]{Hoult:2020eho}
\bibinfo{author}{\bibfnamefont{R.~E.} \bibnamefont{Hoult}} \bibnamefont{and}
  \bibinfo{author}{\bibfnamefont{P.}~\bibnamefont{Kovtun}},
  \bibinfo{journal}{JHEP} \textbf{\bibinfo{volume}{06}}, \bibinfo{pages}{067}
  (\bibinfo{year}{2020}), \eprint{2004.04102}.

\bibitem[{\citenamefont{Noronha et~al.}(2021)\citenamefont{Noronha,
  Spali\'nski, and Speranza}}]{Noronha:2021syv}
\bibinfo{author}{\bibfnamefont{J.}~\bibnamefont{Noronha}},
  \bibinfo{author}{\bibfnamefont{M.}~\bibnamefont{Spali\'nski}},
  \bibnamefont{and} \bibinfo{author}{\bibfnamefont{E.}~\bibnamefont{Speranza}}
  (\bibinfo{year}{2021}), \eprint{2105.01034}.

\bibitem[{\citenamefont{Rocha and Denicol}(2021)}]{Rocha:2021lze}
\bibinfo{author}{\bibfnamefont{G.~S.} \bibnamefont{Rocha}} \bibnamefont{and}
  \bibinfo{author}{\bibfnamefont{G.~S.} \bibnamefont{Denicol}},
  \bibinfo{journal}{Phys. Rev. D} \textbf{\bibinfo{volume}{104}},
  \bibinfo{pages}{096016} (\bibinfo{year}{2021}), \eprint{2108.02187}.

\bibitem[{\citenamefont{Coope and Snider}(1970)}]{coope1970irreducible}
\bibinfo{author}{\bibfnamefont{J.}~\bibnamefont{Coope}} \bibnamefont{and}
  \bibinfo{author}{\bibfnamefont{R.}~\bibnamefont{Snider}},
  \bibinfo{journal}{Journal of Mathematical Physics}
  \textbf{\bibinfo{volume}{11}}, \bibinfo{pages}{1003} (\bibinfo{year}{1970}).

\bibitem[{\citenamefont{Siskens and Van~Weert}(1977)}]{siskens1977transition}
\bibinfo{author}{\bibfnamefont{T.~J.} \bibnamefont{Siskens}} \bibnamefont{and}
  \bibinfo{author}{\bibfnamefont{C.~G.} \bibnamefont{Van~Weert}},
  \bibinfo{journal}{Physica A: Statistical Mechanics and its Applications}
  \textbf{\bibinfo{volume}{89}}, \bibinfo{pages}{163} (\bibinfo{year}{1977}).

\bibitem[{\citenamefont{Nambu and
  Jona-Lasinio}(1961{\natexlab{a}})}]{Nambu:1961tp}
\bibinfo{author}{\bibfnamefont{Y.}~\bibnamefont{Nambu}} \bibnamefont{and}
  \bibinfo{author}{\bibfnamefont{G.}~\bibnamefont{Jona-Lasinio}},
  \bibinfo{journal}{Phys. Rev.} \textbf{\bibinfo{volume}{122}},
  \bibinfo{pages}{345} (\bibinfo{year}{1961}{\natexlab{a}}).

\bibitem[{\citenamefont{Nambu and
  Jona-Lasinio}(1961{\natexlab{b}})}]{Nambu:1961fr}
\bibinfo{author}{\bibfnamefont{Y.}~\bibnamefont{Nambu}} \bibnamefont{and}
  \bibinfo{author}{\bibfnamefont{G.}~\bibnamefont{Jona-Lasinio}},
  \bibinfo{journal}{Phys. Rev.} \textbf{\bibinfo{volume}{124}},
  \bibinfo{pages}{246} (\bibinfo{year}{1961}{\natexlab{b}}).

\end{thebibliography}

\end{document}